\documentclass[preprint,12pt]{elsarticle}

\usepackage{graphicx}
\usepackage{subfigure}
\usepackage{float}
\usepackage{amsmath}
\usepackage{color}

\usepackage{amssymb}

\biboptions{sort&compress}

\begin{document}

\begin{frontmatter}



\title{Numerical assessments of ocean energy extraction from western boundary currents using a quasi-geostrophic ocean circulation model}


\author{Omer San}
\ead{osan@okstate.edu}
\address{School of Mechanical and Aerospace Engineering, \\ Oklahoma State University, Stillwater, OK, USA}

\begin{abstract}
A single-layer, quasi-geostrophic (QG), large-scale ocean circulation model is developed in this paper to study available ocean current energy potentials harnessed by using the ocean current turbines. Power extraction is modeled by adding a parameterized Rayleigh friction term in the barotropic vorticity equation. Numerical assessments are performed by simulating a set of mid-latitude ocean basins in the beta plane, which are standard prototypes of more realistic ocean dynamics considering inter-decadal variability in turbulent equilibrium. The third-order Runge-Kutta scheme for the temporal discretization and the second-order conservative Arakawa scheme for the spatial discretization are utilized to perform Munk scale resolving high-resolution computations. A sensitivity analysis with respect to the turbine parameters is performed for various physical conditions. Results show that the proposed model captures the quasi-stationary ocean dynamics and provides the four-gyre circulation patterns in time mean. After an initial spin-up process, the proposed model reaches a statistically steady state at an average maximum speed between 1.5 m/s and 2.5 m/s, which is close to the observed maximum zonal velocities in the western boundary currents. The probability density function of the available power over a long time period is computed for a wide range of parameters. Numerical results shows that 10 GW mean power can be extracted from the turbines distributed over a length scale of 100 km along the western boundaries. However, it is demonstrated that bigger turbine areas would alter the flow patterns and energetics due to excessive dissipation. An increase in the turbine area results in an increase in the available power ranging from 8 to 22 GW depending on the values of turbine modeling parameters. This first step in the numerical assessment of the proposed QG model shows that the present framework could represent a viable tool for evaluating energy potentials in a highly turbulent flow regime.
\end{abstract}


\end{frontmatter}


\section{Introduction}
\label{}
There is a growing interest in harvesting energy from the oceans. The huge energy potential of moving water in the forms of heat, strong currents, waves and tides has recently attracted a great deal of attention due to increasing demand in alternative renewable energy systems \cite{pelc2002renewable,vanek2008energy,khaligh2009energy,faizal2011ocean,esteban2012current,heath2012review,lopez2013review, wu2015ocean,halder2015high}. Although the biophysical and environmental impacts are under investigation \cite{copping2015environmental}, it is conjectured that the oceans could turn out to be an even more benign clean source of power than wind \cite{tollefson2014blue}. Among the other forms, as discussed by Yang et al. \cite{yang2015national}, strong ocean currents are rich in storing hydrokinetic energy since sea water is about 800 times denser than air. Since the power is proportional to the fluid density and the cube of the flow velocity, ocean currents of about 1/9 the speed of the wind have comparable kinetic power density with wind. It is anticipated that, with its gathering speed, more than 7 \% of the worldwide energy production will be harnessed from the oceans by 2050 \cite{esteban2012current}.

The main drivers of ocean circulation are the Earth's rotation and atmospheric winds. The ocean circulation is characterized by large circulation zones, or gyres, which can be identified with the strong, persistent, sub-tropical and sub-polar western boundary currents such as the Gulf Stream \citep{stommel1972gulf,kelly2010western}. Generally, the circulation is clockwise (CW) in the Northern Hemisphere and counter-clockwise (CCW) in the Southern Hemisphere. These circulation patterns emerge when we average over several years. One of the major similarities between the various ocean basins is the strong western boundary currents and weaker flow in the interior and eastern boundaries. Due to its reliability, persistency and stainability, the Gulf Stream ocean current system, carrying billions of gallons of water per minute, is of paramount interest as a potential clean energy resource for Florida and other coastal states.

The process of hydrokinetic energy conversion due to the underwater turbine systems implies utilization of kinetic energy contained in strong streams and currents for the generation of electricity \cite{khan2009hydrokinetic}. A variety of ocean current energy turbines (OCTs) are currently being proposed and have been tested \cite{tollefson2014blue}. Global energy potentials of ocean currents using turbines have also been identified by VanZwieten et al. \cite{vanzwieten2013global} discussing eight potential locations where ocean current energy could be potentially viable. One of early studies indicated that 10 GB power could be extracted from the Gulf Stream \cite{lissaman1979the,lissamen1979coriolis}. Von Arx et al. \cite{von1974florida} conservatively estimated that the distributed turbine arrays in Florida current can supply 1 GW power without seriously disturbing climate conditions. Interest has grown increasingly over the last few years. An assessment of available ocean current hydrokinetic energy near the North Carolina shore has been performed by Kabir et. al. \cite{kabir2015assessment}. Assessments of the Kurushio current were also discussed by Chen \cite{chen2010kuroshio}. In a recent work by Duerr and Dhanak \cite{duerr2010hydrokinetic, duerr2012assessment}, it is predicted that an amount of 20-25 GW power can be extracted from Gulf Stream system. Yang et. al. \cite{yang2013theoretical} provided a theoretical framework to assess available potential of energy from ocean circulations by using a simplified ocean circulation model, known as the Stommel model \cite{stommel1948westward, vallis2006atmospheric}. Their framework have also been utilized in a more realistic general ocean circulation model \cite{haas2014modeling}. As further explained in \cite{yang2014evaluating}, this approach utilizes a two-dimensional idealized ocean circulation model and represents the presence of turbines as linearized drag force in the form of Rayleigh friction and predicts an average amount of 5 GW available power from the Gulf Stream system corresponding an average of approximately 44 TWh/yr.

Following similar parametrization for representing localized turbines \cite{yang2014evaluating}, the objective of the present work is to study the potential available energy from the western boundary currents using a single layer (two-dimensional), wind-driven, double-gyre, mid-latitude, beta-plane quasi-geostrophic (QG) ocean circulation model. Capturing the inter-annual and inter-decade variability in large-scale ocean basins, this model utilizes the unsteady baratropic vorticity equation. The barotropic vorticity equation (BVE), also known as the single-layer QG model, is one of the most used mathematical models for forced-dissipative large scale ocean circulation problem (i.e., see \cite{bryan1963numerical,holland1975generation,holland1980example,jiang1995multiple,berloff1999large,greatbatch2000four,nadiga2001dispersive,majda2006nonlinear,mcwilliams2006fundamentals,miller2007numerical,san2013an}). Here, the author works in a regime in which the model reaches a state of turbulent equilibrium driven by a double-gyre wind forcing. The time mean circulation patterns are characterized by a four-gyre structure (i.e., southern outer, subtropical, subpolar, and northern outer gyres) when the barotropic vorticity equation is considered under the double-gyre wind forcing in a highly turbulent regime \cite{greatbatch2000four,san2011approximate,san2013coarse}. The two inner gyres circulate in the same directions as the wind stress curl, while two outer gyres at the northern and southern boundaries of the basin circulate in the opposite direction.

Studies of wind-driven circulation using a double-gyre wind forcing have played an important role in understanding various aspects of ocean dynamics, including the role of meso-scale eddies and their effect on the mean circulation, energy transfer, and seasonal and inter-annual oscillations \cite{shen1999wind}. When the barotropic vorticity equation is forced by a double-gyre wind stress and the explicit dissipation mechanism is weak, the instantaneous flow is highly turbulent showing a four-gyre structure in the time mean \cite{nadiga2001dispersive}. In this setting the explicit mechanism plays a minor role and the dominant balance is between the wind forcing and eddy flux of potential vorticity (e.g., see \citep{san2011approximate}). Therefore, the utilized flow setting represents an ideal framework to study the effects of additional turbines to circulation patterns in an unsteady regime. By means of a set of numerical simulations for various ocean basins considering inter-decade variability, this paper investigates the effects of added turbines in the dynamics of circulation zones and energetics of the oceans and computes probability density function of available power in each case. It will be investigated numerically whether the QG model can reproduce the four-gyre flow pattern in the time average when turbines are included by a functional friction term controlling the approximate area of turbine region. Although the present approach does not consider some important factors such as bathymetry, tides and actual wind patters, it extends the study of Stomel's model \citep{yang2014evaluating} to a single-layer QG model with the unsteady barotropic vorticity equation.


The remainder of this paper is organized as follows. Section 2 presents governing equations for the quasi-geostrophic ocean model utilized in our assessments. Section 3 contains a brief description of numerical methods including the second-order energy conserving Arakawa scheme for the nonlinear interaction, and the third-order Runge Kutta scheme for the time integration. The results of carefully selected numerical experiments are provided in Section 4 using a huge set of physical and numerical parameters. Section 5 is devoted to conclusions.

\section{Quasigeostrophic ocean model}
\label{}
Studies of wind-driven circulation using an idealized double-gyre wind forcing have played an important role in understanding various aspects of ocean dynamics, including the role of mesoscale eddies and their effect on mean circulation. In this section, following \cite{san2011approximate,san2013an} and references therein, we present the barotropic vorticity equation (BVE) for forced-dissipative large scale ocean circulation problem. It is also known as single-layer quasigeostrophic model and more details on the physical mechanism and various formulations can be found in geophysical fluid dynamics monographs \cite{gill1982atmosphere,vallis2006atmospheric,mcwilliams2006fundamentals,cushman2011introduction,pedlosky2013geophysical}.

The BVE for one-layer quasigeostrophic ocean model can be written as
\begin{equation}\label{eq:bve}
\frac{\partial \omega}{\partial t} + J(\omega,\psi) -\beta\frac{\partial \psi}{\partial x} = D + R + F,
\end{equation}
where $D$, $R$ and $F$ represent the dissipation, friction and forcing terms, respectively. The three terms in left hand side of Eq.~(\ref{eq:bve}) model local, convective, and rotational effects, respectively. In Eq.~(\ref{eq:bve}), $\omega$ is the kinematic vorticity, the curl of the velocity field, defined as
\begin{equation}\label{eq:vor}
\omega = \frac{\partial v}{\partial x} - \frac{\partial u}{\partial y},
\end{equation}
and $\psi$ is refers the velocity stream function. The kinematic relationship between the vorticity and stream function yields the following Poisson equation:
\begin{equation}\label{eq:poi}
\nabla^2 \psi = -\omega,
\end{equation}
where $\nabla^2$ is the two-dimensional Laplacian operator. The flow velocity components are defined by
\begin{equation}\label{eq:vel}
u = \frac{\partial \psi}{\partial y}, \quad v =- \frac{\partial \psi}{\partial x}.
\end{equation}

\begin{figure}[!t]
\centering
\includegraphics[width=0.8\textwidth]{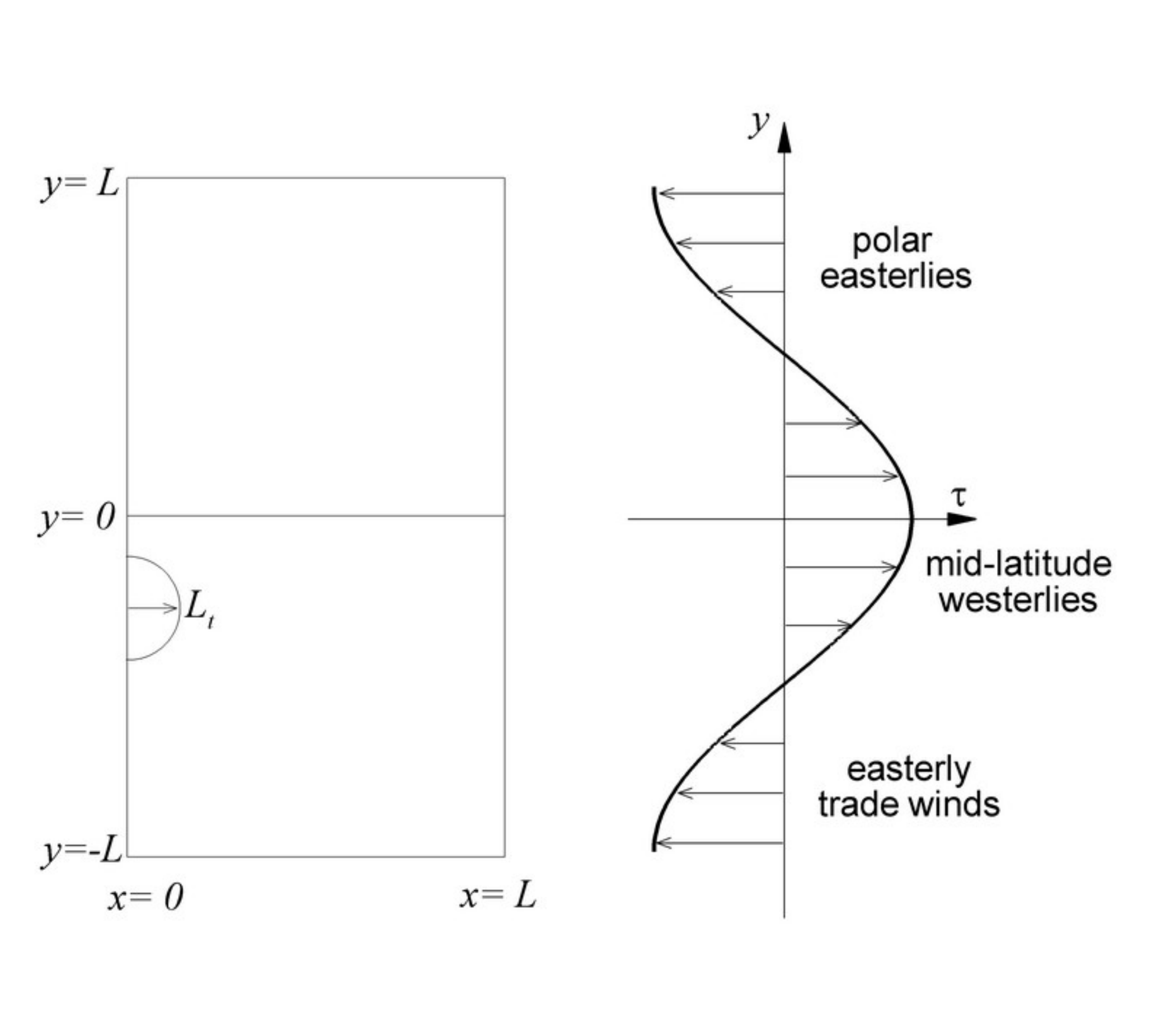}
\caption{Schematics of the ocean basin (left) and the double-gyre wind stress forcing (right). }
\label{f:basin}
\end{figure}

The BVE given by Eq.~(\ref{eq:bve}) uses the beta-plane approximation, which is valid for most of the mid-latitude simplified ocean basins. To account for the Earth's rotational effects, using Taylor series expansion, in the beta-plane approximation the Coriolis parameter is approximated by $f=f_0+\beta y$, where $f_0$ is the constant mean Coriolis parameter at the basin center and $\beta$ is the gradient of the Coriolis parameter (i.e., $\beta=\partial f/\partial y$) at the same location. The convection term in Eq.~(\ref{eq:bve}), called the nonlinear Jacobian, is defined as
\begin{equation}\label{eq:jac}
J(\omega,\psi) =  \frac{\partial \psi}{\partial y}\frac{\partial \omega}{\partial x} - \frac{\partial \psi}{\partial x}\frac{\partial \omega}{\partial y}.
\end{equation}
The viscous dissipation mechanism has the conventional Laplacian form
\begin{equation}\label{eq:vel}
D = \nu \nabla^2 \omega,
\end{equation}
where $\nu$ is the horizontal eddy viscosity coefficient for the ocean basin. The double-gyre wind forcing function in the model is given by
\begin{equation}\label{eq:forc}
F = \frac{1}{\rho H}  \hat{k} \cdot \nabla \times \vec{\tau},
\end{equation}
where $\rho$ is the mean fluid density, and $H$ is the mean depth of the ocean basin, and $\vec{\tau}$ refers the stress vector for the surface wind forcing, and $\hat{k}$ is unit vector in vertical direction. In the equations above, $\nabla$ and $\nabla^2$ are the gradient and Laplacian operators, respectively. In the present model, we use a double-gyre wind forcing only for zonal direction: $\vec{\tau} = (\tau_0\cos(\pi y/L), 0)$, where reference length $L$ is the width of the ocean basin centered at $y=0$, and $\tau_0$ is the maximum amplitude of the wind stress. As shown in Fig. \ref{f:basin}, this form of wind stress represents the meridional profile of easterly trade winds, mid-latitude westerlies, and polar easterlies from South to North over the ocean basin. Taking the curl to the stress field, the forcing term can be written as
\begin{equation}\label{eq:forcb}
F = \frac{\tau_0}{\rho H} \frac{\pi}{L} \sin\Big(\pi \frac{y}{L}\Big).
\end{equation}

\begin{figure}
\centering
\mbox{
\subfigure[$\epsilon= 0.05$]{\includegraphics[width=0.5\textwidth]{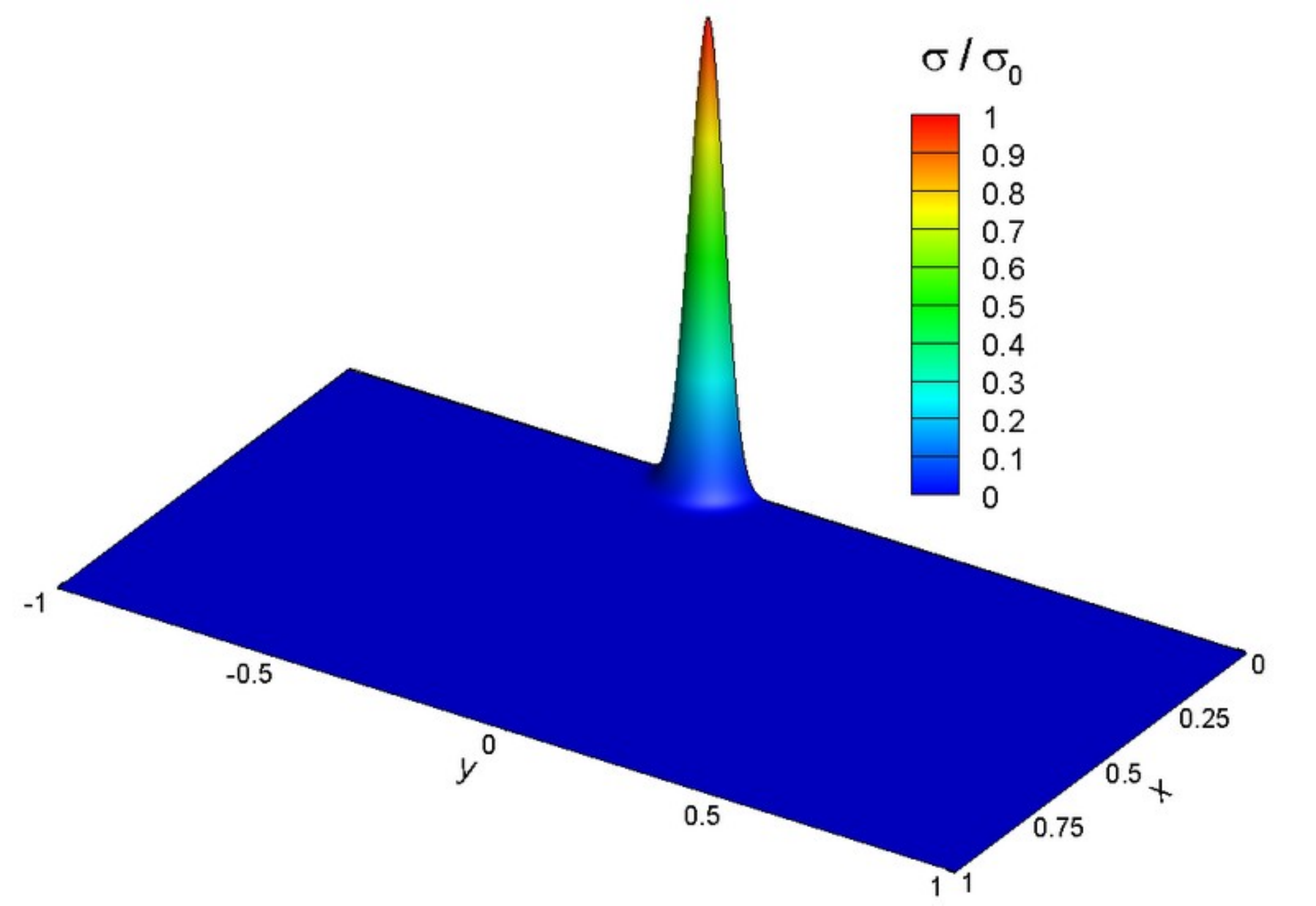}}
\subfigure[$\epsilon= 0.1$]{\includegraphics[width=0.5\textwidth]{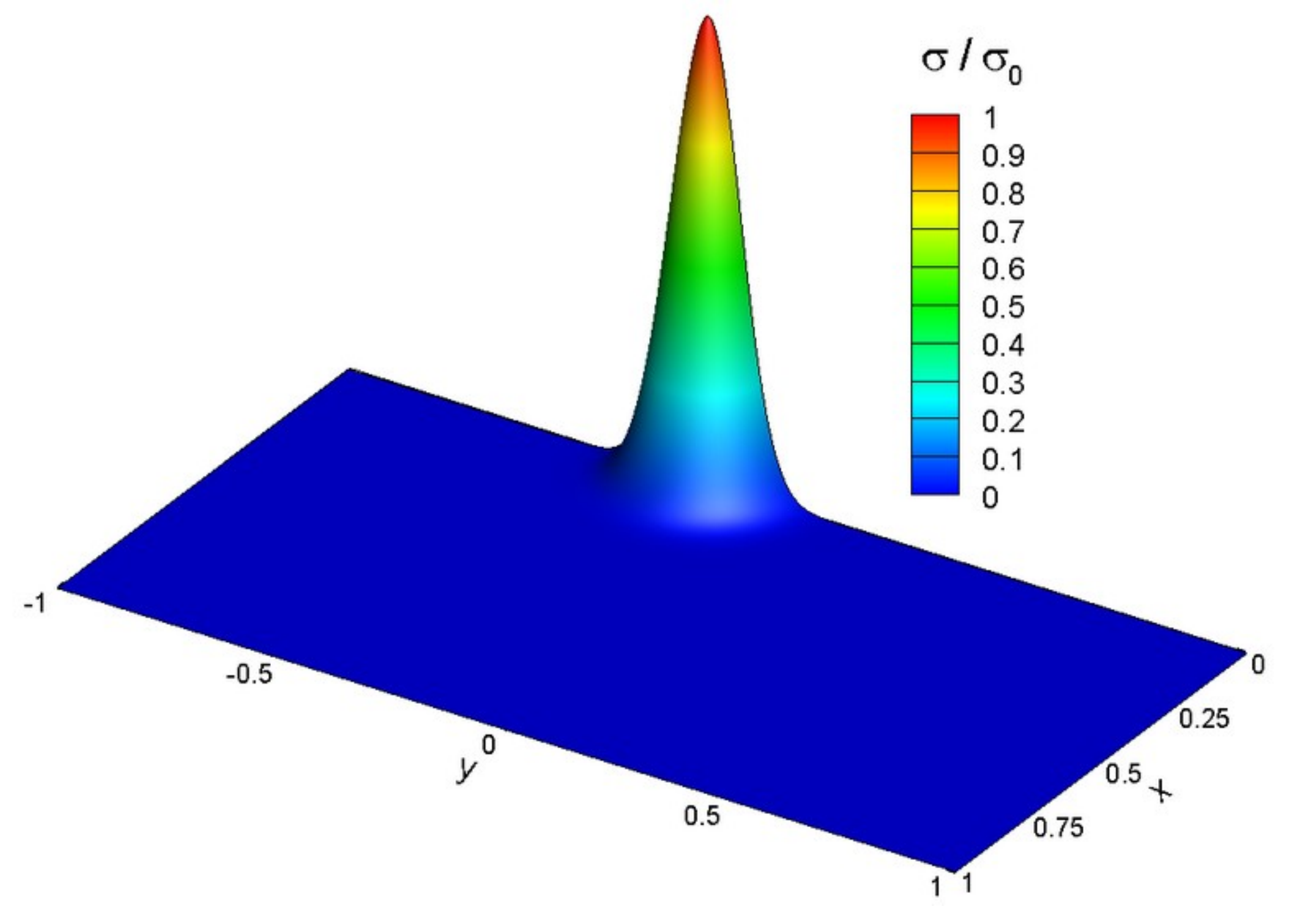}}
}\\
\mbox{
\subfigure[$\epsilon= 0.2$]{\includegraphics[width=0.5\textwidth]{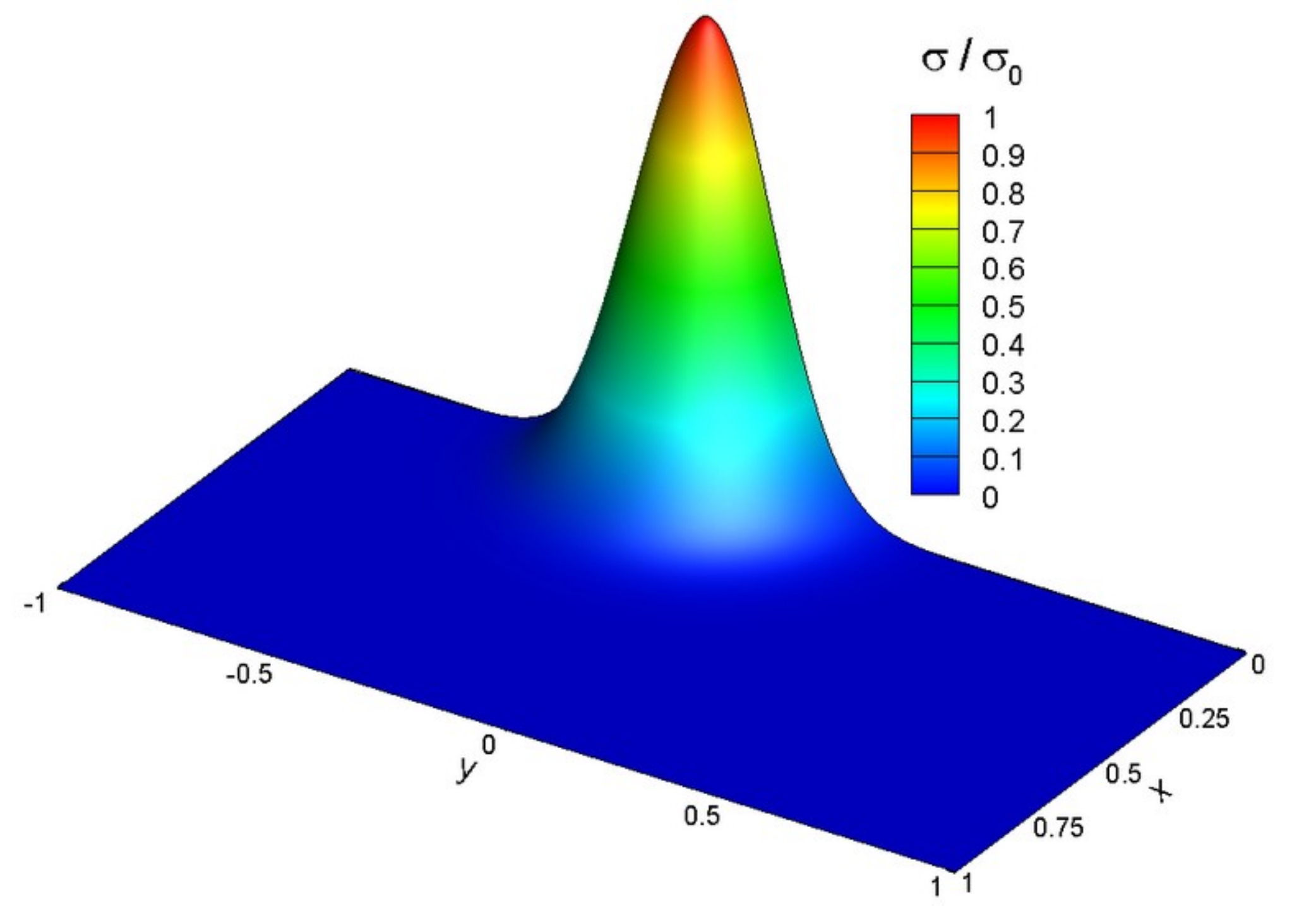}}
}
\caption{Parametrization of dimensionless turbine drag coefficient for various $\epsilon = L_t/L$, where $L_t$ is the length scale with effective turbine distribution, and $L$ is the length scale for the ocean basin.  }
\label{f:turbine}
\end{figure}

In four-gyre problem, the time average of the statistically steady equilibrium state exhibits a four-gyre circulation pattern, in contrast to the standard two-gyre structure associated symmetric double-gyre wind forcing. Finally, the friction term $R$ represents the drag force associated due to the turbines and can be parameterized in the following Rayleigh type linear friction form
\begin{equation}\label{eq:forc}
R = \frac{C_{t}}{H}\omega,
\end{equation}
where $C_{t}$ is the spatially varying drag coefficient profile representing the turbines. Following \cite{yang2014evaluating}, the turbine drag coefficient profile is specified as
\begin{equation}\label{eq:forc}
C_{t}(x,y) = C_{0} \exp(-\frac{(x-x_t)^2 + (y-y_t)^2}{L_{t}^{2}})
\end{equation}
where $C_{0}$ is the peak value of Gaussian turbine drag coefficient centered at $(x_t, y_t)$, and $L_t$ is the length scale parameter controlling the approximate area of the turbine region.  In order to obtain a dimensionless form of the BVE,  we use the following definitions:
\begin{equation}\label{eq:dimless}
\tilde{x} = \frac{x}{L}, \quad \tilde{y} = \frac{y}{L}, \quad \tilde{t} = \frac{t}{L/V}, \quad \tilde{\omega} = \frac{\omega}{V/L}, \quad \tilde{\psi} = \frac{\psi}{V L},
\end{equation}
where the tilde denotes the corresponding nondimensional variables. In the nondimensionalization, $L$ represents the characteristic horizontal length scale (in our study $L$ is the basin dimension in the $x$ direction), and $V$ is the characteristic Sverdrup velocity scale given by the following definition:
\begin{equation}\label{eq:sverdrup}
V = \frac{\tau_0}{\rho H}\frac{\pi}{\beta L}.
\end{equation}
The dimensionless governing equations for the two-dimensional single-layer quasigeostrophic ocean model can be written as
\begin{equation}
\frac{\partial \omega}{\partial t} + J(\omega,\psi) -\frac{1}{Ro}\frac{\partial \psi}{\partial x} = \frac{1}{Re}\nabla^2 \omega + \frac{1}{Ro}\sin(\pi y) + \sigma \omega,
\label{eq:nbve}
\end{equation}
in which,
\begin{equation}
\sigma = \sigma_{0}\exp(-\frac{(x-x_t)^2 + (y-y_t)^2}{\epsilon^{2}})
\label{eq:nbve-3}
\end{equation}
where we omit the tilde over the variables for clarity purposes. Due to the nondimensionalization given by Eq.~(\ref{eq:dimless}), the kinematic relationship given in Eq.~(\ref{eq:poi}) and definition of nonlinear Jacobian given in Eq.~(\ref{eq:jac}) remain the same. In the dimensionless form given in Eq.~(\ref{eq:nbve}), Reynolds and Rossby numbers control quasigeostrophic ocean dynamics, which are related to the physical parameters in the following way:
\begin{equation}\label{eq:ReRo}
Re = \frac{V L}{\nu}, \quad Ro = \frac{V}{\beta L^2};
\end{equation}
and the nondimensional paramaters $\sigma_{0}$ and $\epsilon$ control added turbine effects, which are defined as:
\begin{equation}\label{eq:epsi}
\quad \sigma_0 = \frac{C_{0} L}{V H}, \quad \epsilon = \frac{L_t}{L}.
\end{equation}
As shown in Fig. \ref{f:turbine}, turbine drag coefficient profile is designed such that $C_{t}$ peaks in the western boundary at $(x_t,y_t) = (0, -0.2)$ in order to realistically represent the energy extraction from the Gulf Stream system. It should be noted that large $\epsilon$ values correspond to larger turbine areas in the ocean basin.

In order to completely specify the mathematical model, boundary and initial conditions need to be prescribed. In many theoretical studies of large scale ocean circulation models, slip or no-slip boundary conditions are used in simplified Cartesian oceanic basins. Following \cite{cummins1992inertial,ozgokmen1998emergence,greatbatch2000four,holm2003modeling,san2011approximate,san2013approximate,san2013an},
we use slip boundary conditions for the velocity, which translate into homogeneous Dirichlet boundary conditions for the vorticity: $\omega|_{\Gamma} = 0$, where $\Gamma$ symbolizes all the Cartesian boundaries. The corresponding impermeability boundary condition is imposed as $\psi|_{\Gamma} = 0$. For the initial condition, we start our computations from a quiescent state (i.e., $\omega = 0$, and $\psi=0$) and integrate Eq.~(\ref{eq:nbve}) until we obtain a statistically steady state in which the wind forcing, dissipation, friction, and nonlinear Jacobian balance each other.

\section{Numerical methods}
\label{}
In many physically relevant ocean circulation models, such as the QG models, the solutions do not converge to a steady state as time goes to infinity \cite{medjo2000numerical}. Rather they remain time dependent by producing statistically steady state. One of the main purpose of the present study is to investigate available energy potentials from the western boundary currents considering this quasi-stationary flow regime. Therefore, numerical schemes designed for numerical integration of such phenomena should be suited for such behavior of the solutions and for the long-time integration. In this section, we provide a brief description of the numerical methods employed in this study.

Semi-discrete ordinary differential equations are obtained after a spatial discretization of the partial differential equations \cite{moin2010fundamentals}. To implement the Runge-Kutta scheme for the time integration, we cast the governing equation given by Eq.~(\ref{eq:nbve}) in the following form
\begin{equation}
\frac{d\omega_{i,j}}{dt} = \pounds_{i,j} ,
\label{eq:rk}
\end{equation}
where subscripts $i$ and $j$ represent the discrete spatial indices in $x-$ and $y-$directions, respectively. Here, $\pounds_{i,j}$ denotes the discrete spatial derivative operators, including the convective nonlinear Jacobian, $\beta-$plane approximation of the Coriolis force, the linear Laplacian diffusive term, the double-gyre wind forcing stress term, and the Rayleigh friction term due to additional turbines. We assume that the numerical approximation for time level $n$ is known, and we seek the numerical approximation for time level $n+1$, after the time step $\Delta t$. The optimal third-order accurate total variation diminishing Runge-Kutta (TVDRK3) scheme is then given as \cite{gottlieb1998total}
\begin{eqnarray}
\omega_{i,j}^{(1)} &=& \omega_{i,j}^{(n)} + \Delta t \pounds_{i,j}^{(n)}, \nonumber \\
\omega_{i,j}^{(2)} &=& \frac{3}{4} \omega_{i,j}^{(n)} + \frac{1}{4} \omega_{i,j}^{(1)} + \frac{1}{4}\Delta t \pounds_{i,j}^{(1)}, \nonumber \\
\omega_{i,j}^{(n+1)} &=& \frac{1}{3} \omega_{i,j}^{(n)} + \frac{2}{3} \omega_{i,j}^{(2)} + \frac{2}{3}\Delta t \pounds_{i,j}^{(2)}.
\label{eq:TVDRK}
\end{eqnarray}
where $\Delta t$ is the adaptive time step, which can be computed at the end of each time step by specifying the Courant-Friedrichs-Lewy (CFL) number satisfying the numerical stability criteria (e.g., $CFL \leq 1$ for the TVDRK3 scheme \cite{san2013an}). The CFL number is set to 0.8 in the present study by ensuring numerical stability. The source term, $\pounds_{i,j}$, is written as
\begin{equation}
\pounds_{i,j} = - J(\omega_{i,j},\psi_{i,j}) + \frac{1}{Ro}\frac{\partial \psi_{i,j}}{\partial x} + \frac{1}{Re}\nabla^2 \omega_{i,j} + \frac{1}{Ro}\sin(\pi y) + \sigma \omega_{i,j} ,
\label{eq:ssww}
\end{equation}
where we use standard second-order central finite difference schemes in linear terms. Therefore, the derivative operators in the third and fourth terms of Eq.~(\ref{eq:ssww}) can be written as:
\begin{equation}
\frac{\partial \psi_{i,j}}{\partial x} = \frac{\psi_{i+1,j}-\psi_{i-1,j}}{2\Delta x} ,
\label{eq:dfd1}
\end{equation}
\begin{equation}
\nabla^2 \omega_{i,j} = \frac{\omega_{i+1,j}-2\omega_{i,j}+\omega_{i-1,j}}{\Delta x^2} + \frac{\omega_{i,j+1}-2\omega_{i,j}+\omega_{i,j-1}}{\Delta y^2} ,
\label{eq:dfd2}
\end{equation}
where $\Delta x$ and $\Delta y$ are the mesh sizes in $x-$ and $y-$directions, respectively.

Modeling nonlinear term, Arakawa \cite{arakawa1966computational} suggested that the conservation of energy, enstrophy, and skew-symmetry is sufficient to avoid computational instabilities stemming from nonlinear interactions. The following second-order Arakawa scheme for the Jacobian is written
\begin{equation}
J(\omega_{i,j},\psi_{i,j}) =\frac{1}{3}\bigl(J_{1}+J_{2}+J_{3}\bigr) ,
\label{eq:ja1}
\end{equation}
where the discrete Jacobians have the following forms:

\begin{eqnarray}
J_{1}&=& \frac{1}{4\Delta x \Delta y}\Big[(\omega_{i+1,j}-\omega_{i-1,j})(\psi_{i,j+1}-\psi_{i,j-1}) \nonumber \\
&-&(\omega_{i,j+1}-\omega_{i,j-1})(\psi_{i+1,j}-\psi_{i-1,j})\Big] ,
\label{eq:j1} \\
J_{2} &=& \frac{1}{4\Delta x \Delta y}\Big[\omega_{i+1,j}(\psi_{i+1,j+1}-\psi_{i+1,j-1}) -\omega_{i-1,j}(\psi_{i-1,j+1}-\psi_{i-1,j-1}) \nonumber \\
&-&\omega_{i,j+1}(\psi_{i+1,j+1}-\psi_{i-1,j+1}) +\omega_{i,j-1}(\psi_{i+1,j-1}-\psi_{i-1,j-1}) \Big] ,
\label{eq:j2} \\
J_{3} &=& \frac{1}{4\Delta x \Delta y}\Big[\omega_{i+1,j+1}(\psi_{i,j+1}-\psi_{i+1,j}) -\omega_{i-1,j-1}(\psi_{i-1,j}-\psi_{i,j-1}) \nonumber \\
&-&\omega_{i-1,j+1}(\psi_{i,j+1}-\psi_{i-1,j}) +\omega_{i+1,j-1}(\psi_{i+1,j}-\psi_{i,j-1}) \Big] .
\label{eq:j3}
\end{eqnarray}
Note that $J_{1}$, which corresponds to the central second-order difference scheme, is not sufficient for the conservation of energy, enstrophy, and skew-symmetry by the numerical discretization.
Arakawa \cite{arakawa1966computational} showed that the judicious combination of $J_{1}, J_{2}$, and $J_{3}$ in Eq.~\eqref{eq:ja1} achieves the above discrete conservation properties.

Most of the demand on computing resources posed by QG models comes in the solution of the elliptic Poisson equation \citep{miller2007numerical}. This is also true for our study to find stream function values from updated vorticity values at each substep in time integration. However, taking advantage of the simple Cartesian domain, an efficient fast Fourier transform (FFT) method is utilized for solving the kinematic relationship given in Eq.~(\ref{eq:poi}). Specifically, the discrete form of Eq.~(\ref{eq:poi}) is given by
\begin{equation}
\frac{\psi_{i+1,j}-2\psi_{i,j}+\psi_{i-1,j}}{\Delta x^2} + \frac{\psi_{i,j+1}-2\psi_{i,j}+\psi_{i,j-1}}{\Delta y^2} = -\omega_{i,j} ,
\label{eq:dfd2}
\end{equation}
and boundary conditions suggest the use of a fast sine transform.
The procedure to solve Eq.~(\ref{eq:dfd2}) involves three steps. First, an inverse sine transform for the source term is given by:
\begin{equation}
\hat{\omega}_{k,l}=\frac{2}{N_x}\frac{2}{N_y} \sum_{i=1}^{N_x-1} \sum_{j=1}^{N_y-1} \omega_{i,j} \sin \left(\frac{\pi k i}{N_x}\right) \sin \left(\frac{\pi l j}{N_y}\right) ,
\label{eq:ifft-1b}
\end{equation}
where $N_x$ and $N_y$ are the total number of grid points in $x$ and $y$ directions. Here the symbol hat is used to represent the corresponding Fourier coefficient of the physical grid data with a subscript pair $i,j$, where $i=0,1, ... N_x$ and $j=0,1, ... N_y$. As a second step, we directly solve Eq.~(\ref{eq:dfd2}) in Fourier space:
\begin{equation}
\hat{\psi}_{k,l}= -\frac{\hat{\omega}_{k,l}}{\frac{2}{\Delta x^2}\left(\cos(\frac{\pi k }{N_x}) -1\right)+\frac{2}{\Delta y^2}\left(\cos(\frac{\pi l }{N_y})-1\right) },
\label{eq:isub-1}
\end{equation}
Finally, the stream function values are found by performing a forward sine transform:
\begin{equation}
\psi_{i,j}= \sum_{k=1}^{N_x-1} \sum_{l=1}^{N_y-1} \hat{\psi}_{k,l} \sin \left(\frac{\pi k i}{N_x}\right) \sin \left(\frac{\pi l j}{N_y}\right) .
\label{eq:ffft-1}
\end{equation}
The computational cost of this elliptic solver is $\displaystyle \mathcal{O}\left(N_x \, N_y \, \log(N_x) \, \log(N_y) \right)$. The FFT algorithm given by \cite{press1992numerical} is used for forward and inverse sine transforms.

\begin{table}[!t]
\centering
\caption{Physical parameter sets used in the numerical experiments representing mid-latitude ocean basins. }
\label{tab:sets1}       
\begin{tabular}{llll}
\noalign{\smallskip}\hline\noalign{\smallskip}
Variable (unit) & Basin I & Basin II & Basin III \\
\noalign{\smallskip}\hline\noalign{\smallskip}
QG ocean modeling parameters \\
  $L$ (km)   & 2000 & 1600 & 3000 \\
  $H$ (km)   & 1 & 0.8 & 1 \\
  $\nu$ (m$^{2}$s$^{-1}$)    & 100    & 100 & 200  \\
  $\beta$ (m$^{-1}$s$^{-1}$) & $1.62\times10^{-11}$ & $1.62\times10^{-11}$ & $1.62\times10^{-11}$ \\
  $\tau_0$ (N m$^{-2}$)    & 0.2    & 0.2 & 0.2  \\
  $\rho$ (km m$^{-3}$)    & 1030    & 1030 & 1030  \\
  Ro   & $2.9\times10^{-4}$   & $7.1\times10^{-4}$ & $8.6\times10^{-4}$ \\
  Re   & 376.5  & 470.7 & 188.3 \\
Turbine modeling parameters (base model) \\
$C_0$ (m s$^{-1}$)    & 0.001    & 0.001 & 0.001  \\
$L_t$ (km)    & 100    & 100 & 150  \\
$\sigma_0$     & 106.2    & 68 & 239  \\
$\epsilon$     & 0.05    & 0.0625 & 0.05  \\
\noalign{\smallskip}\hline
\end{tabular}
\end{table}

\begin{figure}
\centering
\mbox{
\subfigure[simulation window]{\includegraphics[width=0.5\textwidth]{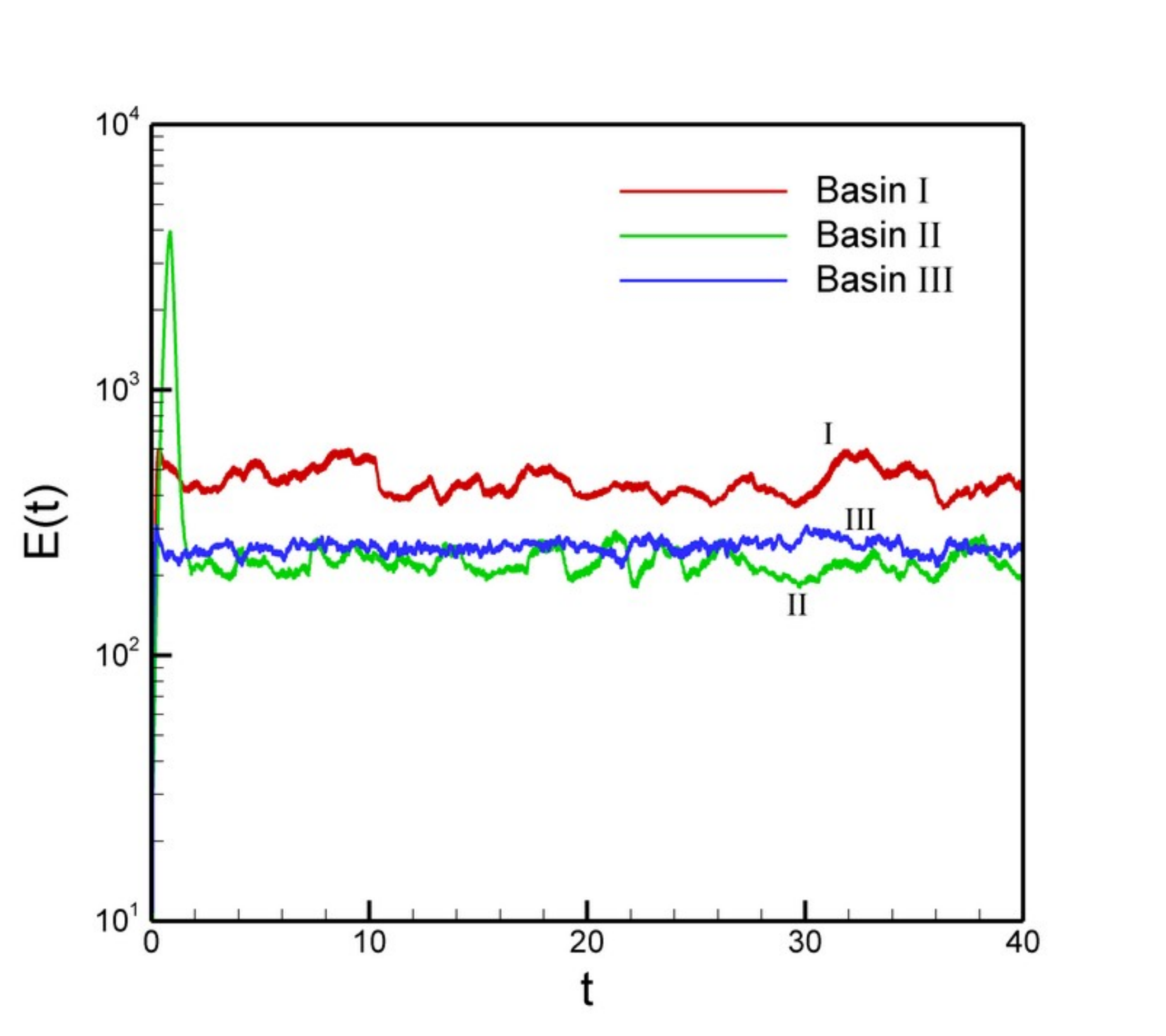}}
\subfigure[data collection window]{\includegraphics[width=0.5\textwidth]{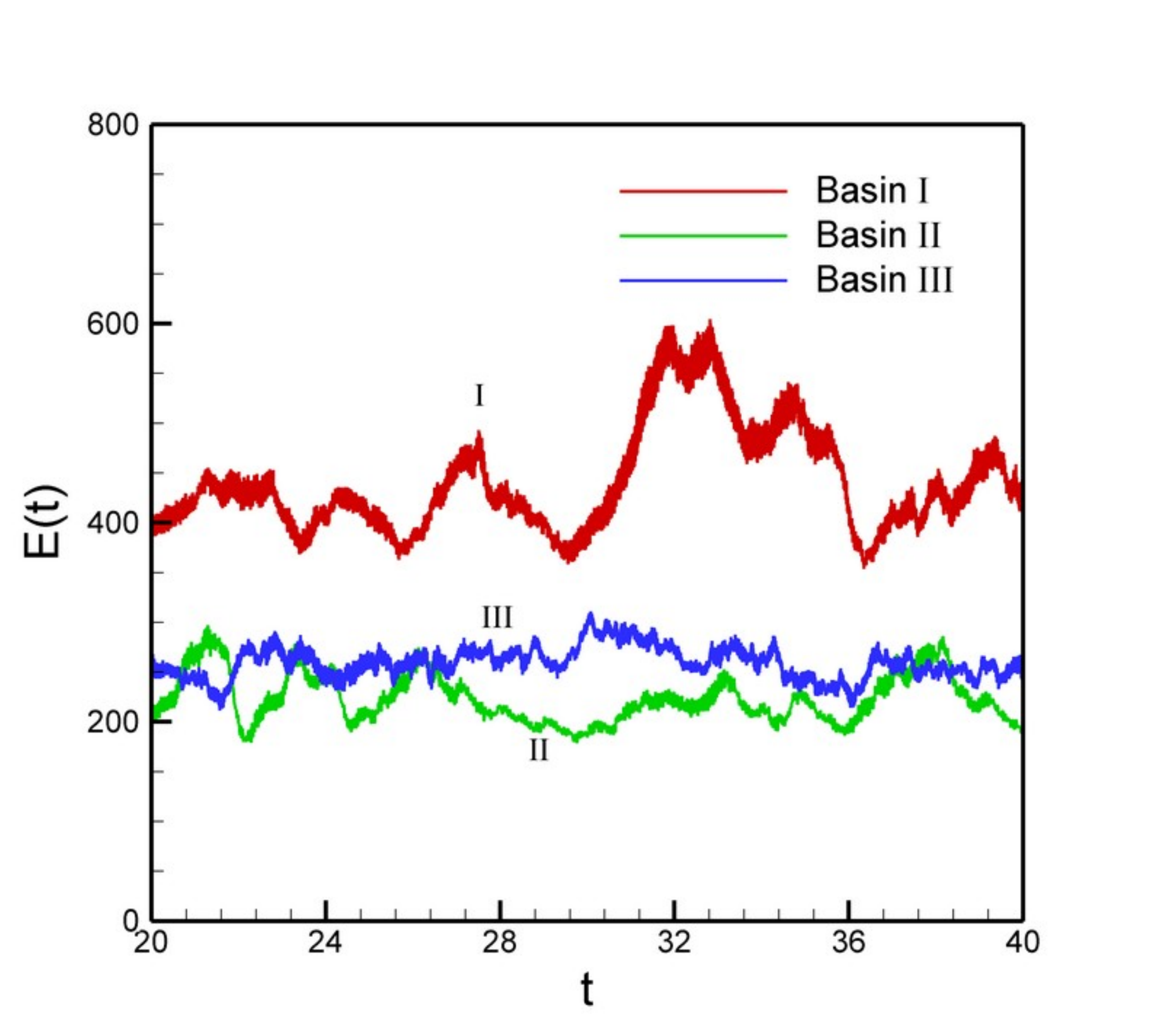}}
}\\
\mbox{
\subfigure[dimensionally scaled window for 5 years]{\includegraphics[width=0.5\textwidth]{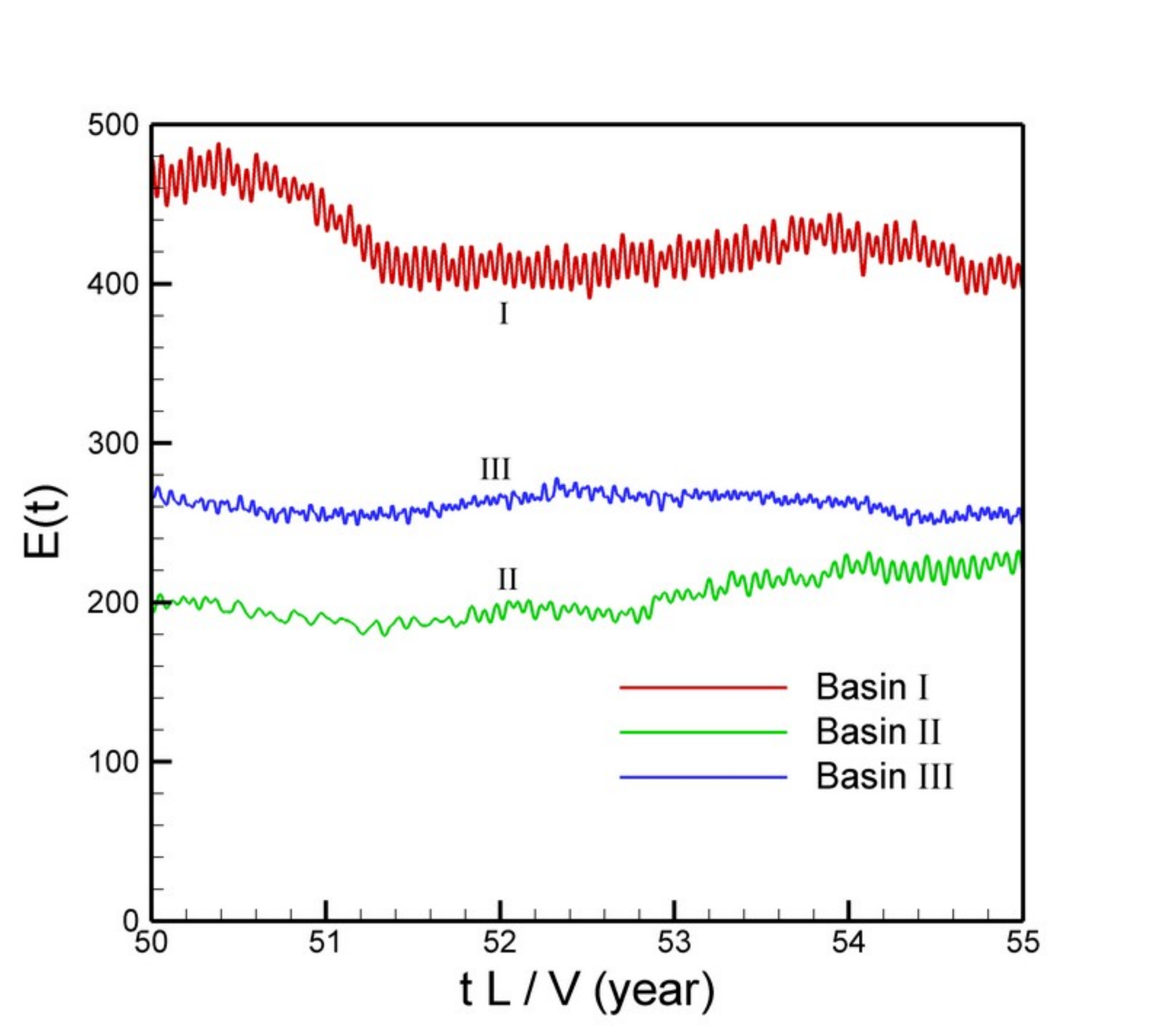}}
}
\caption{Time histories of the basin integrated total energy showing reference (undisturbed) quasistationary flow dynamics for three representative ocean basins.}
\label{f:energy}
\end{figure}

\section{Results}
\label{}
In this section, the results of QG ocean model described in Section 2 will be analyzed considering three representative mid-latitude ocean basins given by Table~\ref{tab:sets1}. Following \cite{greatbatch2000four,san2011approximate,san2015stabilized}, these physical parameters correspond to a volume transport of approximately 37.65 Sv (e.g., see Eq.~(\ref{eq:sverdrup})) demonstrating the robustness of the four-gyre structure with respect to the three different basin length scales with different Reynolds and Rossby numbers. After an initial transient period, the system with these parameters reaches a statistically steady state at a maximum current speed between 1.5 m/s and 2.5 m/s, which is agree well with the observed maximum zonal velocities (e.g., see \cite{dijkstra2005nonlinear} for 2 m/s maximum velocity observed at 68$^{\circ}$W). Reference turbine modeling parameters are gathered from Yang et. al. \cite{yang2013theoretical,yang2014evaluating}. In each ocean basin, a detailed sensitivity analysis with respect to turbine modeling parameters will be also presented to evaluate the characteristics of ocean energy potential for western boundary currents. The QC ocean model employs the barotropic vorticity equation driven by a symmetric double-gyre wind forcing given by Eq.~(\ref{eq:forcb}), which can yield four-gyre mean circulation in the time mean, depending on the physical modeling parameters corresponding to the ocean basin. The proposed model reaches a state of turbulent equilibrium such that the eddy flux of potential vorticity dominates the explicit dissipation and balances the vorticity from wind stress forcing. This test problem in the highly turbulent regime has been used in numerous studies (e.g., see \cite{cummins1992inertial,scott1998small,greatbatch2000four,nadiga2001dispersive,holm2003modeling,san2011approximate,san2013an,san2015stabilized}) and represents an ideal test bed for the numerical assessment of ocean energy potential of Gulf Stream system.
Starting from quiescent state, all numerical experiments conducted here are solved for a maximum dimensionless time of $T_{max}=40$. This value corresponds to the dimensional times of $134.7$, $68.9$ and $303.1$ years for Basin I, Basin II and Basin III, respectively, which are long enough to capture statistically steady states. All numerical computations are performed by using a resolution of $256 \times 512$ grid points, which is enough to perform Munk layer resolving simulations in our computations (e.g., please see \cite{san2013an} for a detailed sensitivity analysis with respect to the grid resolution).

\begin{figure}
\centering
\mbox{
\subfigure[Basin I]{\includegraphics[width=0.33\textwidth]{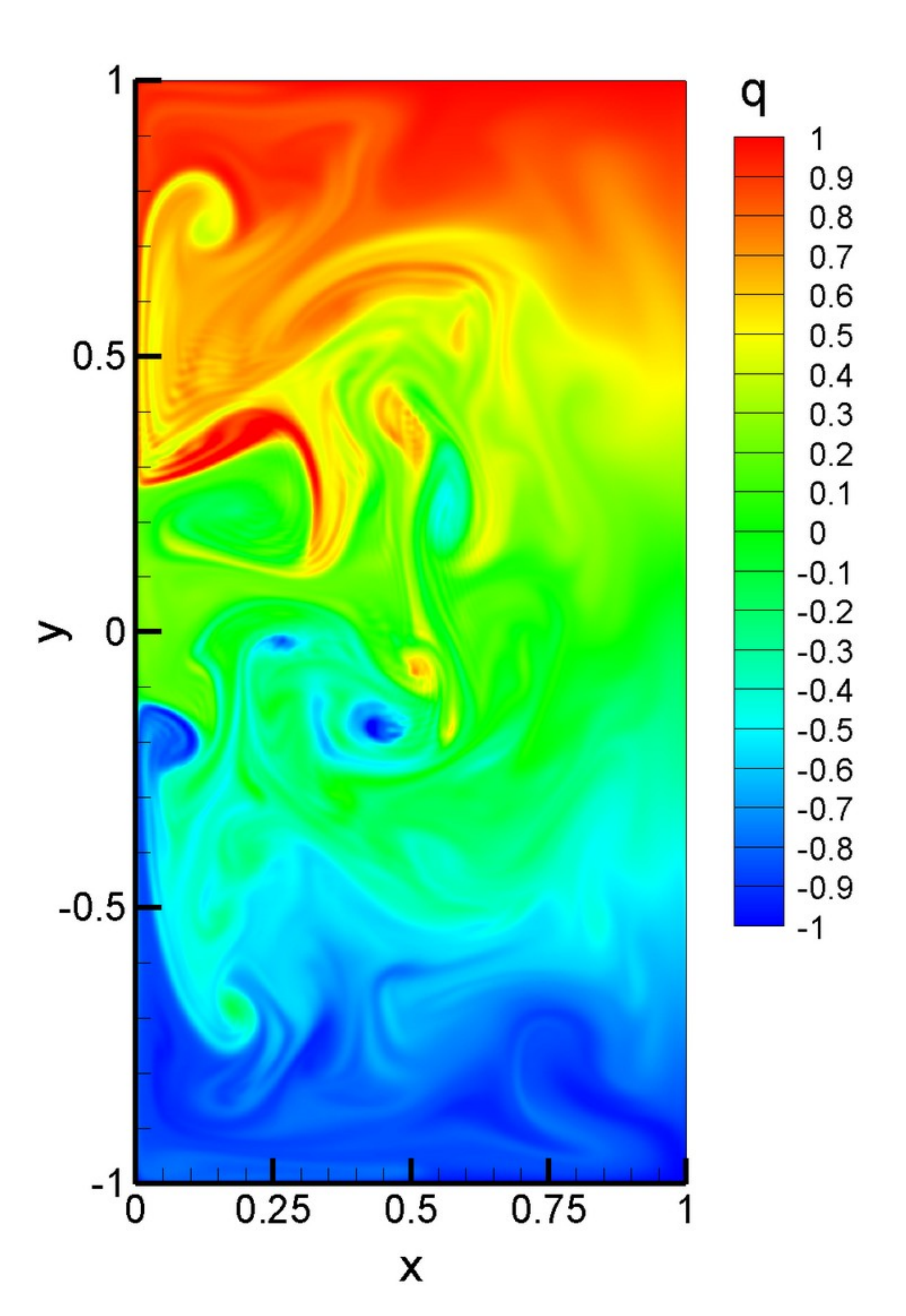}}
\subfigure[Basin II]{\includegraphics[width=0.33\textwidth]{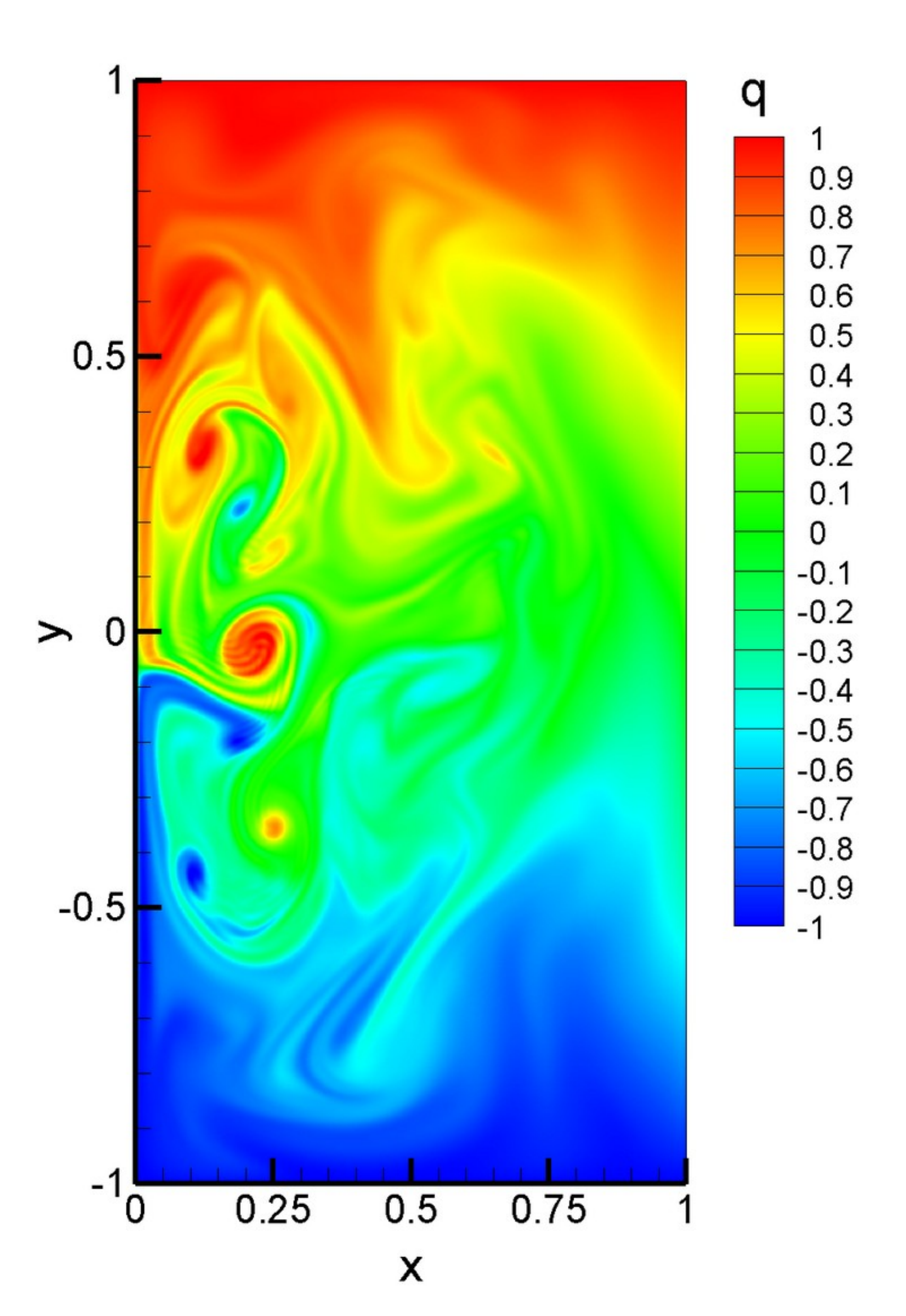}}
\subfigure[Basin III]{\includegraphics[width=0.33\textwidth]{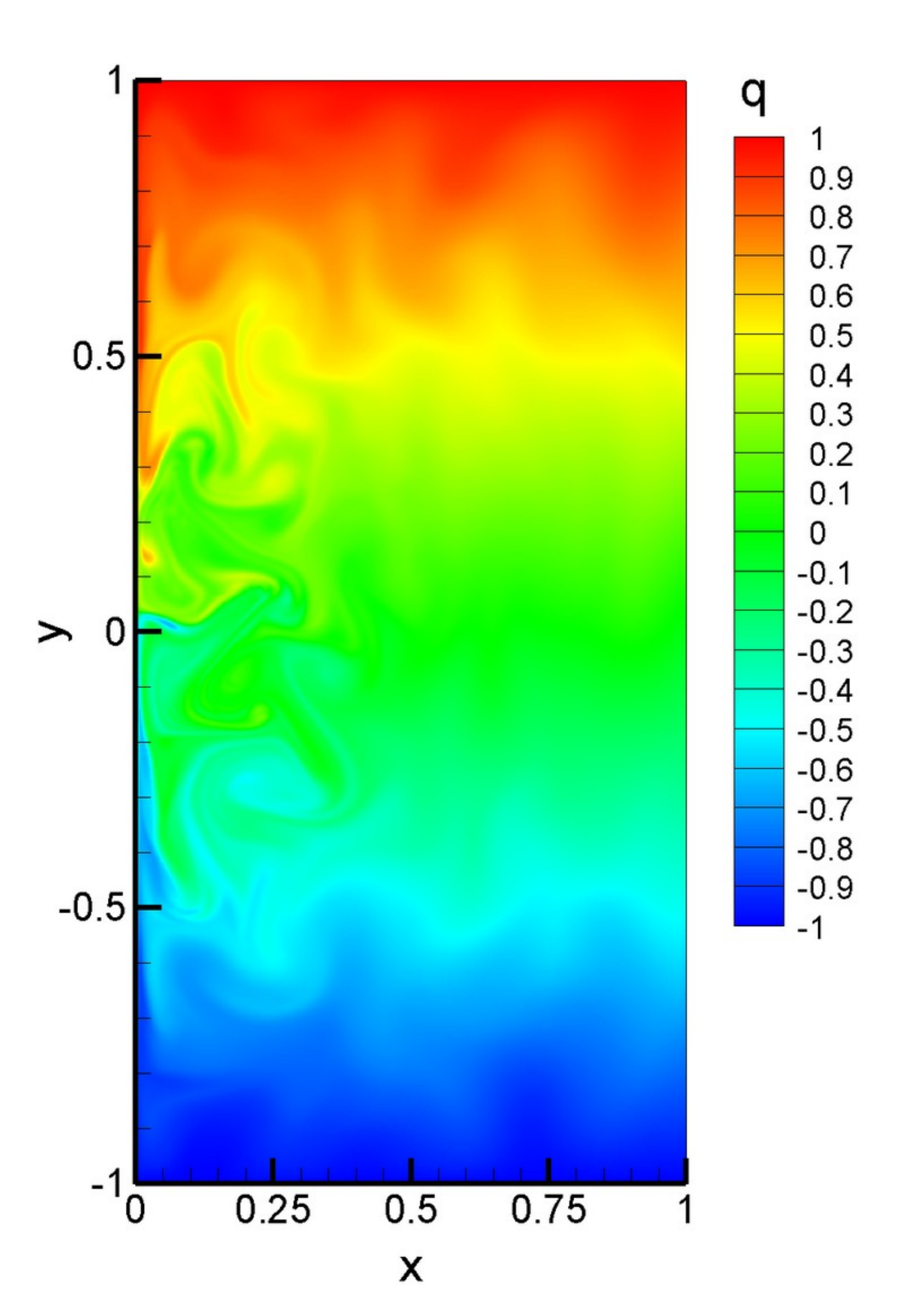}}
}
\caption{Instantaneous potential vorticity fields for the undisturbed ocean basins.}
\label{f:ins}
\end{figure}

In Fig.~\ref{f:energy}, we plot the time evolution of the basin integrated total kinetic energy given by,
\begin{equation}\label{eq:hist}
    E(t) = \frac{1}{2}\int\int \Big(\frac{\partial \psi}{\partial x}\Big)^2 + \Big(\frac{\partial \psi}{\partial y}\Big)^2  dxdy,
\end{equation}
for each ocean basin when there is no turbine (i.e., $\sigma = 0$). As shown in this figure, after an initial transition period, a quasi-stationary regime is achieved in each experiment. Statistically steady state data sets are collected between $t=20$ and $t=40$ for all the assessments presented in this study. The inter-annual and inter-decade variability of the ocean dynamics can also be seen from the figure. For undisturbed ocean basins, instantaneous flow fields at time $t=40$ are shown in Fig.~\ref{f:ins} for each ocean basin illustrating the potential vorticity field, defined as $q = \omega + Ro~y$ in QG models.

\begin{figure}
\centering
\mbox{
\subfigure[undisturbed]{\includegraphics[width=0.25\textwidth]{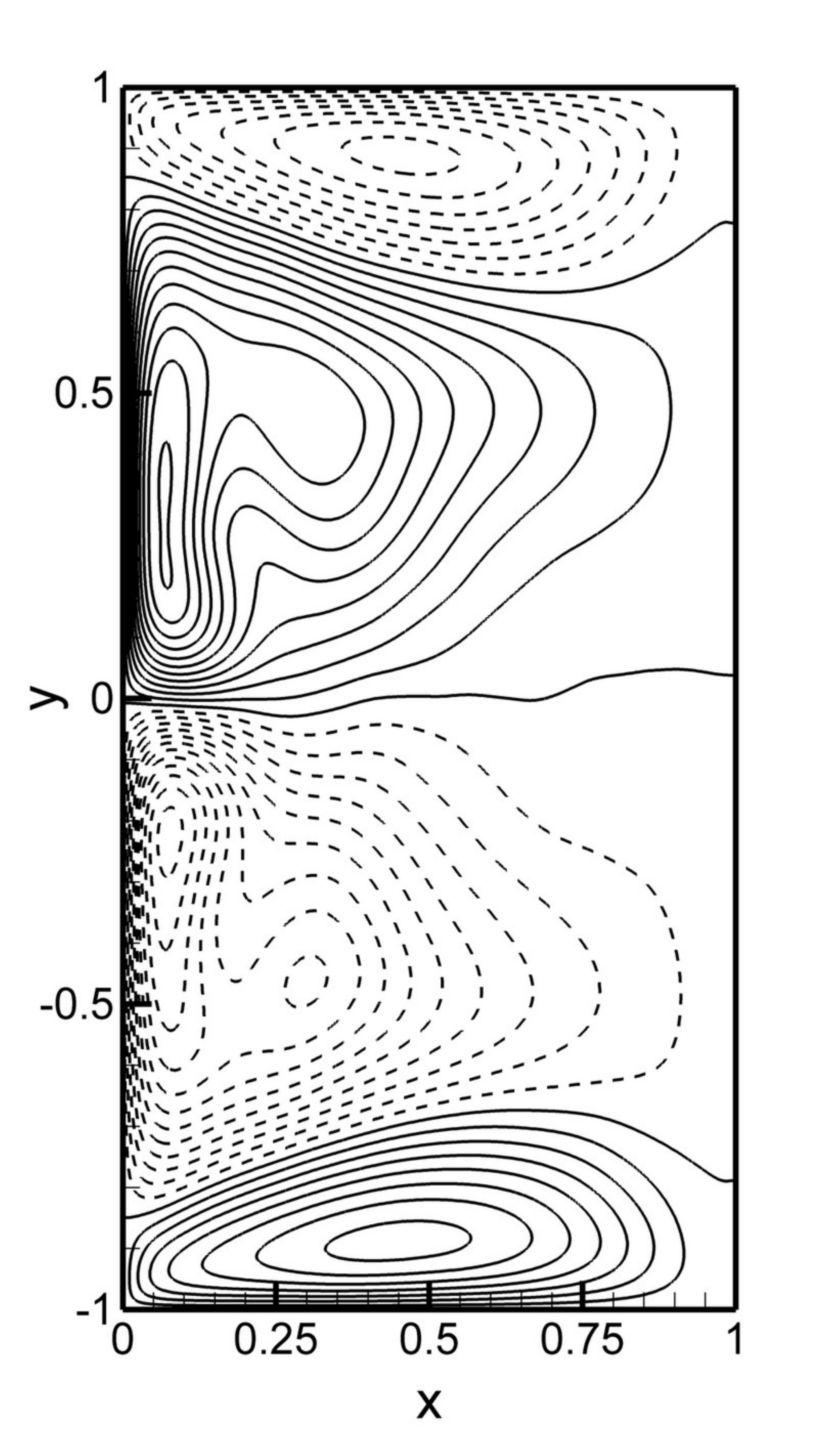}}
\subfigure[$\epsilon = 0.05$]{\includegraphics[width=0.25\textwidth]{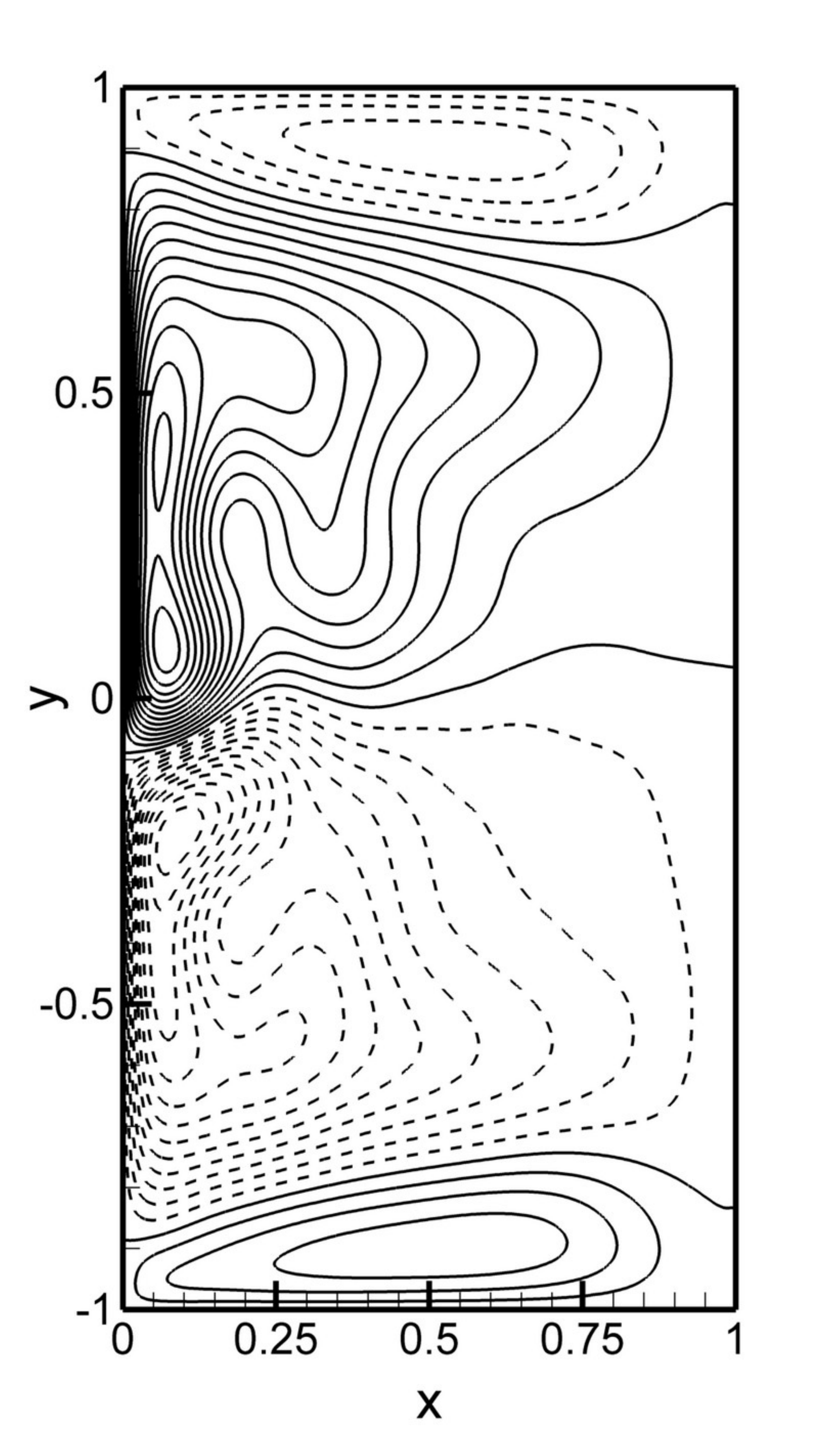}}
\subfigure[$\epsilon = 0.1$]{\includegraphics[width=0.25\textwidth]{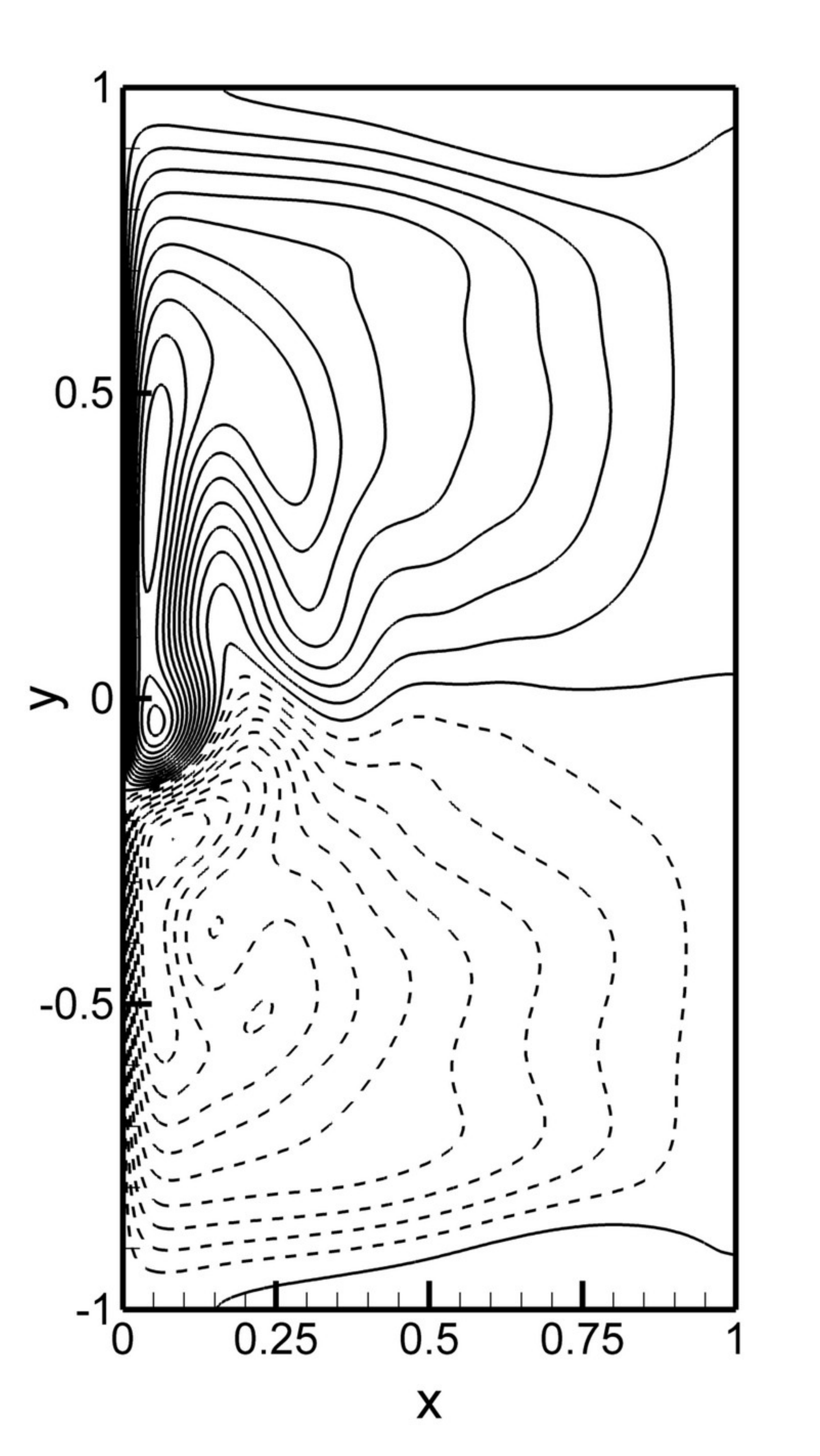}}
\subfigure[$\epsilon = 0.2$]{\includegraphics[width=0.25\textwidth]{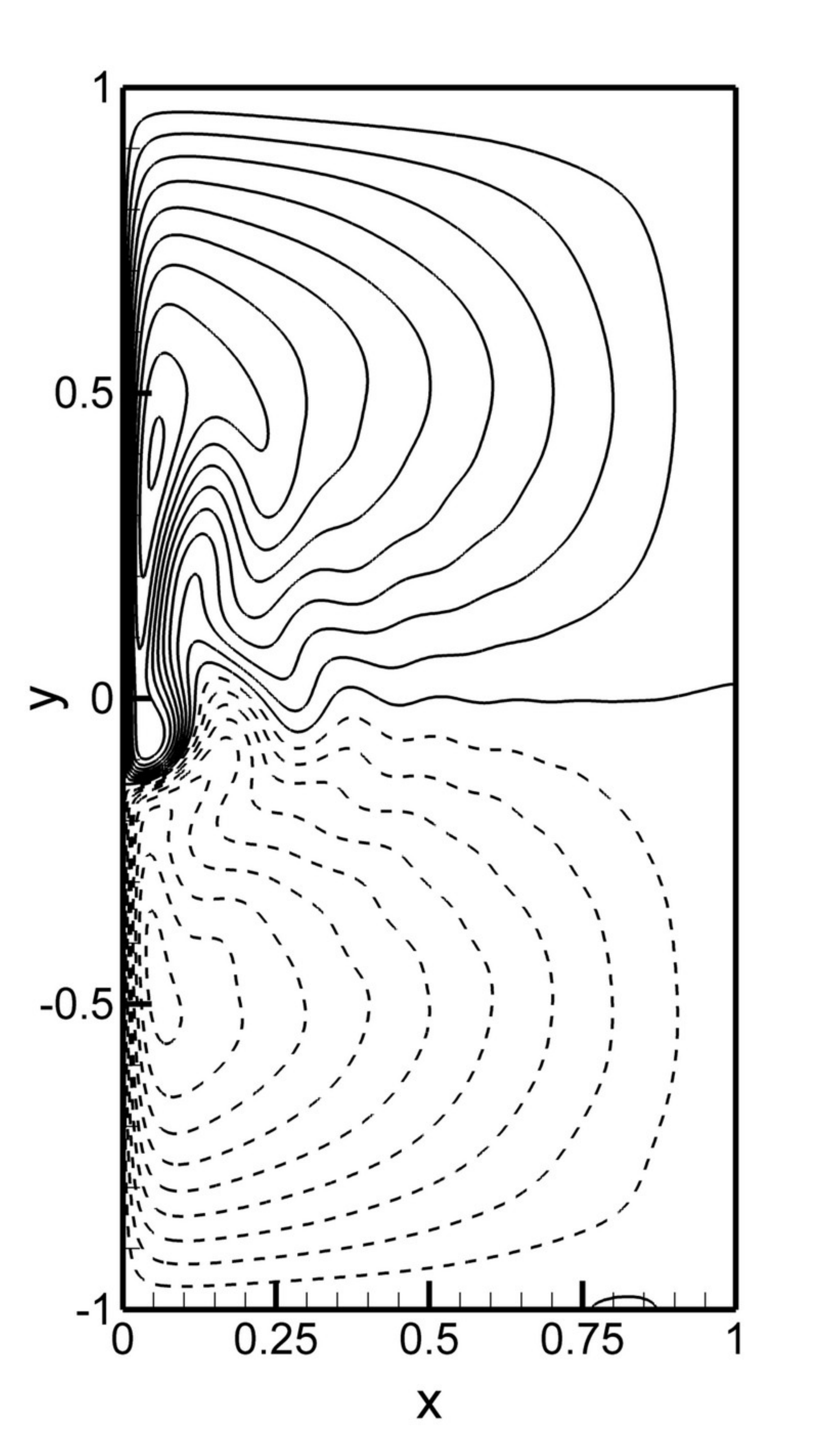}}
}
\caption{Comparison of the mean stream functions varying for $\epsilon$ for the ocean Basin I with $C_0 = 0.001$ m/s. Solid and dashed lines represent CCW and CW circulations with $0.1$ increments, respectively.}
\label{f:a-1}
\end{figure}

\begin{figure}
\centering
\mbox{
\subfigure[undisturbed]{\includegraphics[width=0.25\textwidth]{g-e1-l.pdf}}
\subfigure[$C_0 = 0.0008$ m/s]{\includegraphics[width=0.25\textwidth]{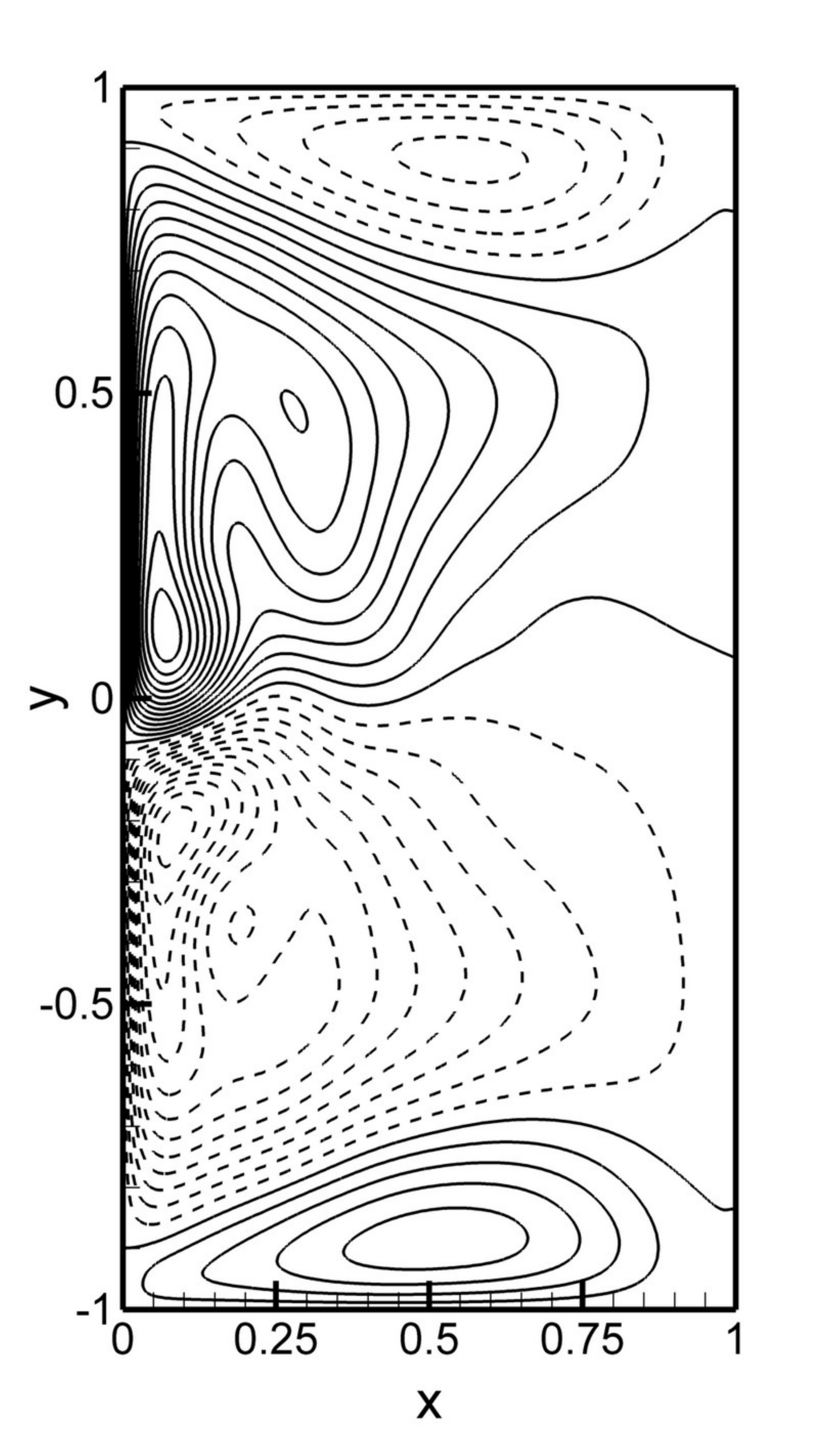}}
\subfigure[$C_0 = 0.0012$ m/s]{\includegraphics[width=0.25\textwidth]{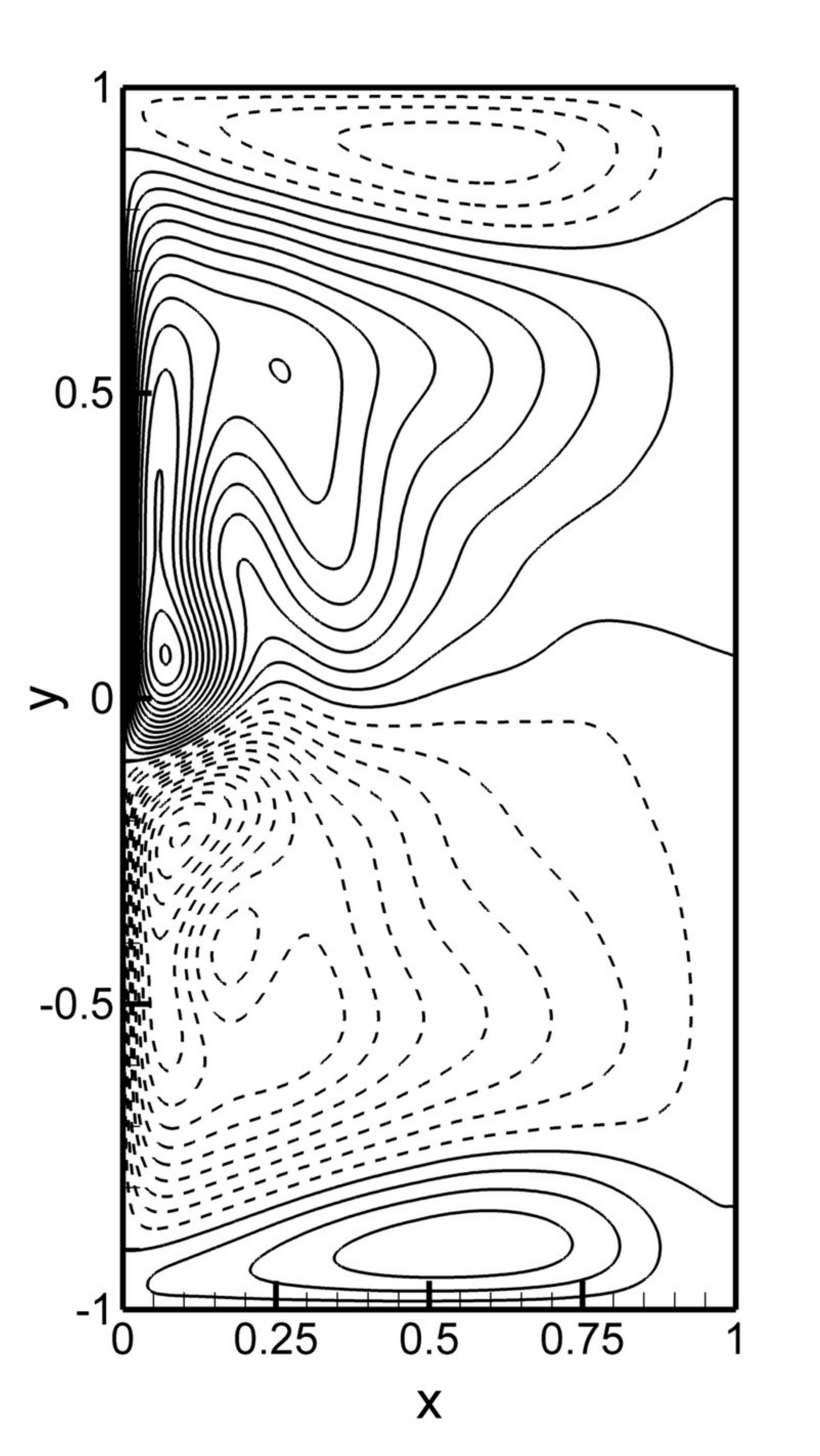}}
\subfigure[$C_0 = 0.0016$ m/s]{\includegraphics[width=0.25\textwidth]{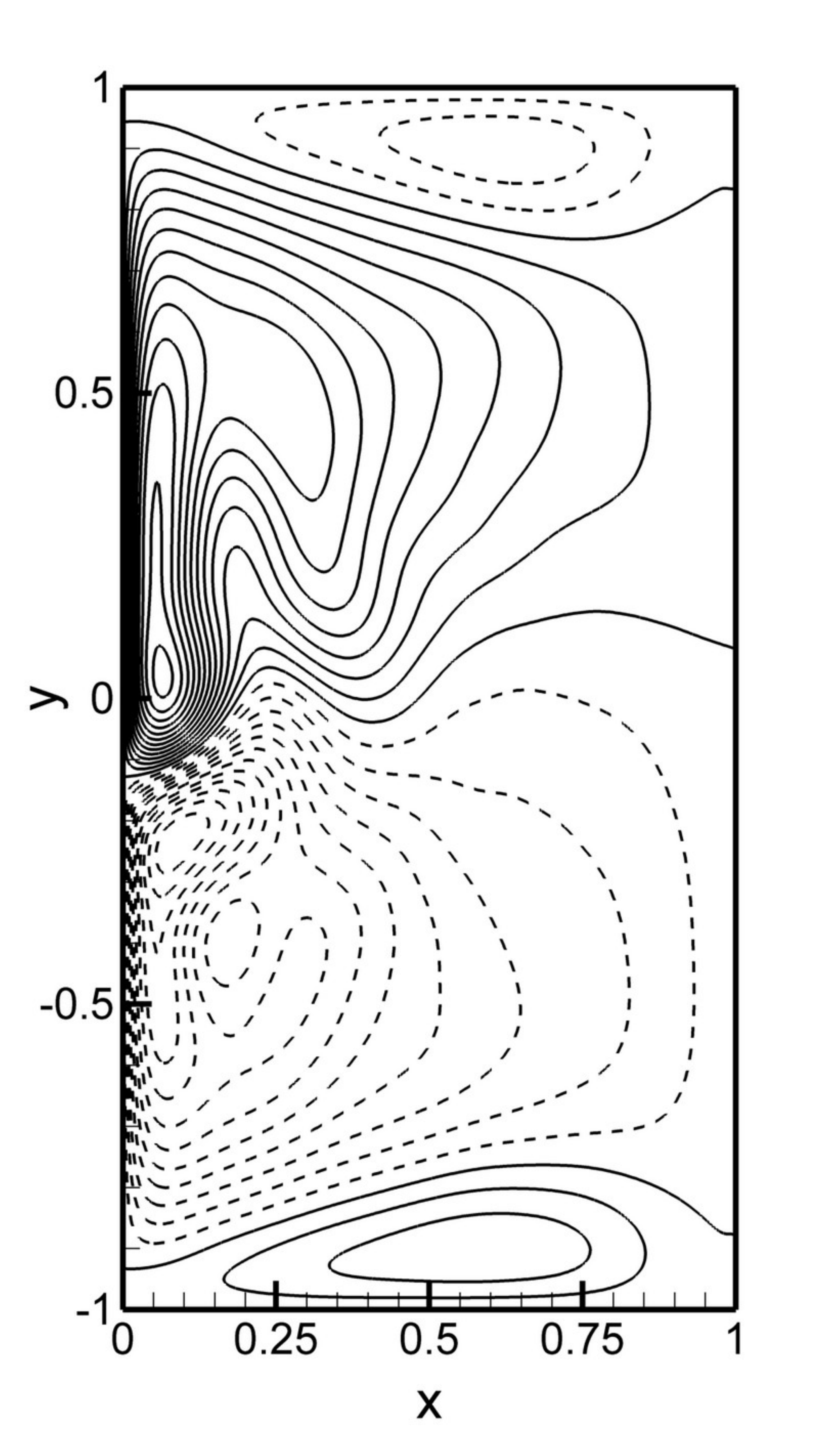}}
}
\caption{Comparison of the mean stream functions varying for $\sigma_0$ for the ocean Basin I with $\epsilon = 0.05$. Solid and dashed lines represent CCW and CW circulations with $0.1$ increments, respectively.}
\label{f:c-1}
\end{figure}

\begin{figure}
\centering
\mbox{
\subfigure[undisturbed]{\includegraphics[width=0.25\textwidth]{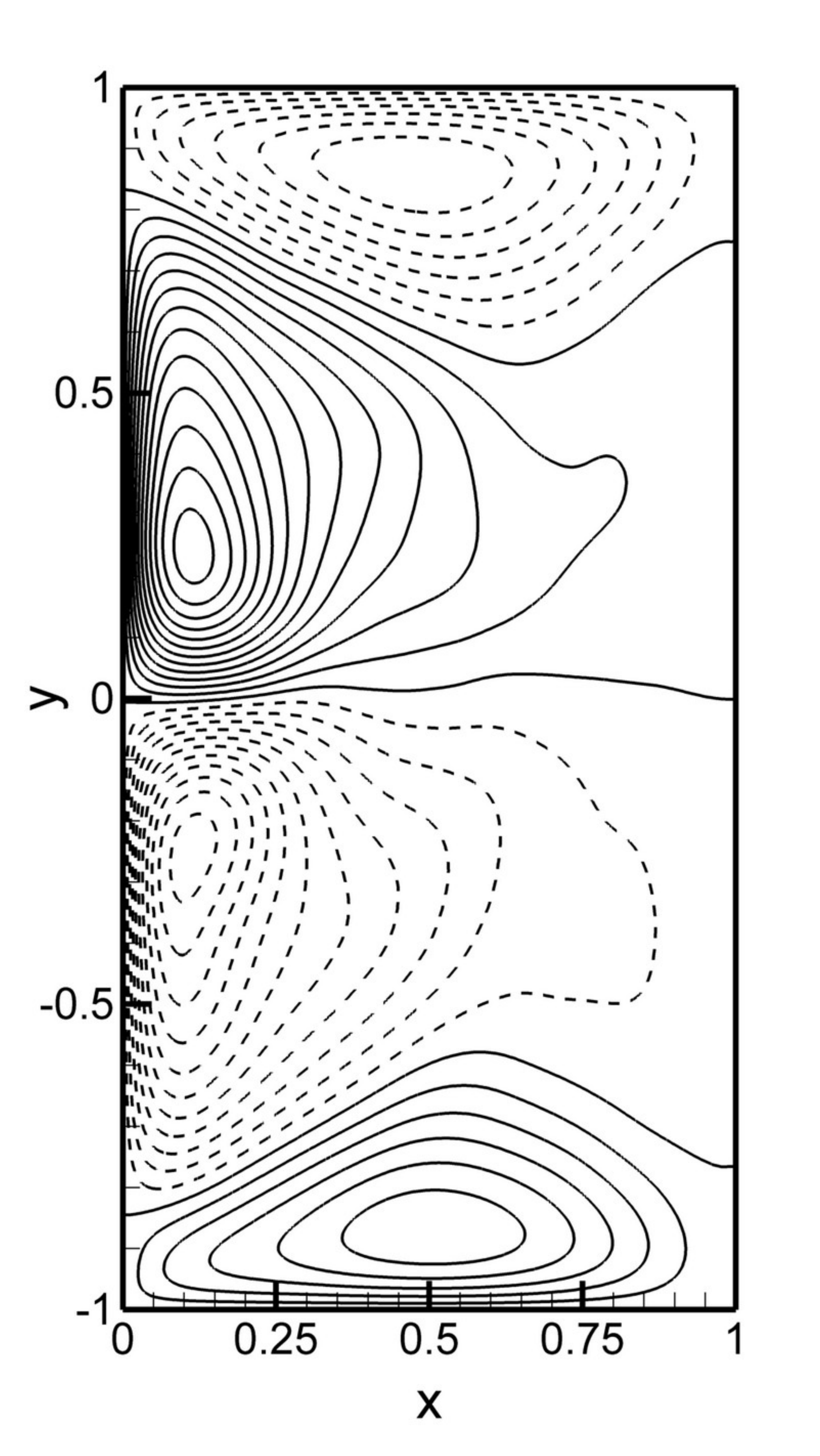}}
\subfigure[$\epsilon = 0.0625$]{\includegraphics[width=0.25\textwidth]{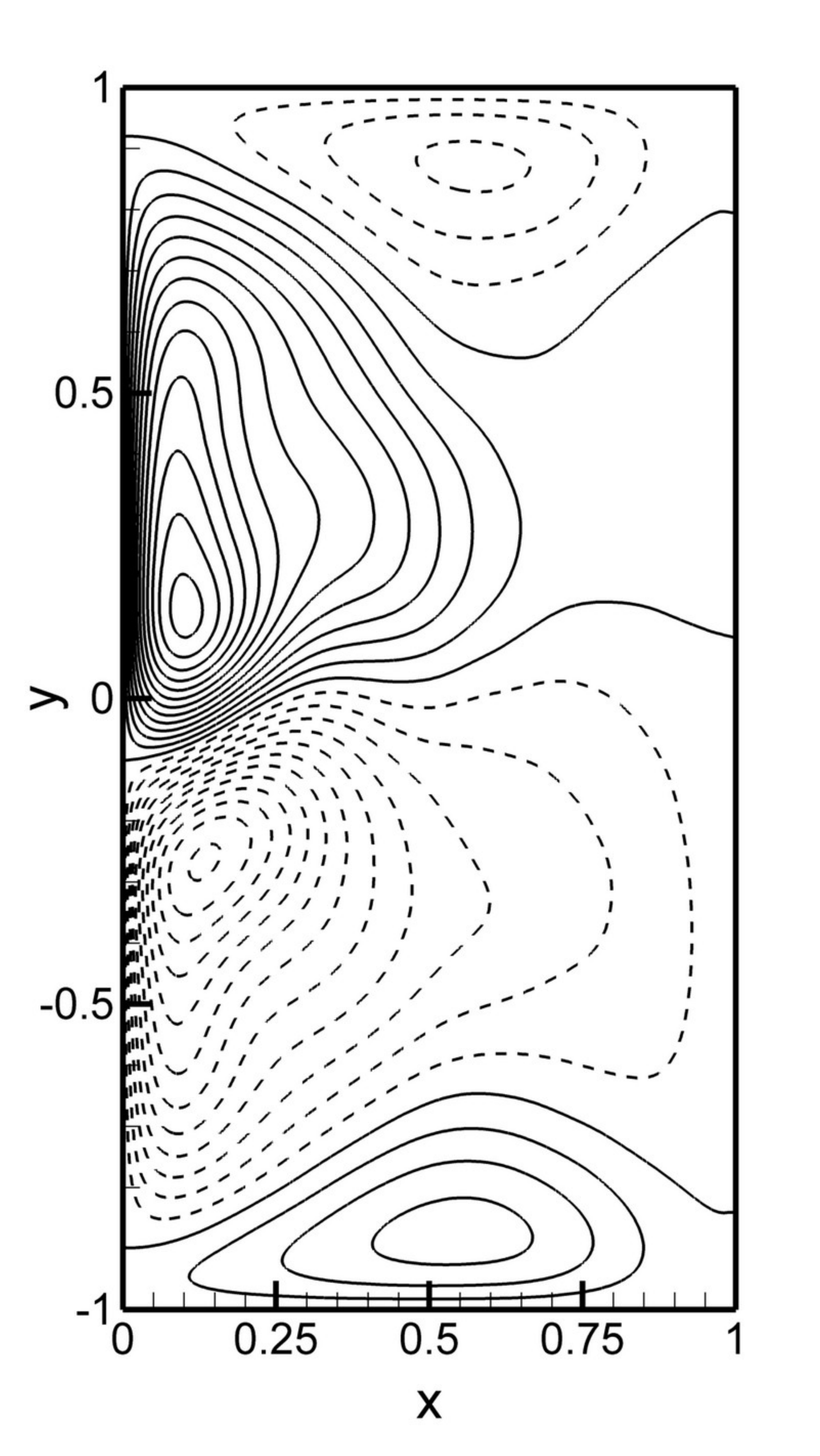}}
\subfigure[$\epsilon = 0.125$]{\includegraphics[width=0.25\textwidth]{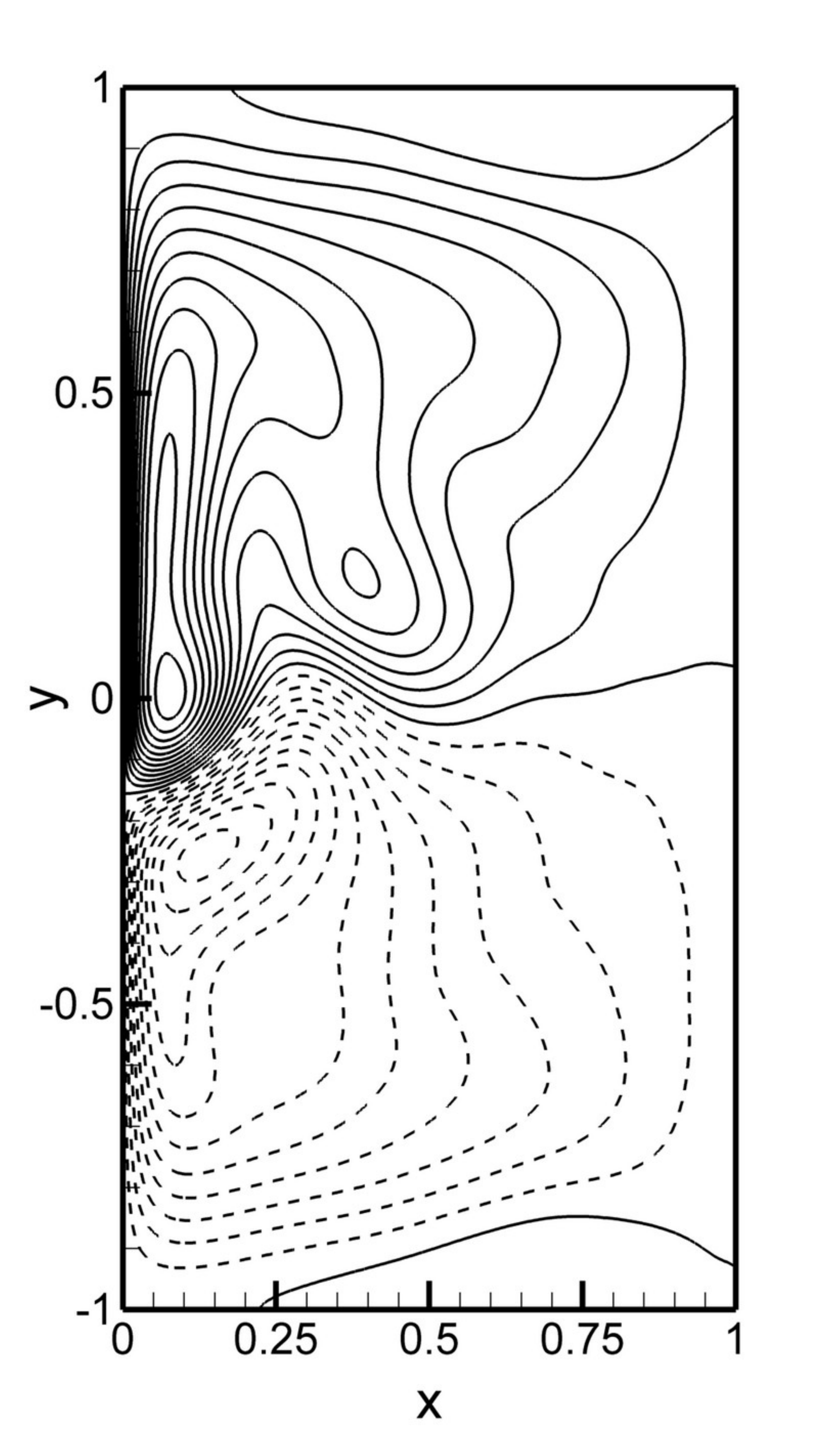}}
\subfigure[$\epsilon = 0.25$]{\includegraphics[width=0.25\textwidth]{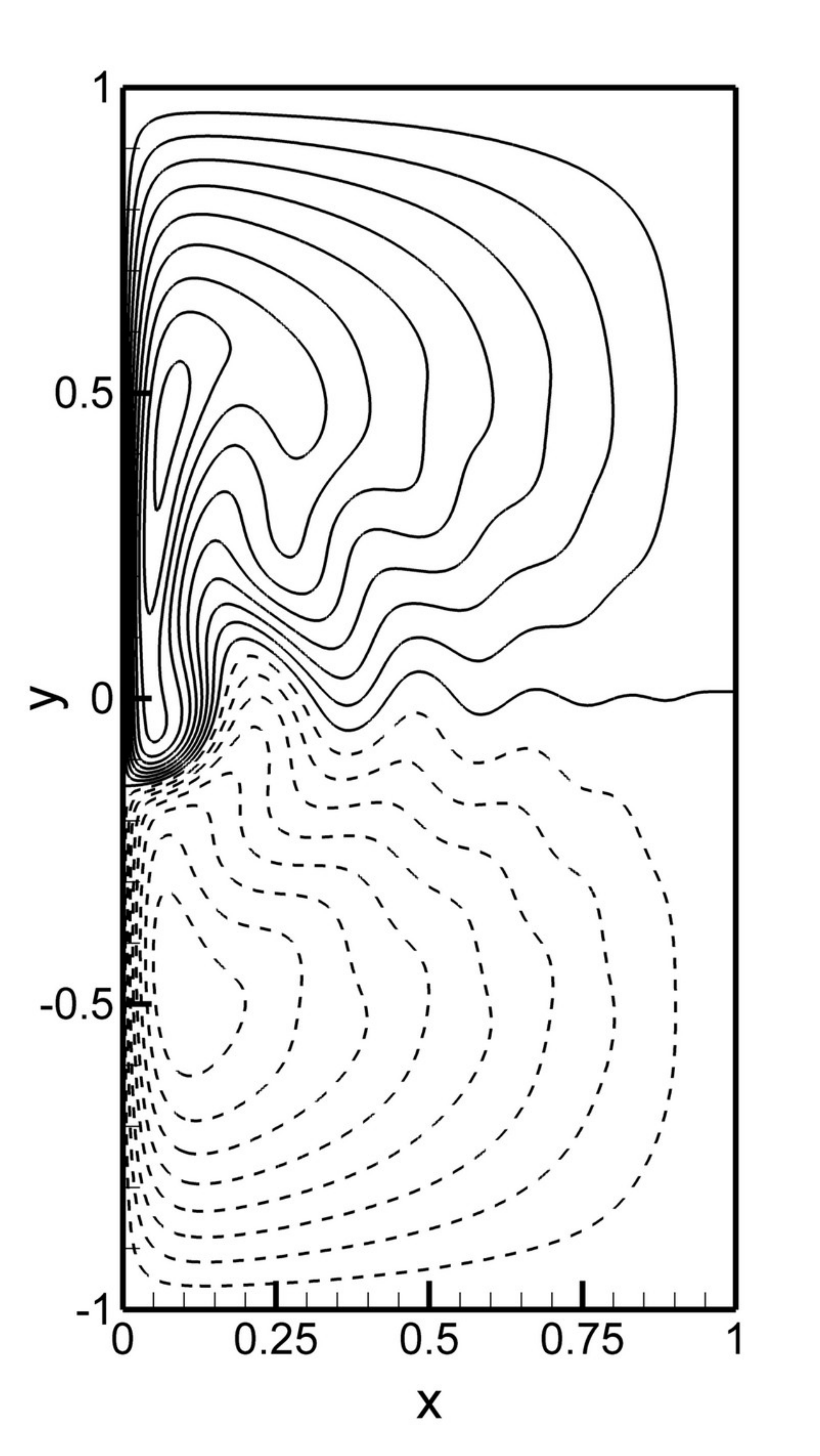}}
}
\caption{Comparison of the mean stream functions varying for $\epsilon$ for the ocean Basin II with $C_0 = 0.001$ m/s. Solid and dashed lines represent CCW and CW circulations with $0.1$ increments, respectively.}
\label{f:a-2}
\end{figure}

\begin{figure}
\centering
\mbox{
\subfigure[undisturbed]{\includegraphics[width=0.25\textwidth]{g-e2-l.pdf}}
\subfigure[$C_0 = 0.0008$ m/s]{\includegraphics[width=0.25\textwidth]{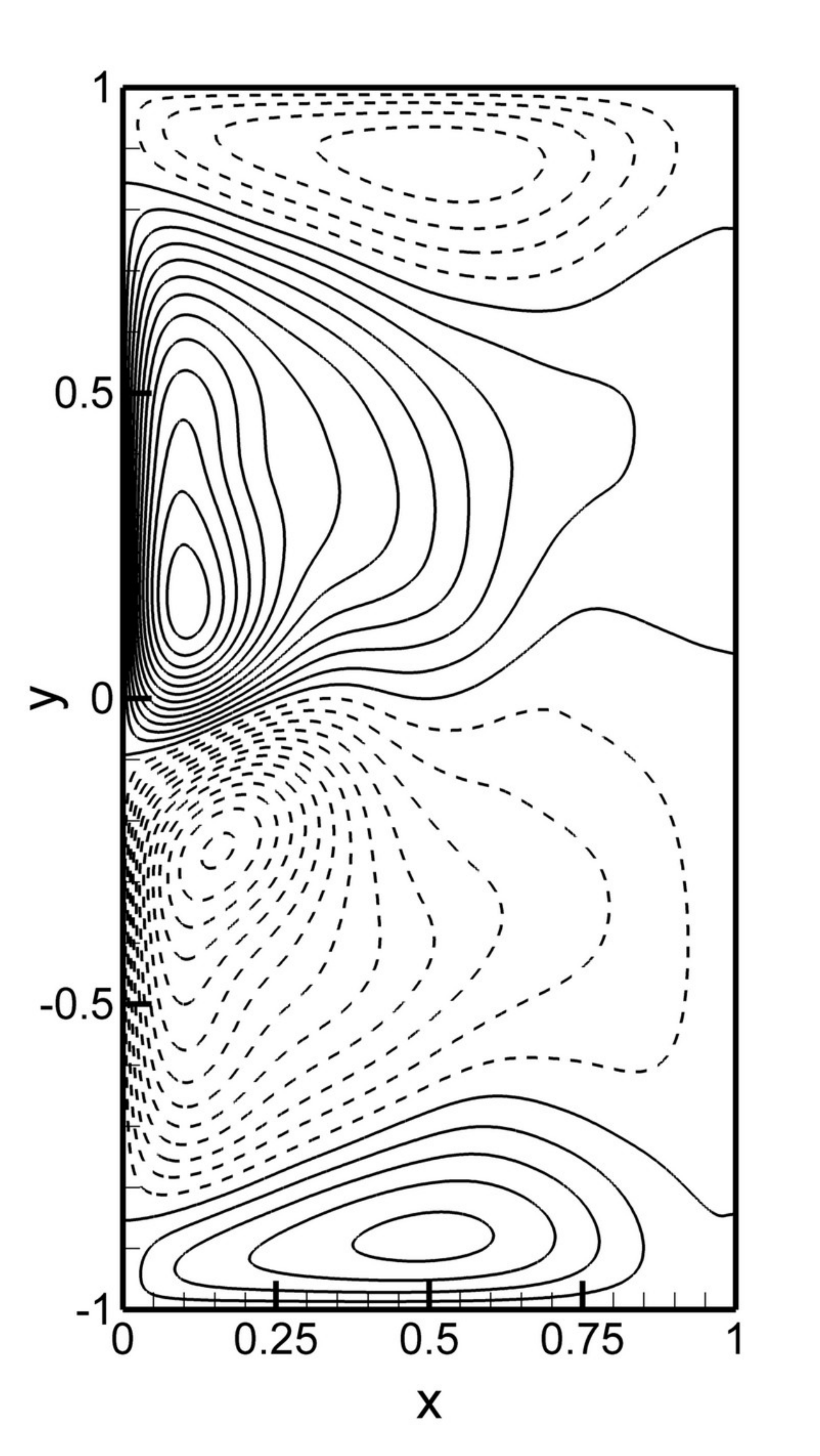}}
\subfigure[$C_0 = 0.0012$ m/s]{\includegraphics[width=0.25\textwidth]{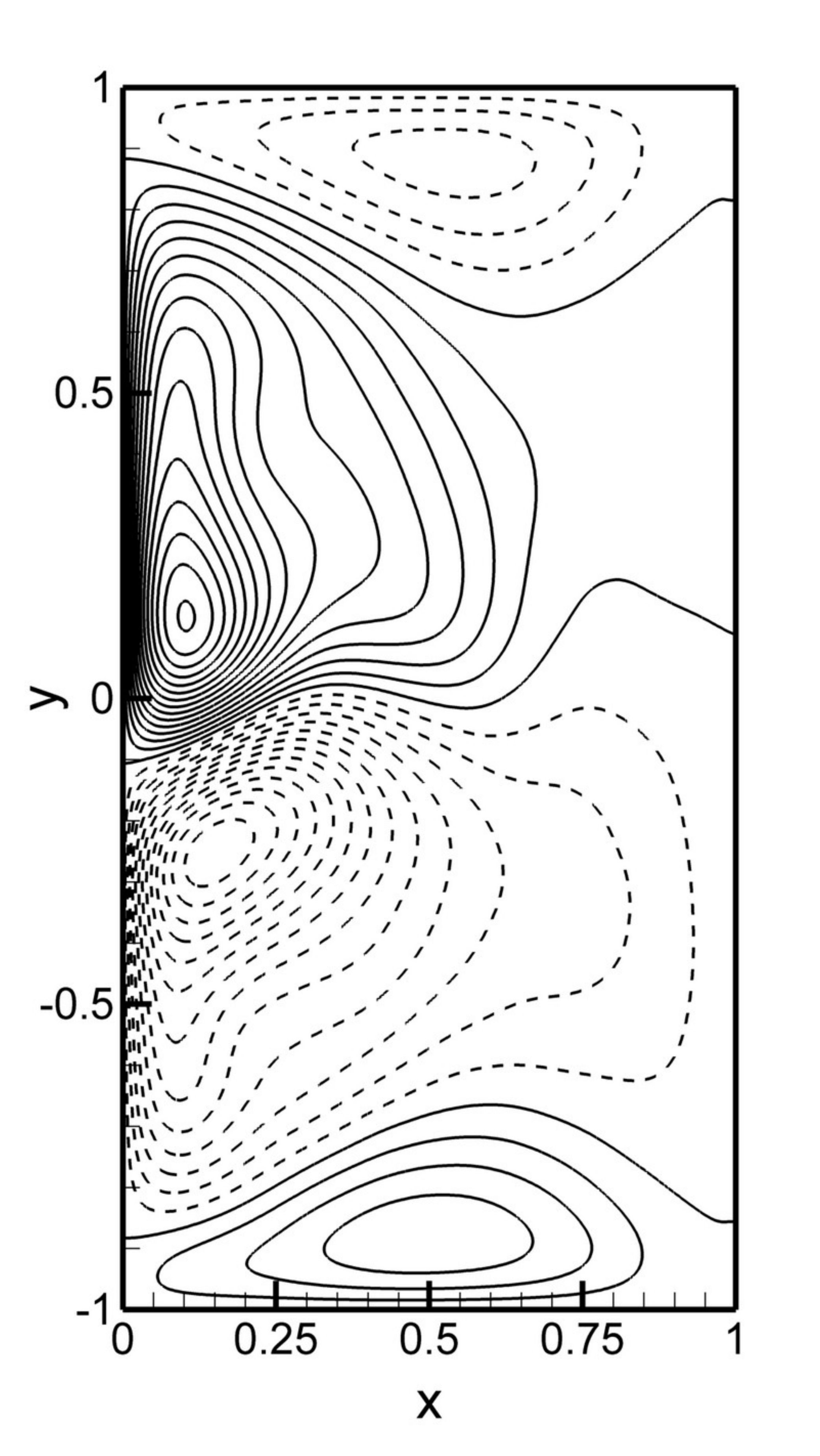}}
\subfigure[$C_0 = 0.0016$ m/s]{\includegraphics[width=0.25\textwidth]{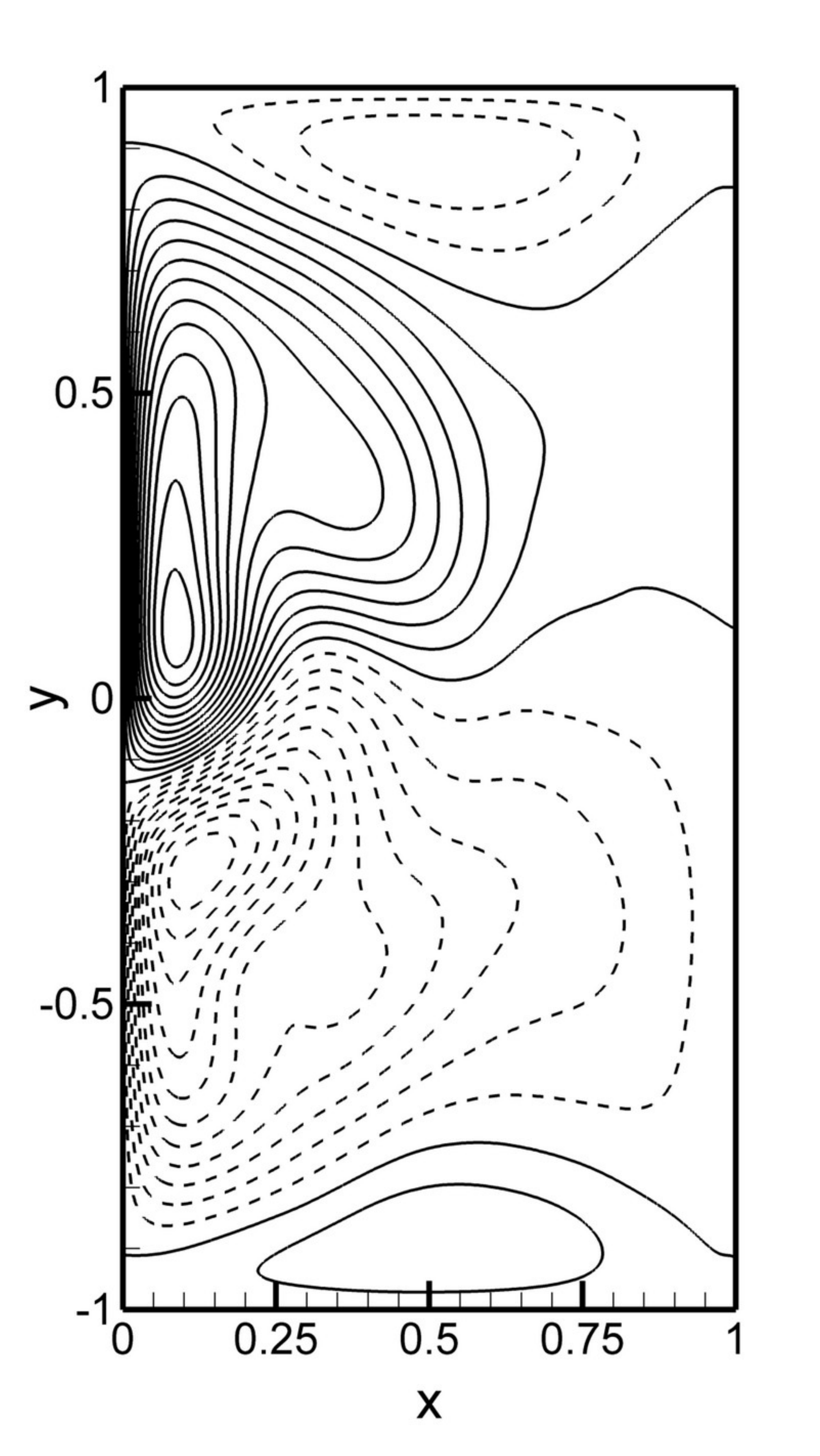}}
}
\caption{Comparison of the mean stream functions varying for $C_0$ for the ocean Basin II with $\epsilon = 0.0625$. Solid and dashed lines represent CCW and CW circulations with $0.1$ increments, respectively.}
\label{f:c-2}
\end{figure}

\begin{figure}
\centering
\mbox{
\subfigure[undisturbed]{\includegraphics[width=0.25\textwidth]{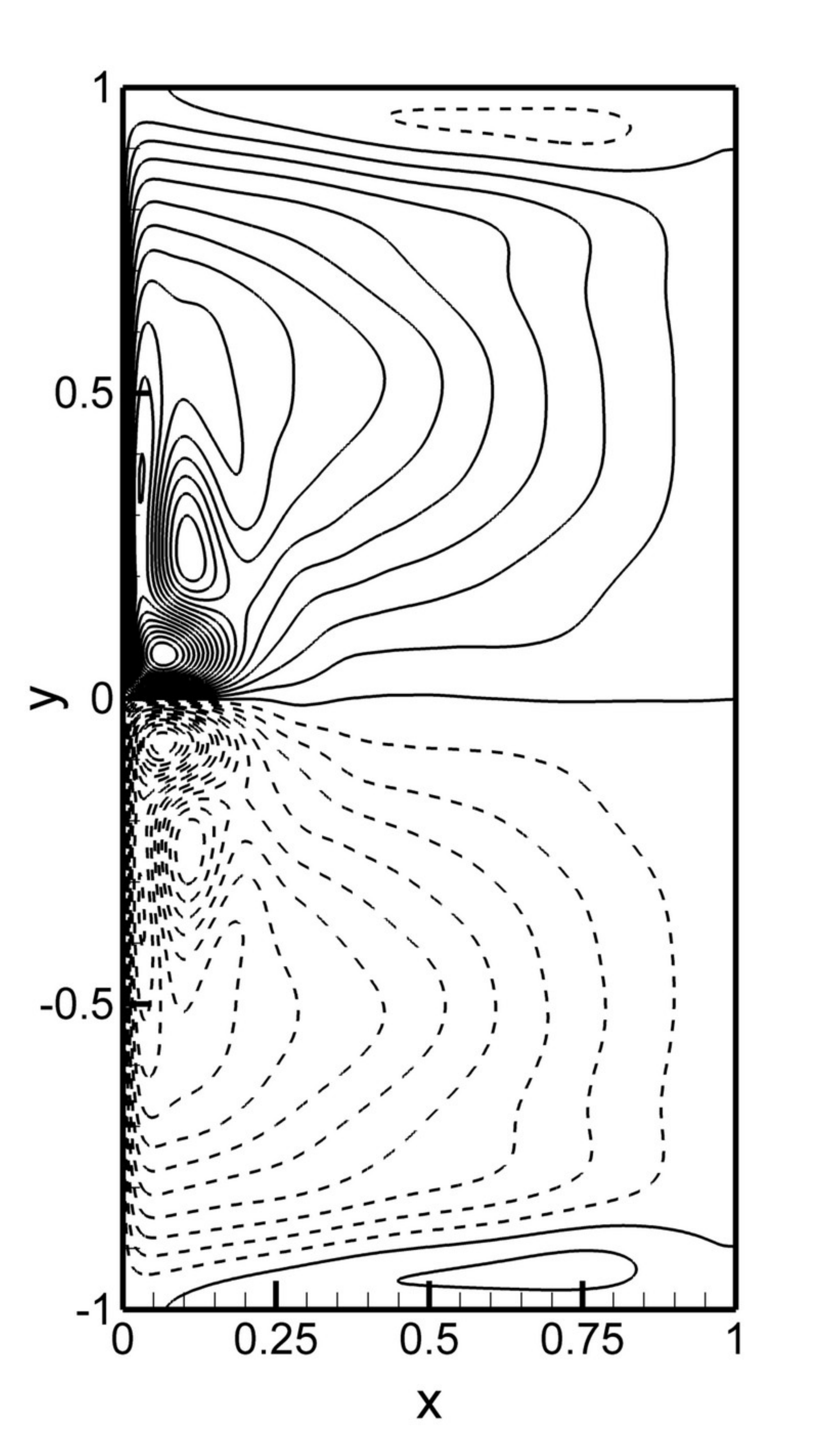}}
\subfigure[$\epsilon = 0.05$]{\includegraphics[width=0.25\textwidth]{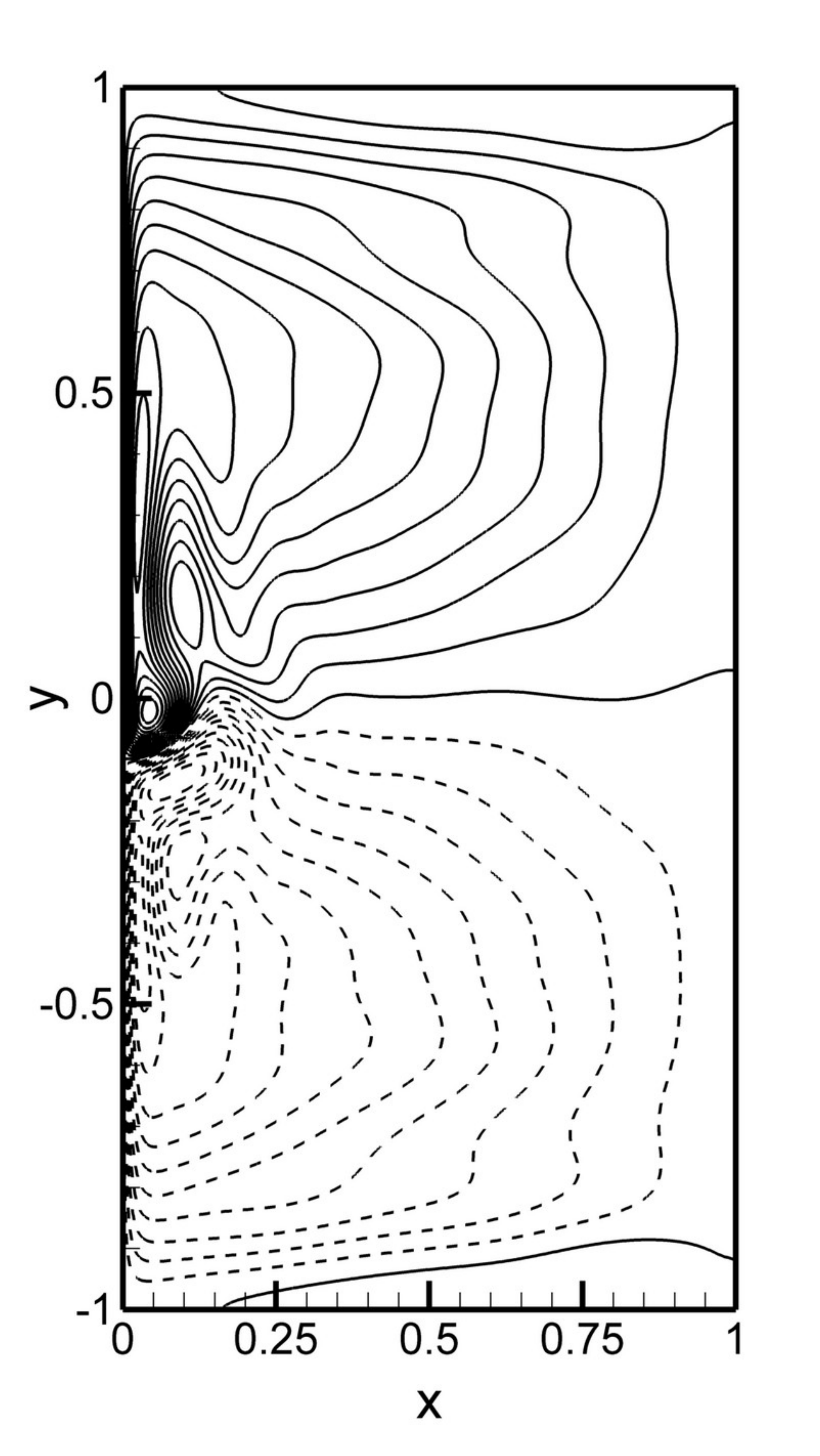}}
\subfigure[$\epsilon = 0.1$]{\includegraphics[width=0.25\textwidth]{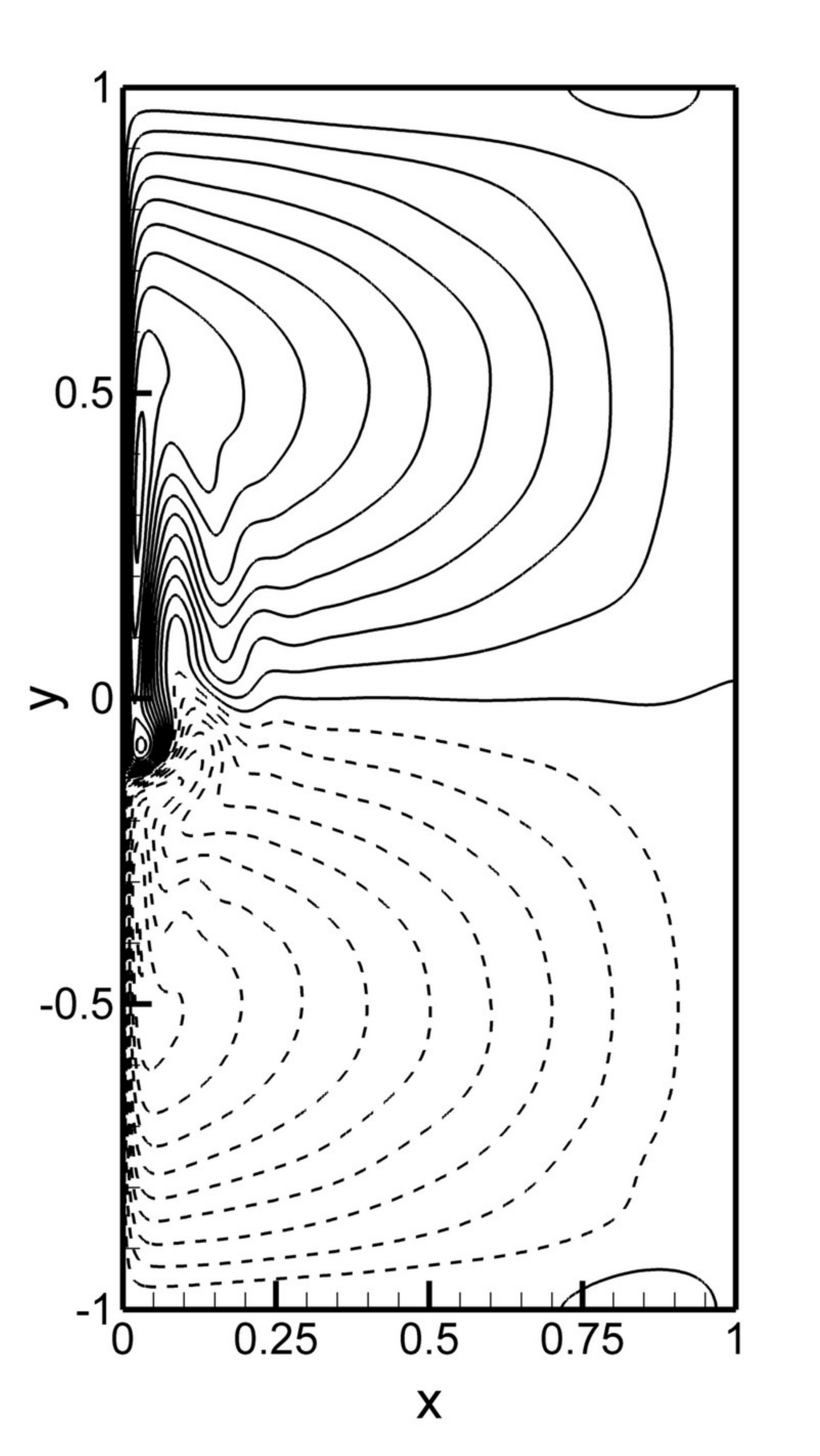}}
\subfigure[$\epsilon = 0.2$]{\includegraphics[width=0.25\textwidth]{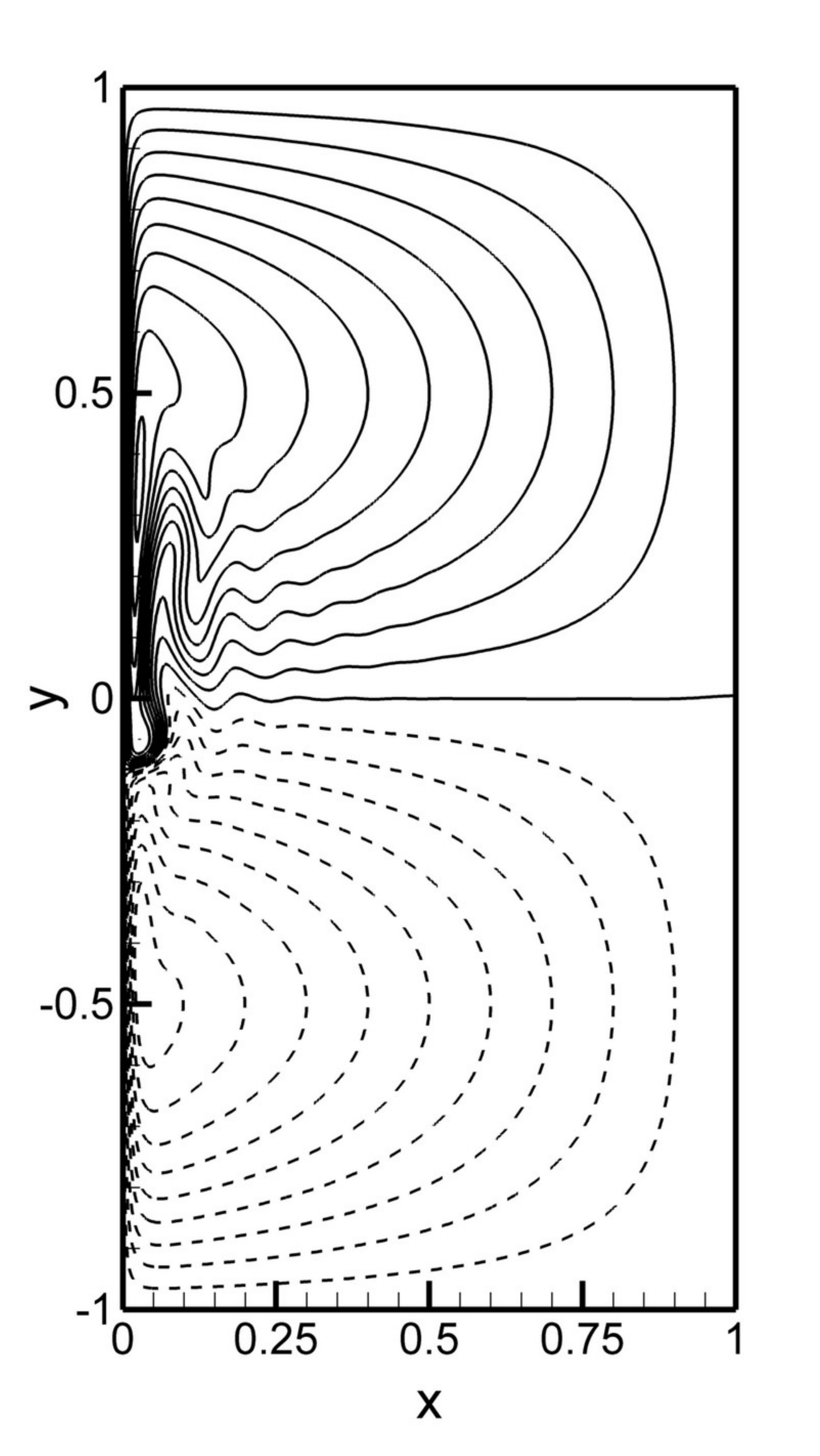}}
}
\caption{Comparison of the mean stream functions varying for $\epsilon$ for the ocean Basin III with $C_0 = 0.001$. Solid and dashed lines represent CCW and CW circulations with $0.1$ increments, respectively.}
\label{f:a-3}
\end{figure}

\begin{figure}
\centering
\mbox{
\subfigure[undisturbed]{\includegraphics[width=0.25\textwidth]{g-e3-l.pdf}}
\subfigure[$C_0 = 0.0008$ m/s]{\includegraphics[width=0.25\textwidth]{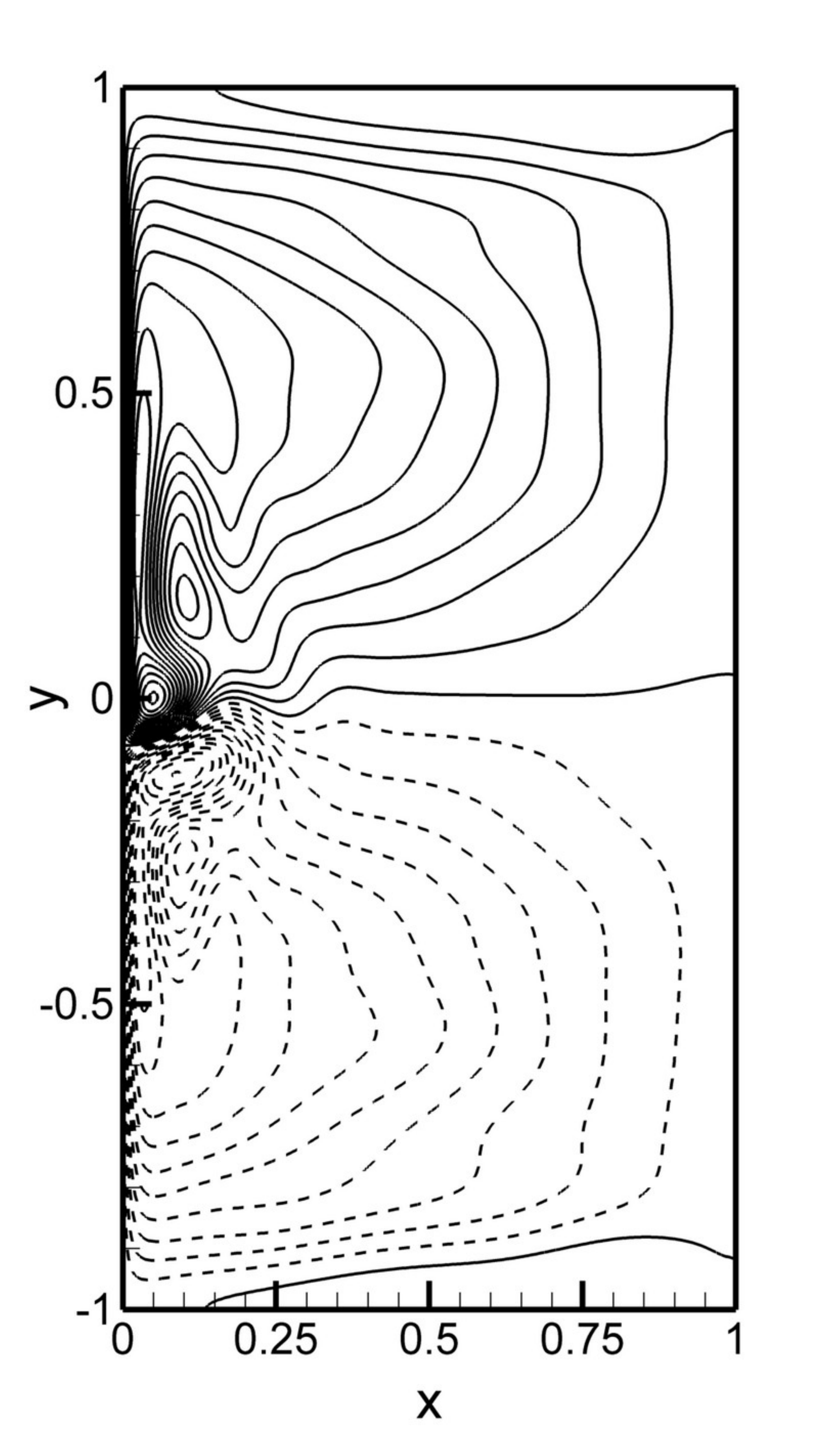}}
\subfigure[$C_0 = 0.0012$ m/s]{\includegraphics[width=0.25\textwidth]{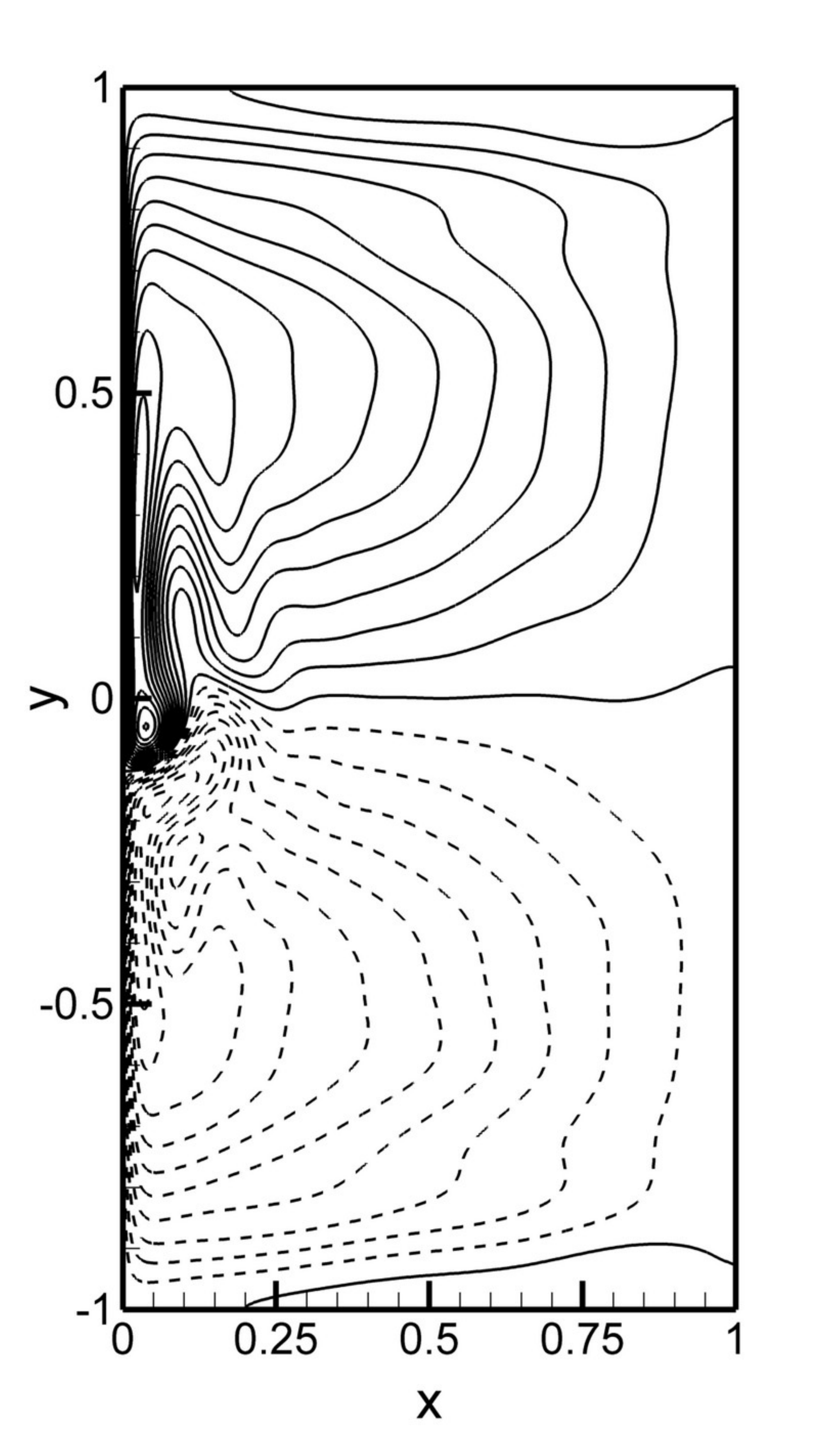}}
\subfigure[$C_0 = 0.0016$ m/s]{\includegraphics[width=0.25\textwidth]{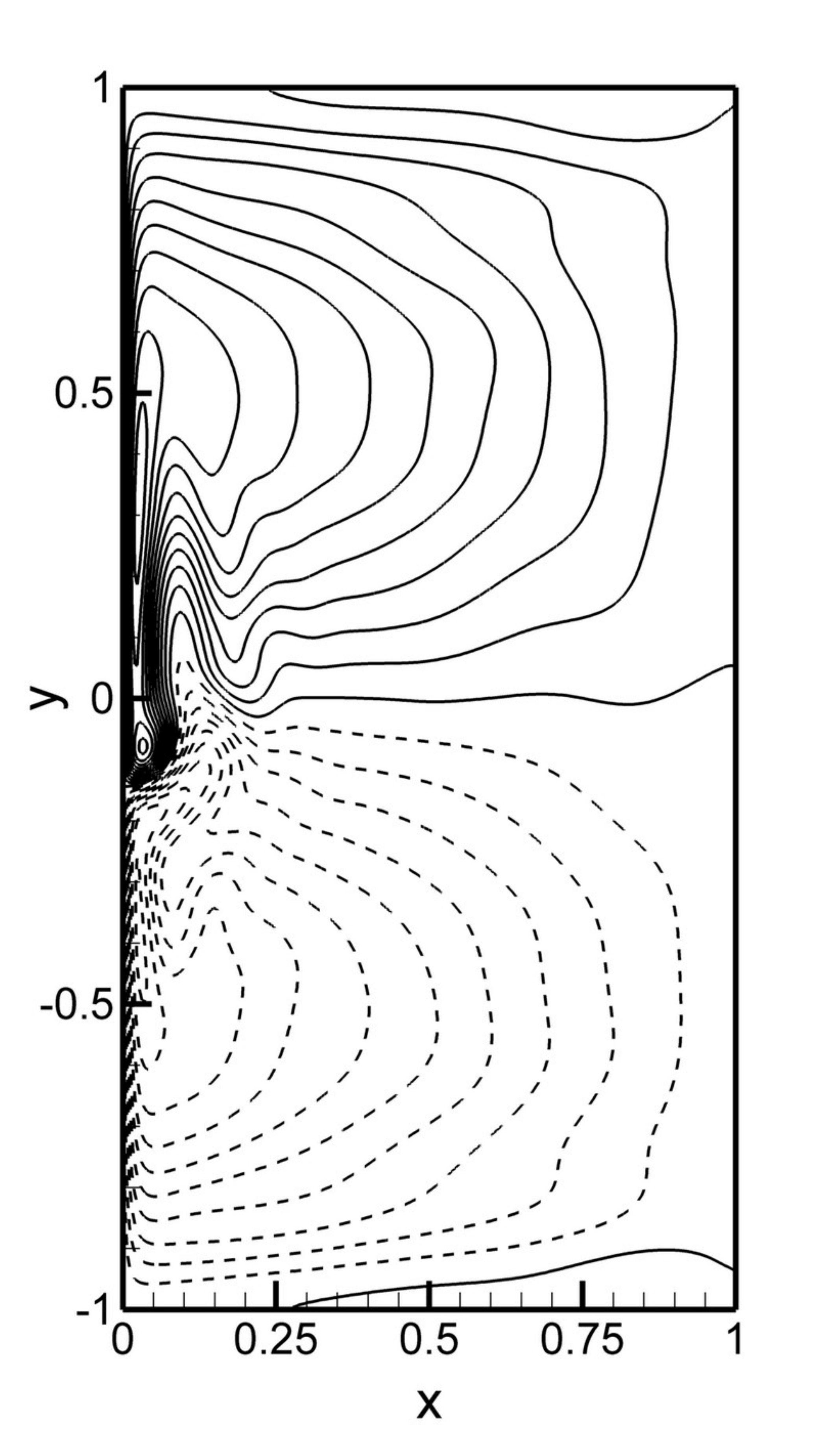}}
}
\caption{Comparison of the mean stream functions varying for $C_0$ for the ocean Basin III with $\epsilon = 0.05$. Solid and dashed lines represent CCW and CW circulations with $0.1$ increments, respectively.}
\label{f:c-3}
\end{figure}

Next, we investigate the effect of the added turbines to the ocean circulation in each ocean basin. We will investigate numerically whether we can reproduce the four gyre time average circulation field by adding turbine profile which is parameterized by $\sigma$ (i.e., $C_0$ in dimensional form) and $\epsilon$ parameters. Using $C_0=0.001$ m/s, Fig.~\ref{f:a-1} compares mean stream function for the Basin I in various effective turbine areas parameterized by $\epsilon$.  Solid and dashed lines in this figure represent CCW and CW circulations with $0.1$ increments, respectively. Mean circulation field due to the undisturbed simulation is also included for comparison purpose. As shown in this figure, although the long time average yields a four gyre pattern for undisturbed and slightly disturbed cases (i.e., when the effective scale for the added turbine $L_t =100$ km), an increase in $\epsilon$ value may result in a change the mean flow dynamics. Four-gyre pattern may change the double-gyre mean circulation pattern because an increase of effective turbine area. Since mean flow date is averaged over several decade, this result may suggest that adding excessive turbines in the western boundary currents may result in a significant change in time average flow pattern. Without an explicit relationship between the turbine array distribution and corresponding turbine area, we found that turbine should be distributed over a length scale less than $L_t=100$ km in order to keep similar gyre structures in the time average of statistically steady state. Mean stream function contour plots are also illustrated in Fig.~\ref{f:c-1} for varying turbine drag coefficient $C_0$ at $\epsilon = 0.05$. With decrease in the turbine drag coefficient (i.e., possibly due to a less denser turbine distribution per area) asymptote to the undisturbed flow dynamics demonstrating the four-gyre flow pattern. Similar analyses are shown in Figs.~\ref{f:a-2}-\ref{f:c-2} and Figs.~\ref{f:a-3}-\ref{f:c-3} for Basin II and Basin III, respectively, illustrating the same trend with respect to turbine parameters.

\begin{figure}
\centering
\mbox{
\subfigure[$\epsilon = 0.05$, $C_0 = 0.001$ m/s]{\includegraphics[width=0.34\textwidth]{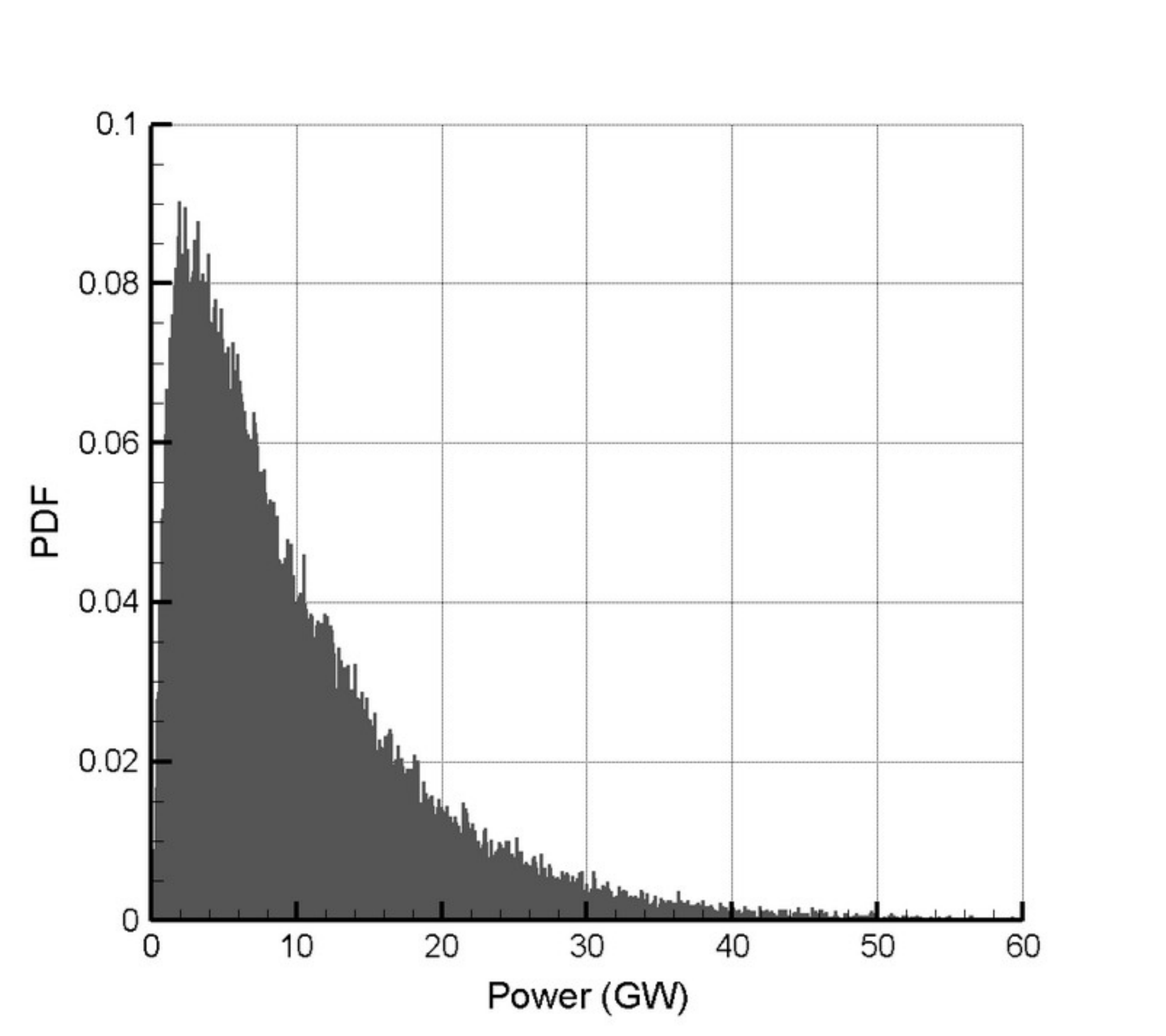}}
\subfigure[$\epsilon = 0.1$, $C_0 = 0.001$ m/s]{\includegraphics[width=0.34\textwidth]{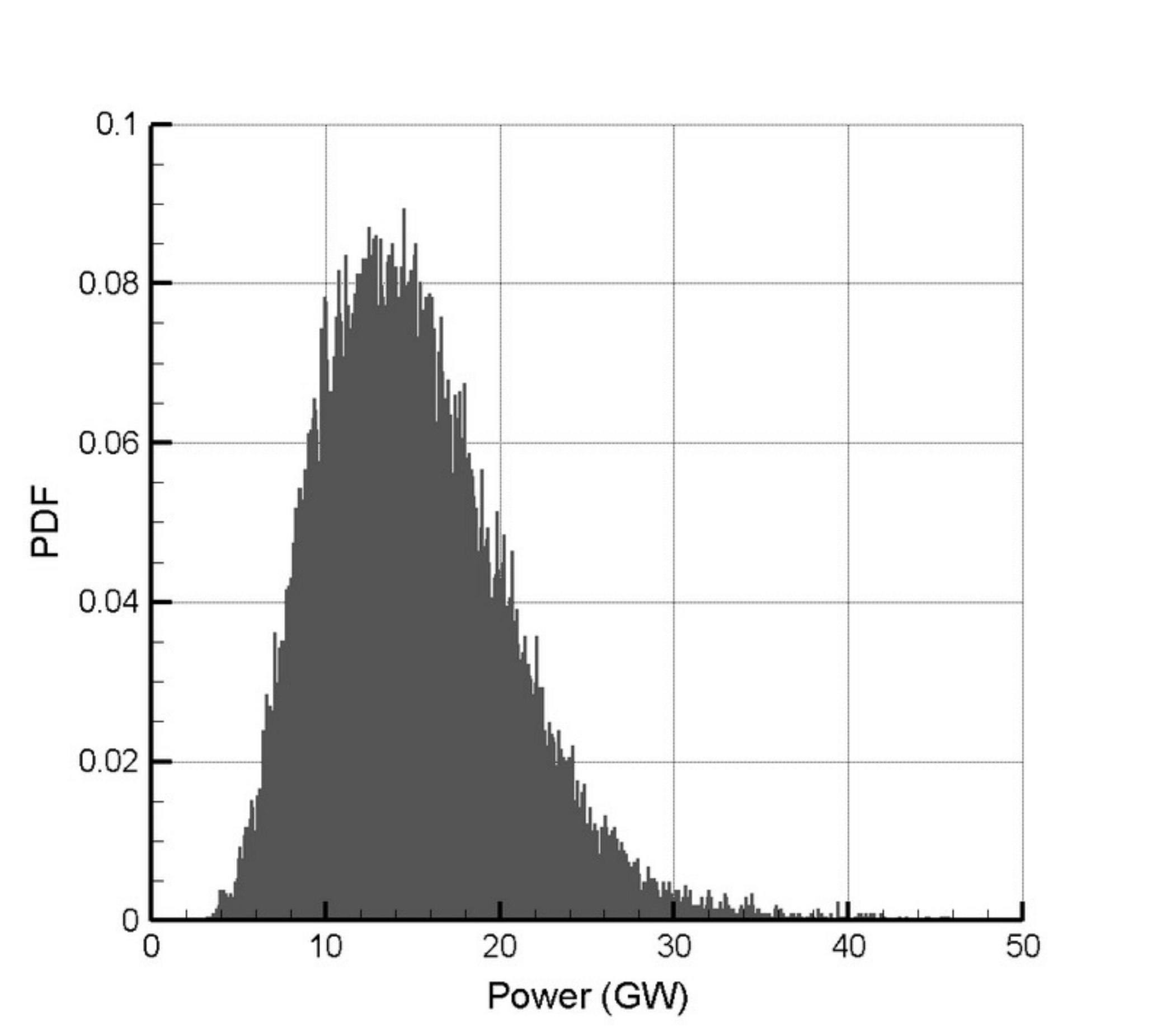}}
\subfigure[$\epsilon = 0.2$, $C_0 = 0.001$ m/s]{\includegraphics[width=0.34\textwidth]{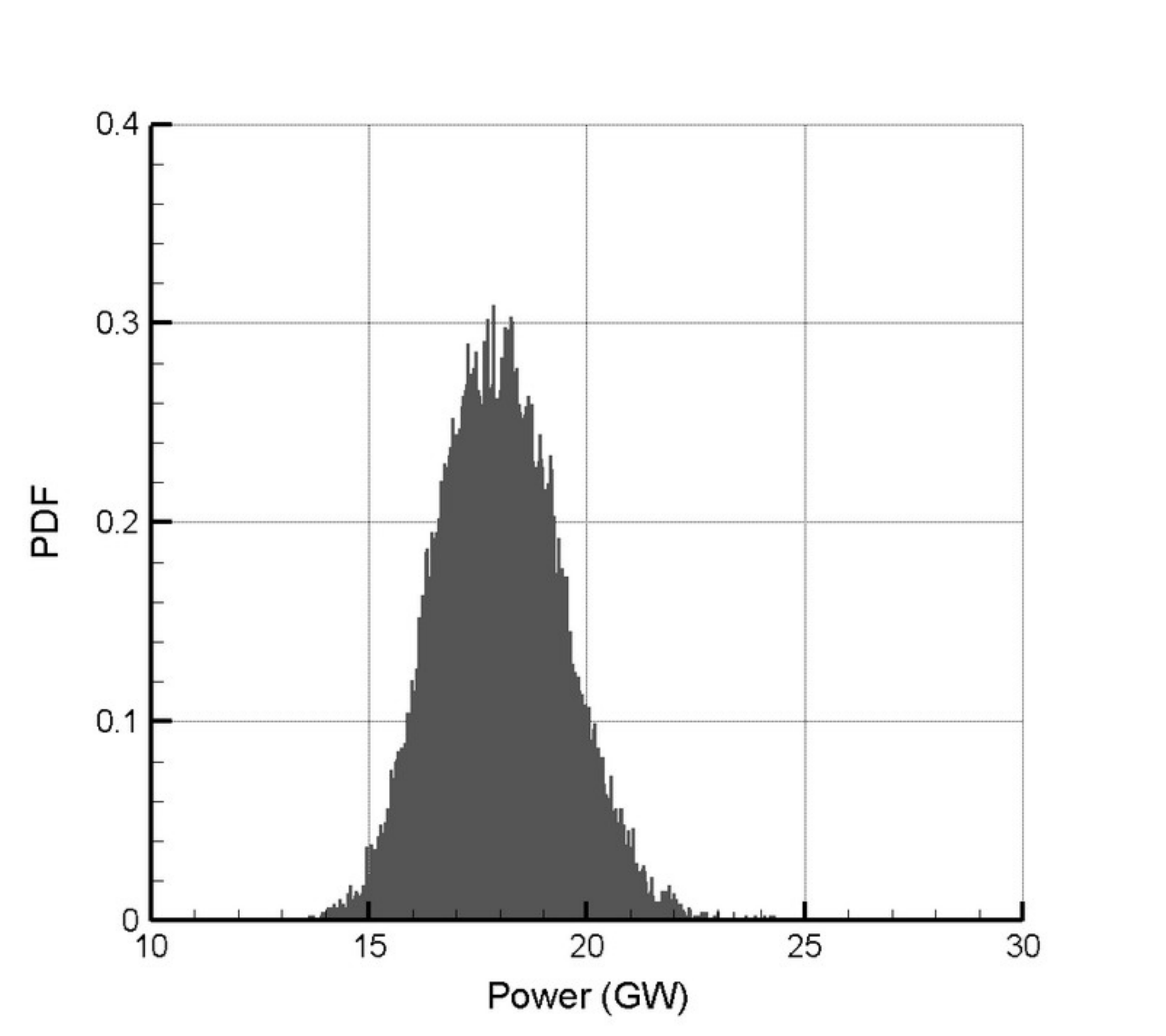}}
}\\
\mbox{
\subfigure[$\epsilon = 0.05$, $C_0 = 0.0008$ m/s]{\includegraphics[width=0.34\textwidth]{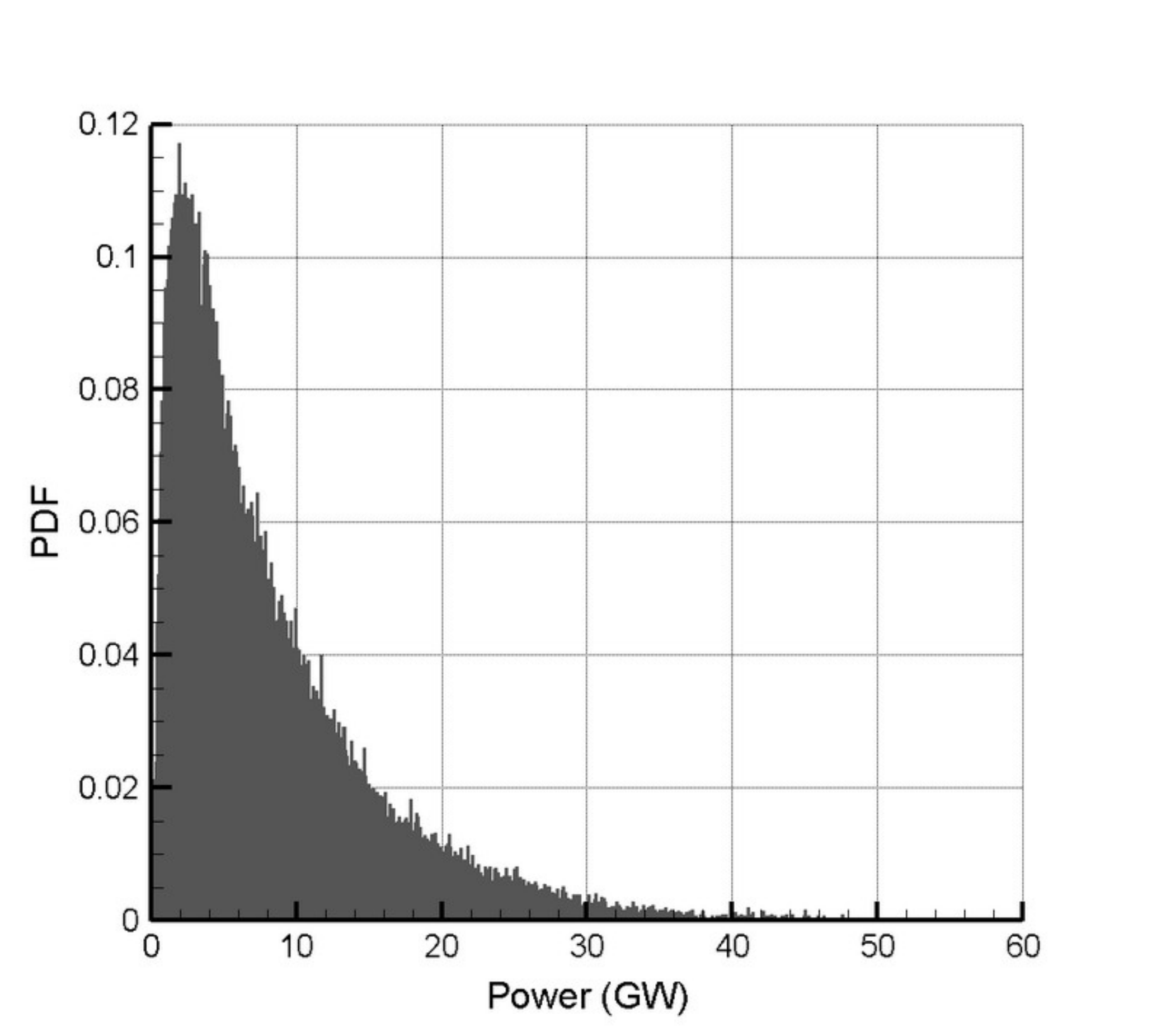}}
\subfigure[$\epsilon = 0.05$, $C_0 = 0.0012$ m/s]{\includegraphics[width=0.34\textwidth]{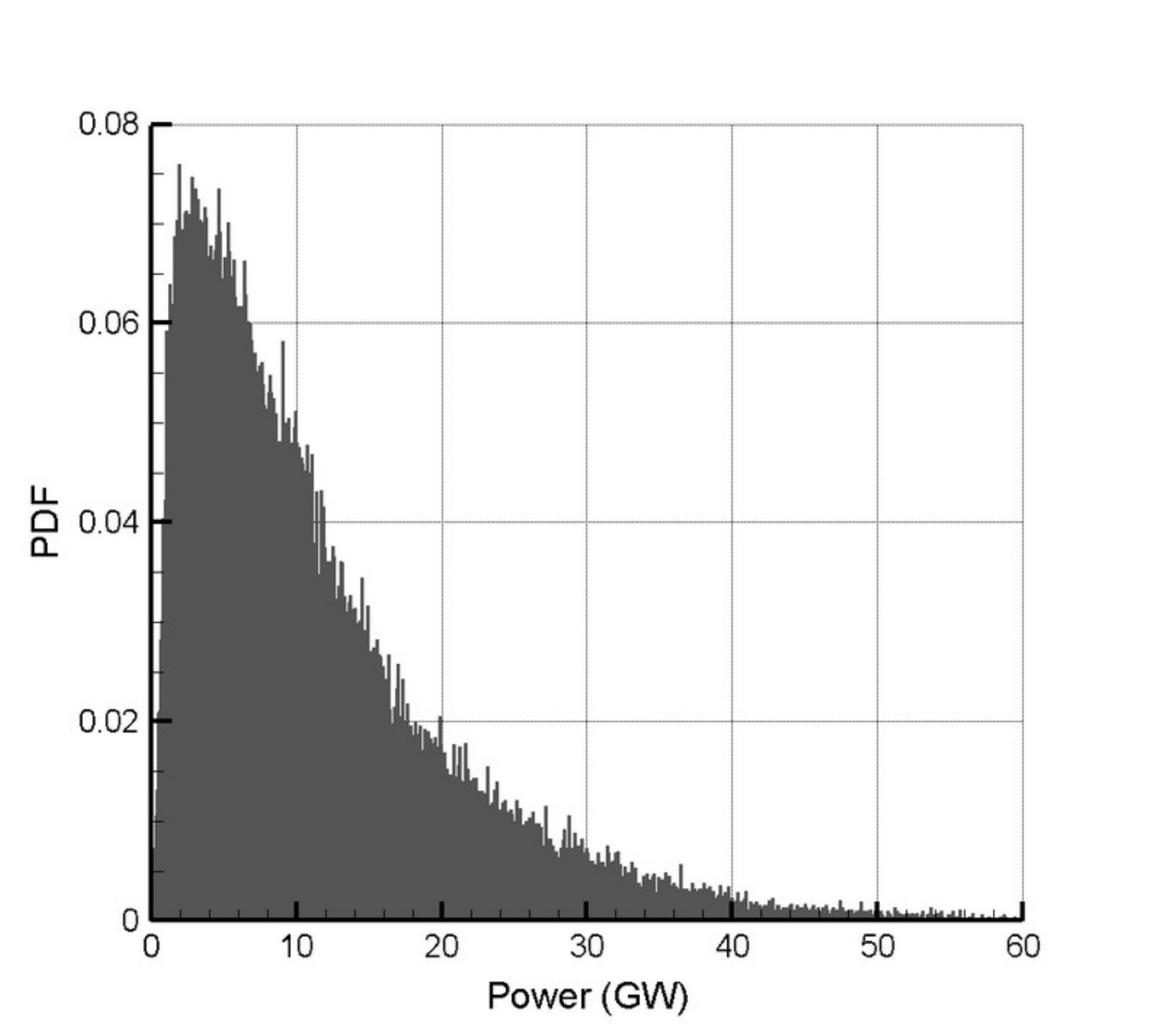}}
\subfigure[$\epsilon = 0.05$, $C_0 = 0.0016$ m/s]{\includegraphics[width=0.34\textwidth]{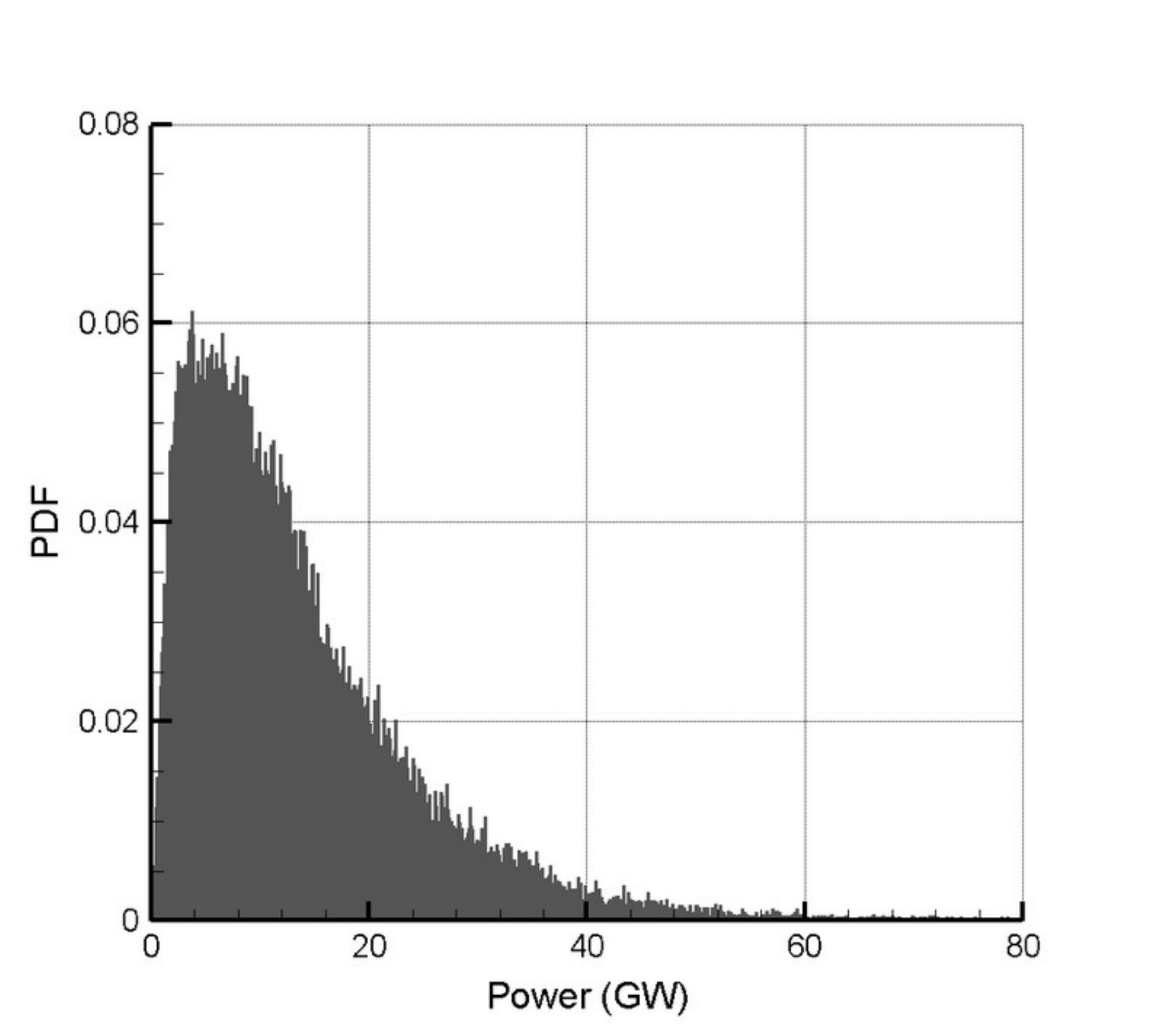}}
}
\caption{Probability density function (PDF) of the extracted power from the ocean Basin I.}
\label{f:pdf-p-1}
\end{figure}

\begin{figure}
\centering
\mbox{
\subfigure[$\epsilon = 0.0625$, $\sigma_0 = 0.001$]{\includegraphics[width=0.33\textwidth]{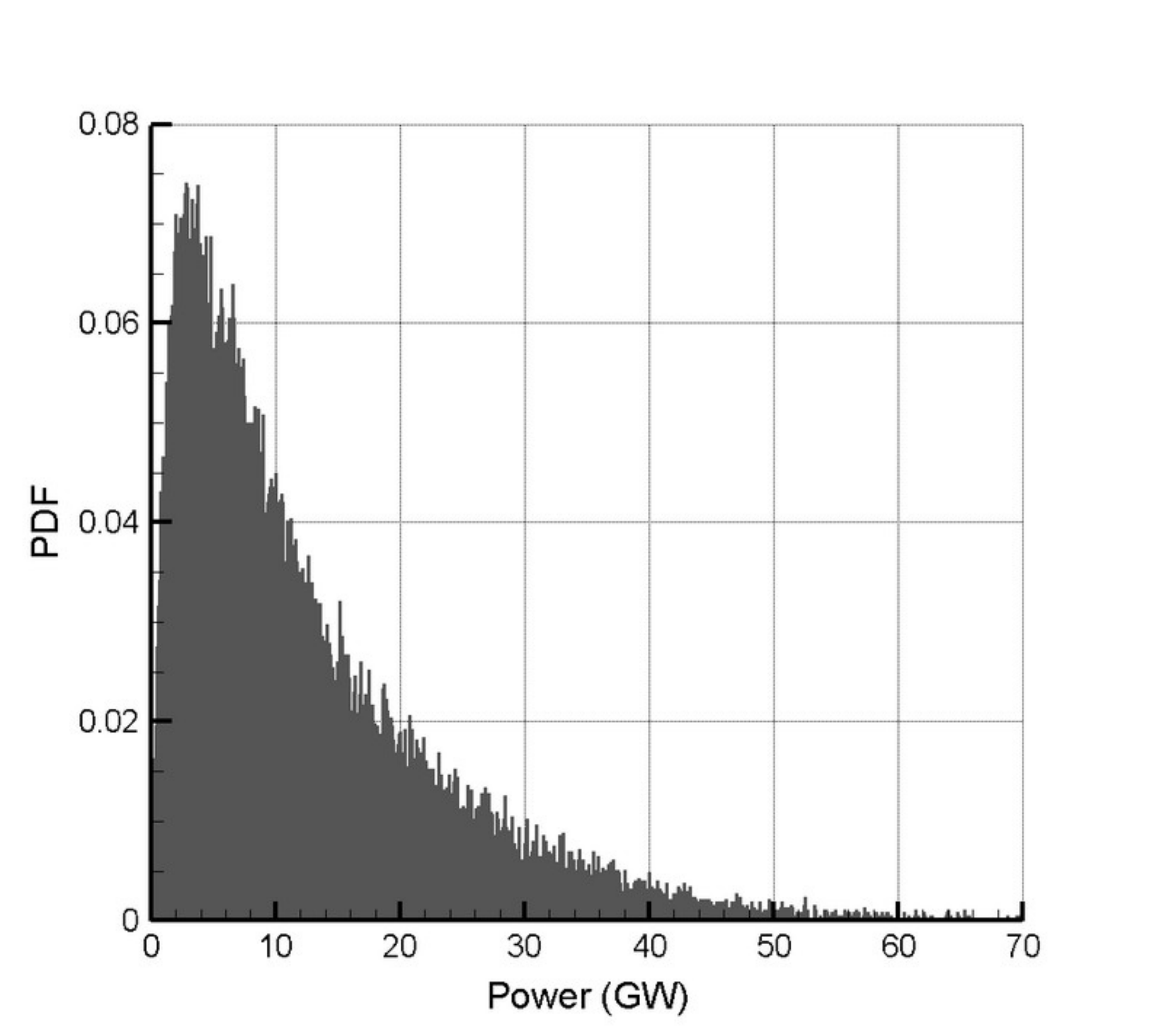}}
\subfigure[$\epsilon = 0.125$, $\sigma_0 = 0.001$]{\includegraphics[width=0.33\textwidth]{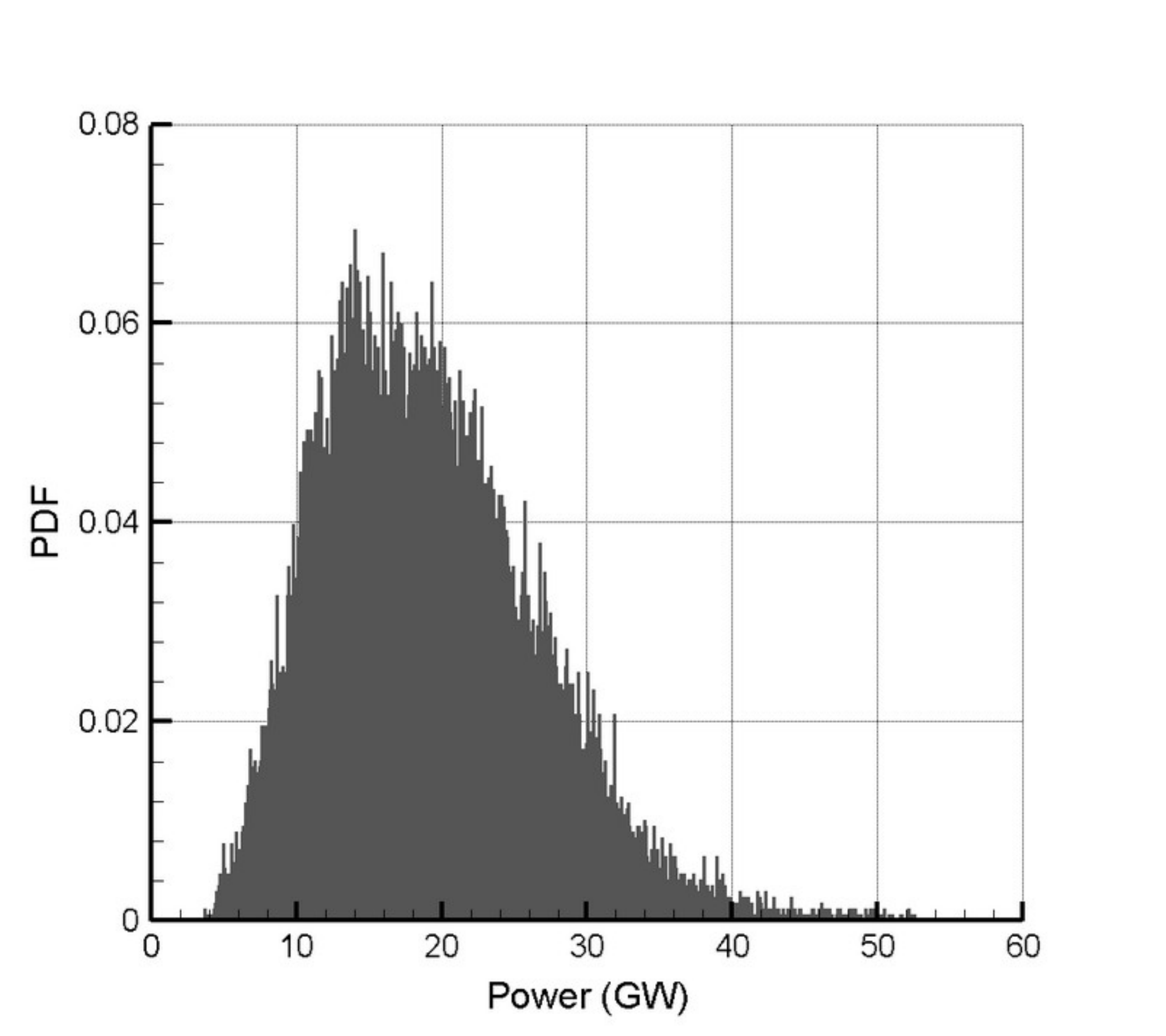}}
\subfigure[$\epsilon = 0.25$, $\sigma_0 = 0.001$]{\includegraphics[width=0.33\textwidth]{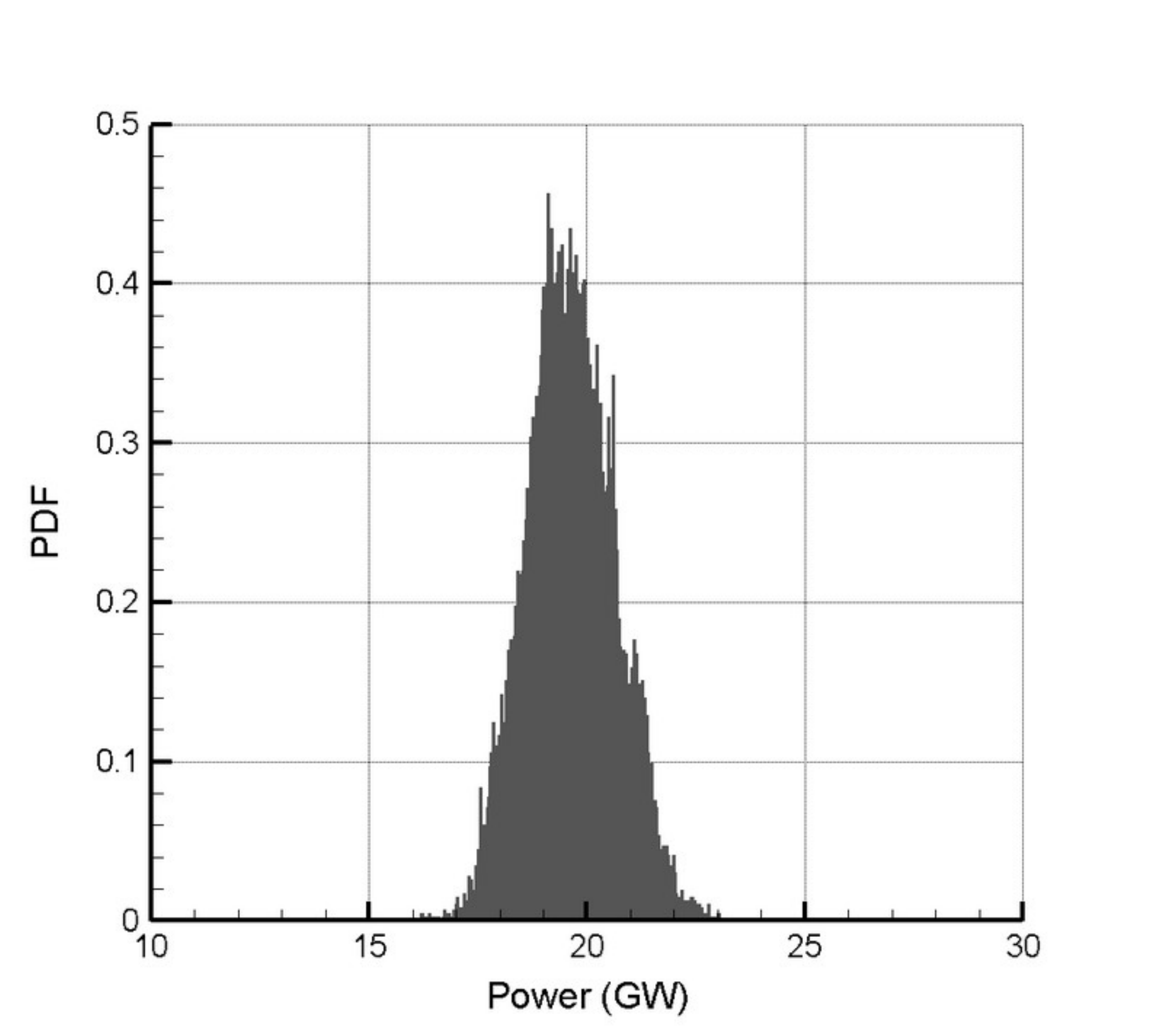}}
}\\
\mbox{
\subfigure[$\epsilon = 0.0625$, $\sigma_0 = 0.0008$]{\includegraphics[width=0.33\textwidth]{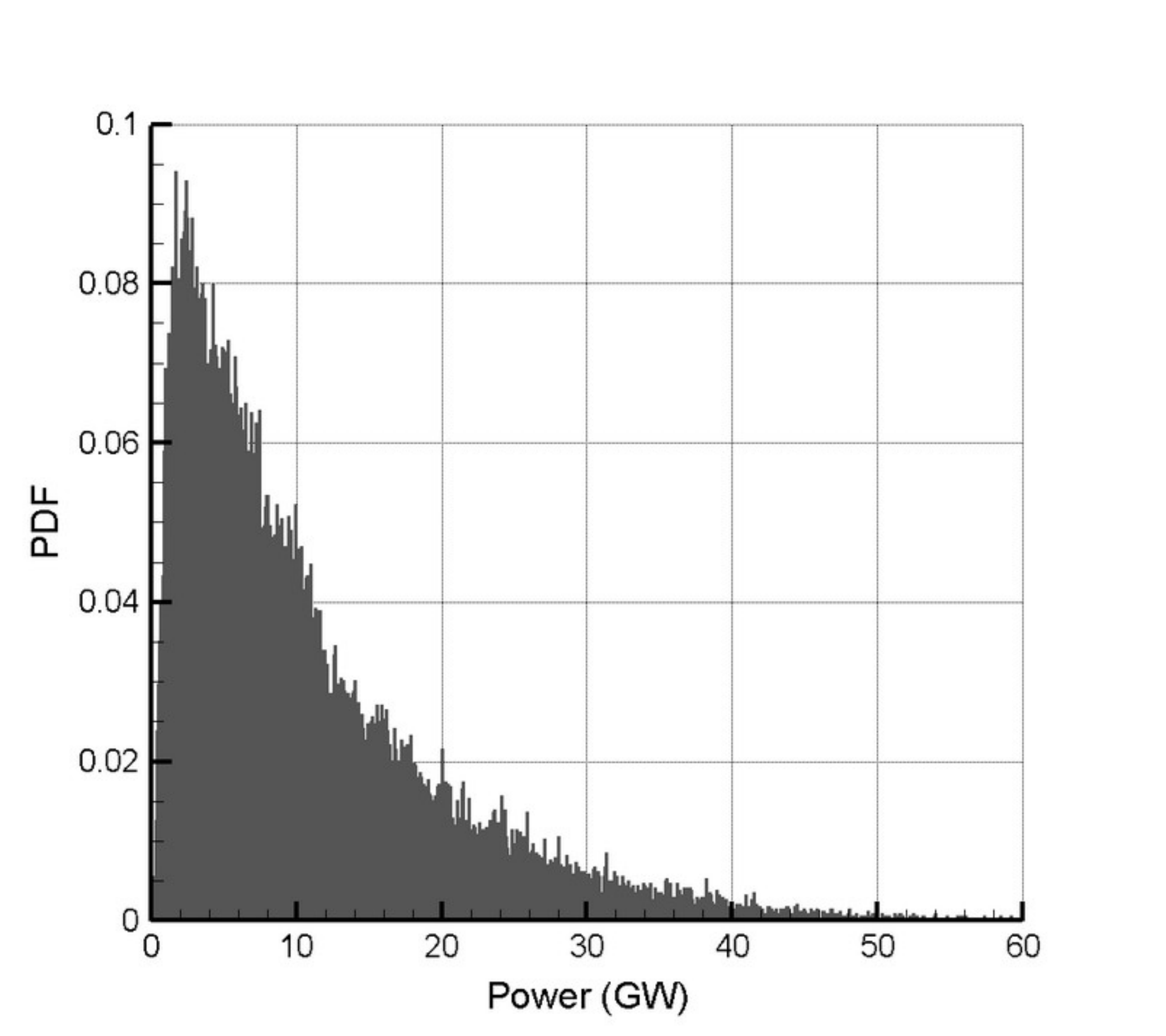}}
\subfigure[$\epsilon = 0.0625$, $\sigma_0 = 0.0012$]{\includegraphics[width=0.33\textwidth]{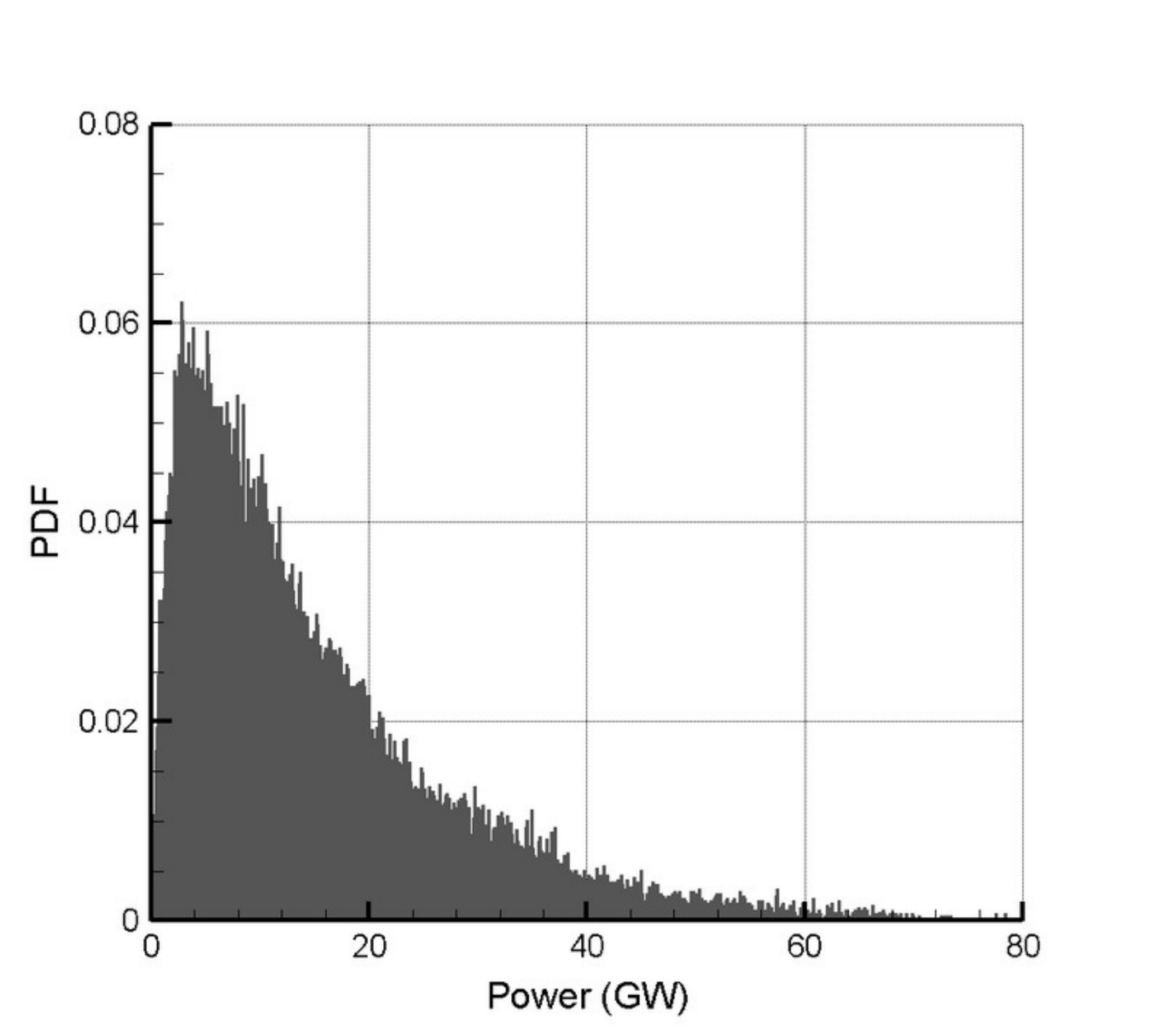}}
\subfigure[$\epsilon = 0.0625$, $\sigma_0 = 0.0016$]{\includegraphics[width=0.33\textwidth]{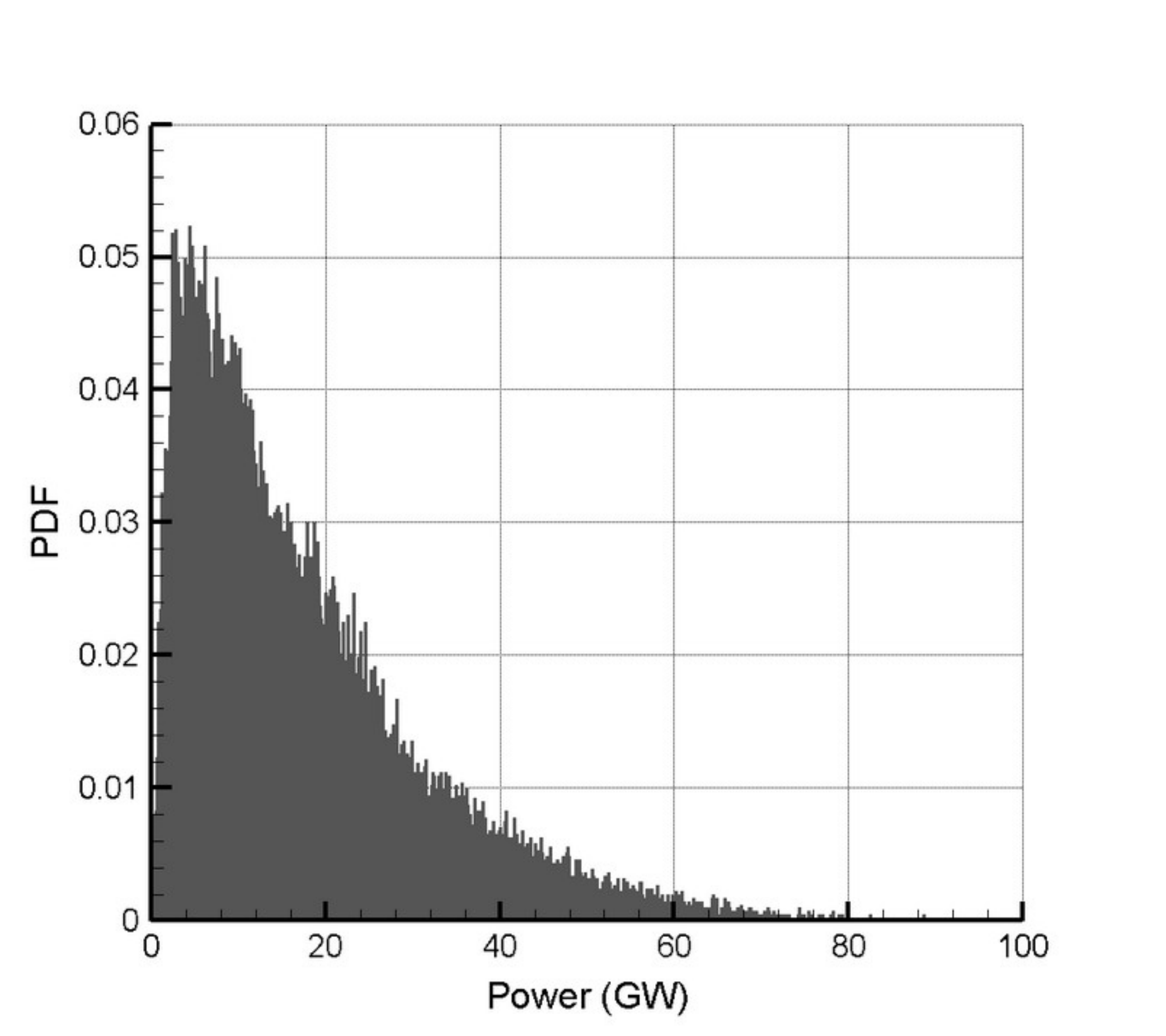}}
}
\caption{Probability density function (PDF) of the extracted power from the ocean Basin II.}
\label{f:pdf-p-2}
\end{figure}

\begin{figure}
\centering
\mbox{
\subfigure[$\epsilon = 0.05$, $\sigma_0 = 0.001$]{\includegraphics[width=0.33\textwidth]{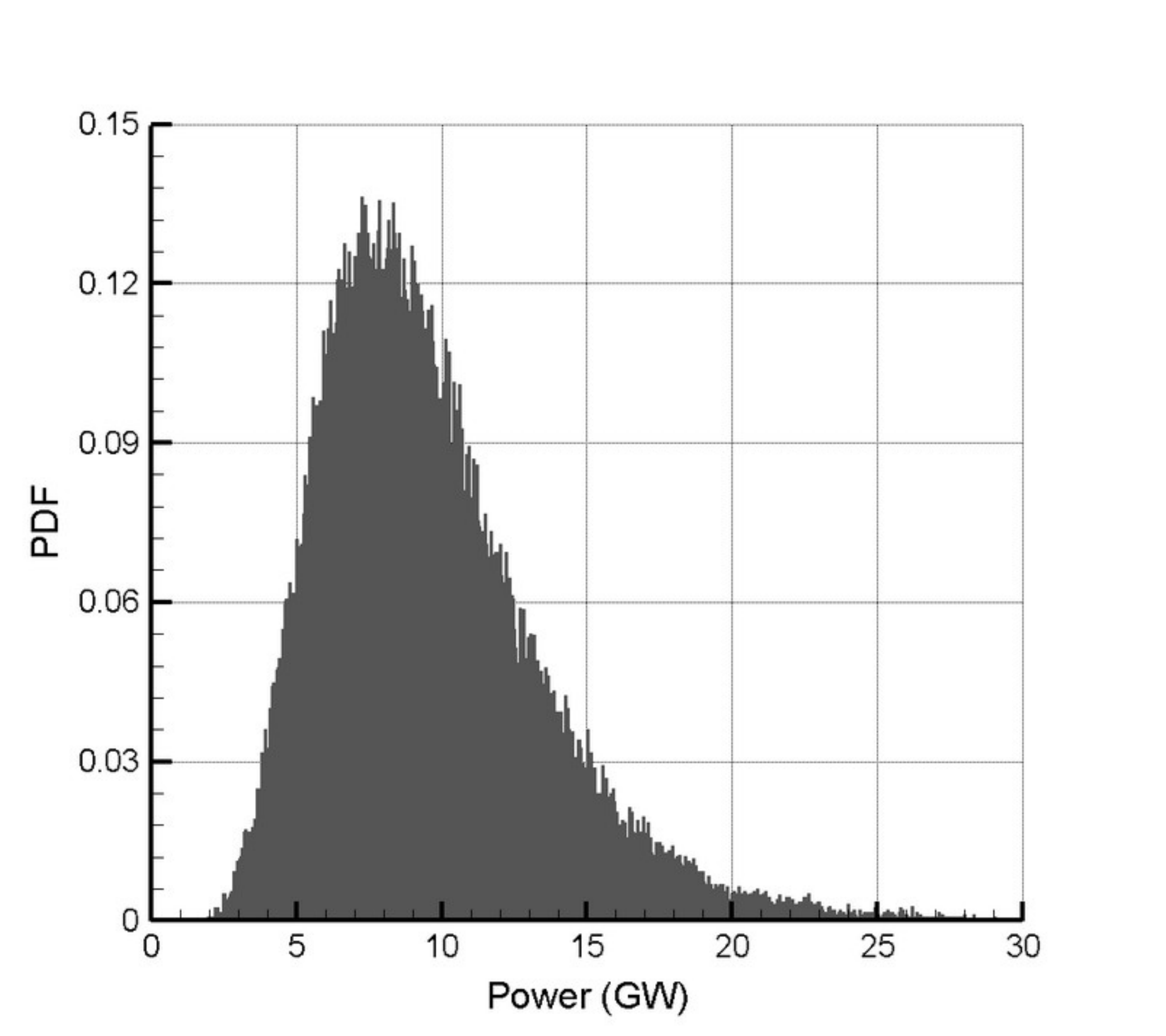}}
\subfigure[$\epsilon = 0.1$, $\sigma_0 = 0.001$]{\includegraphics[width=0.33\textwidth]{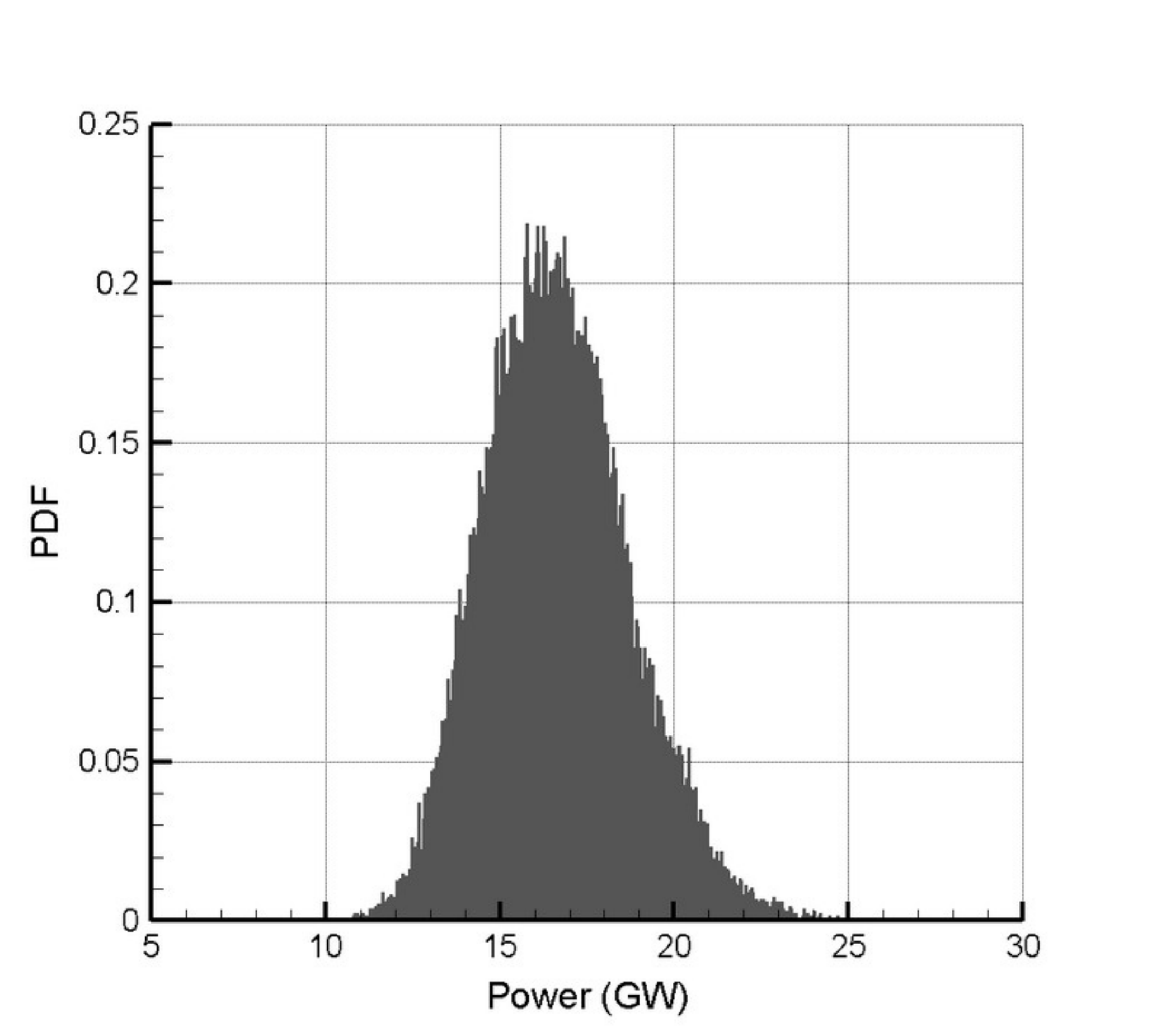}}
\subfigure[$\epsilon = 0.2$, $\sigma_0 = 0.001$]{\includegraphics[width=0.33\textwidth]{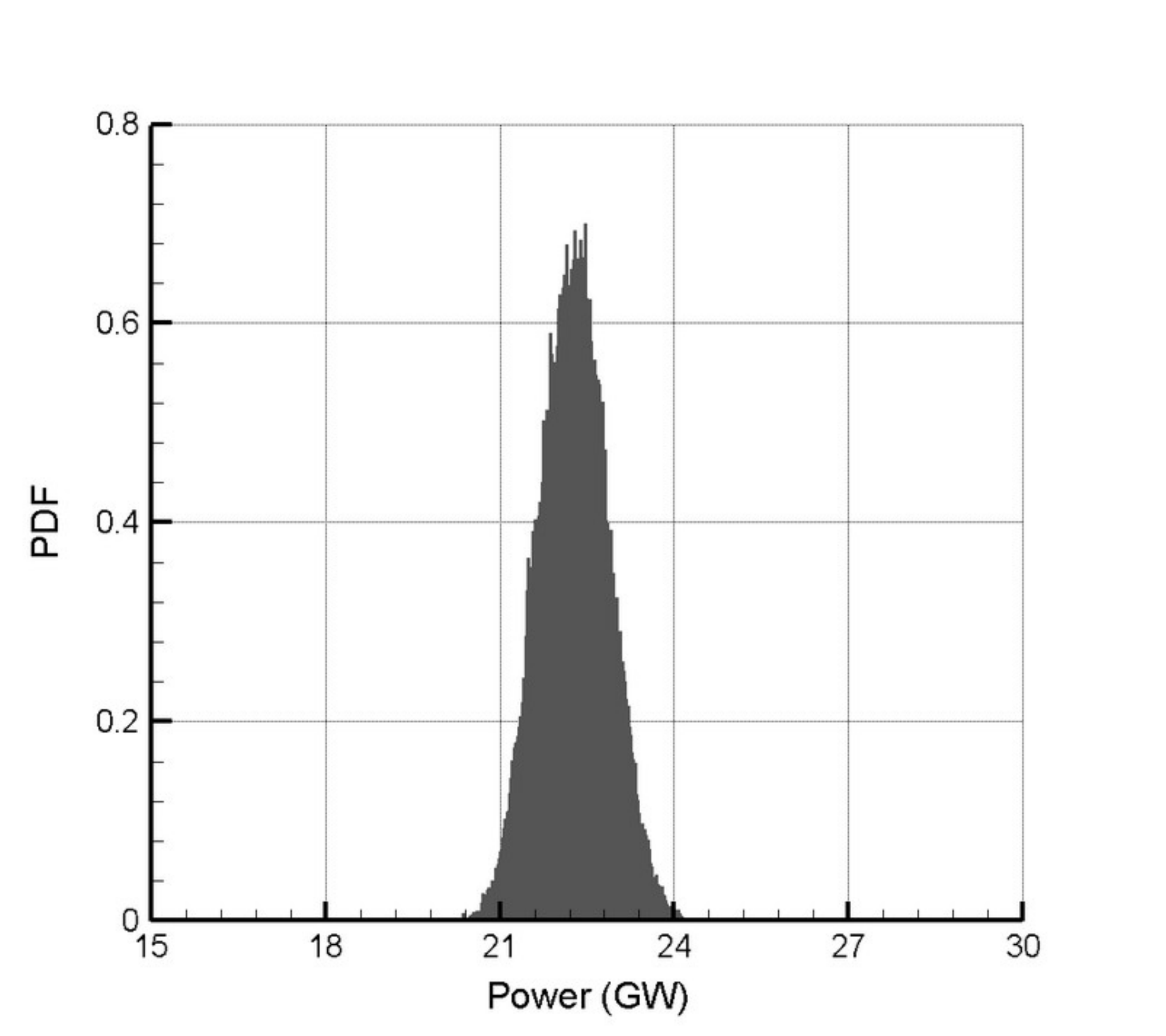}}
}\\
\mbox{
\subfigure[$\epsilon = 0.05$, $\sigma_0 = 0.0008$]{\includegraphics[width=0.33\textwidth]{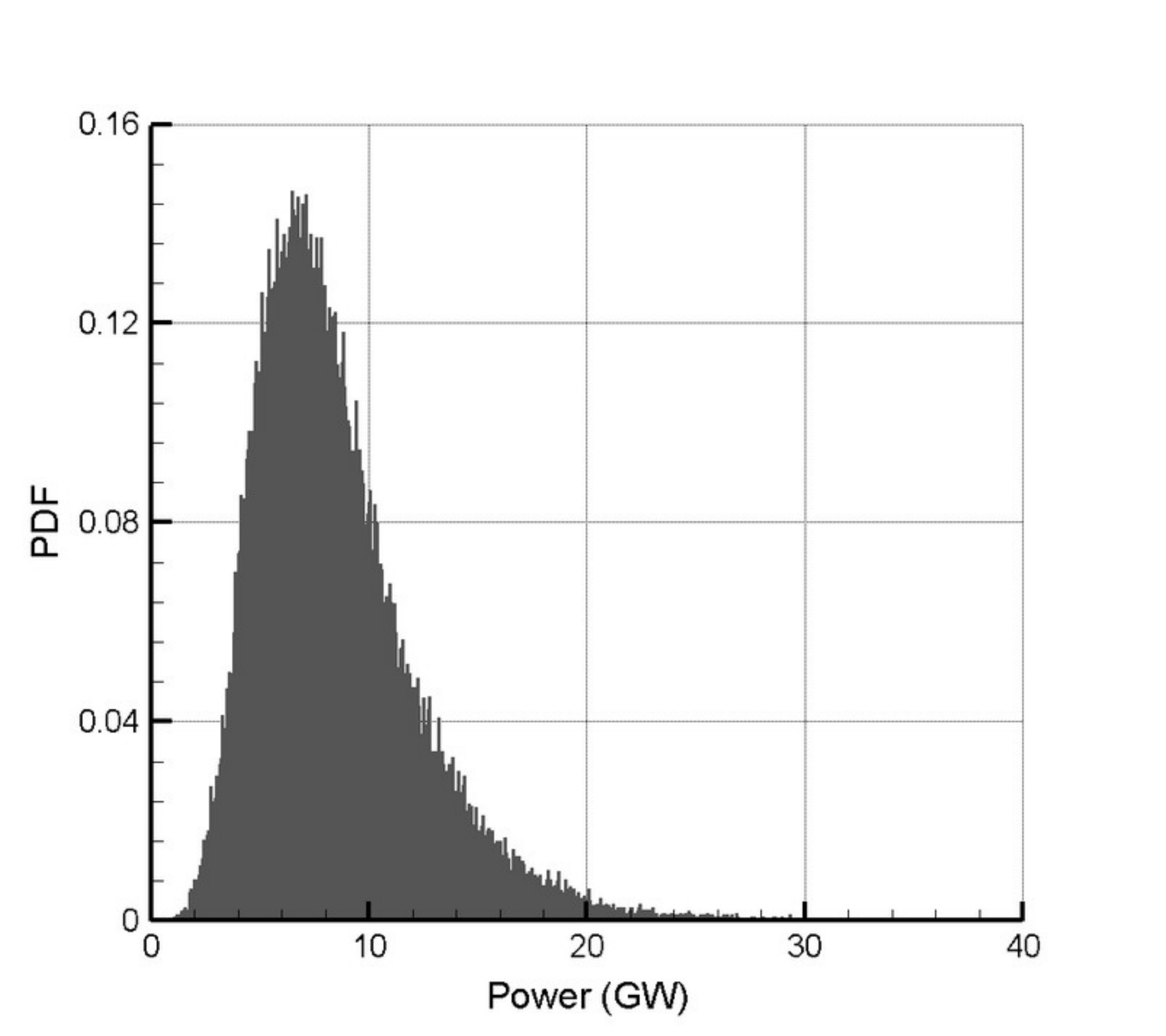}}
\subfigure[$\epsilon = 0.05$, $\sigma_0 = 0.0012$]{\includegraphics[width=0.33\textwidth]{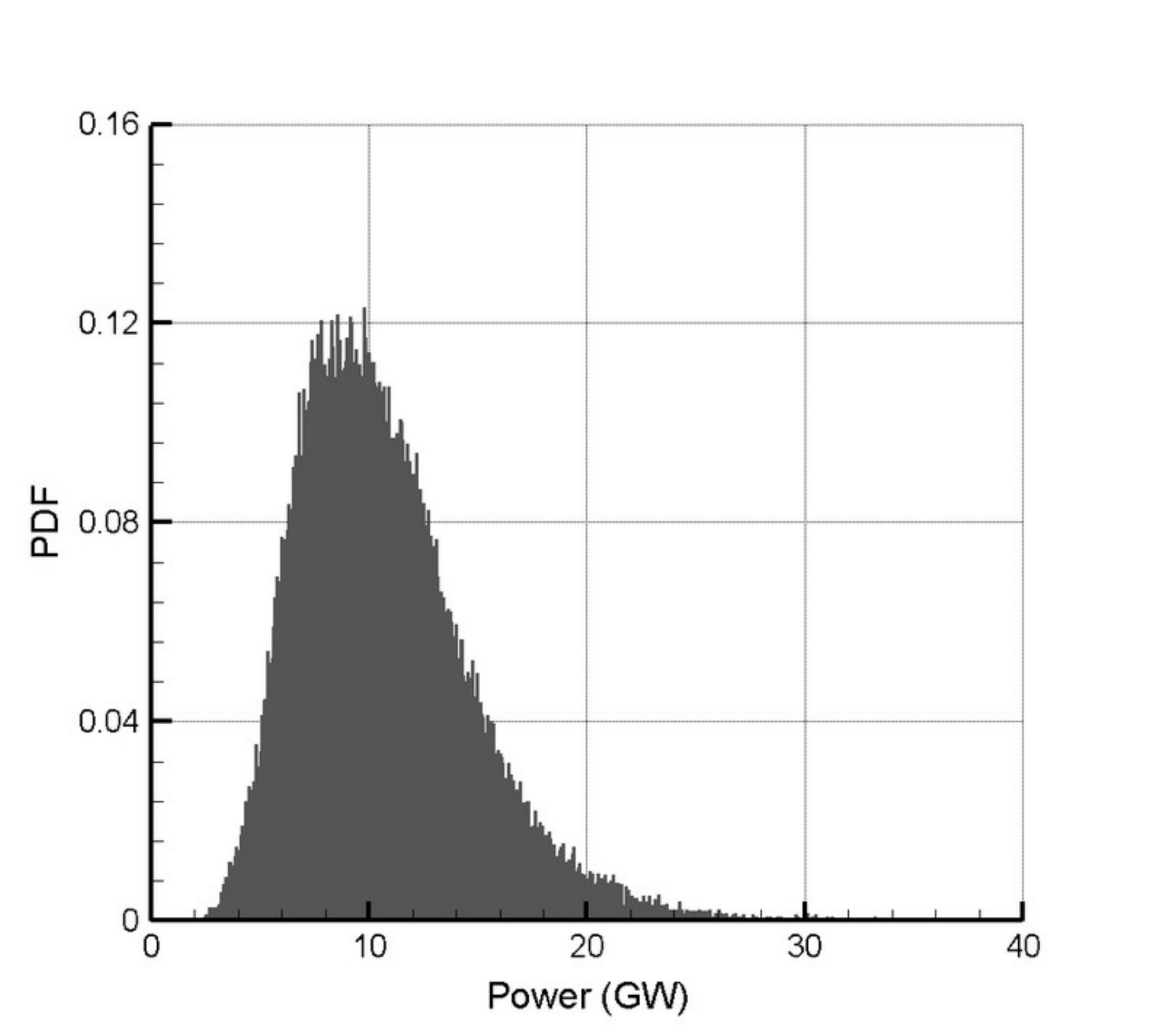}}
\subfigure[$\epsilon = 0.05$, $\sigma_0 = 0.0016$]{\includegraphics[width=0.33\textwidth]{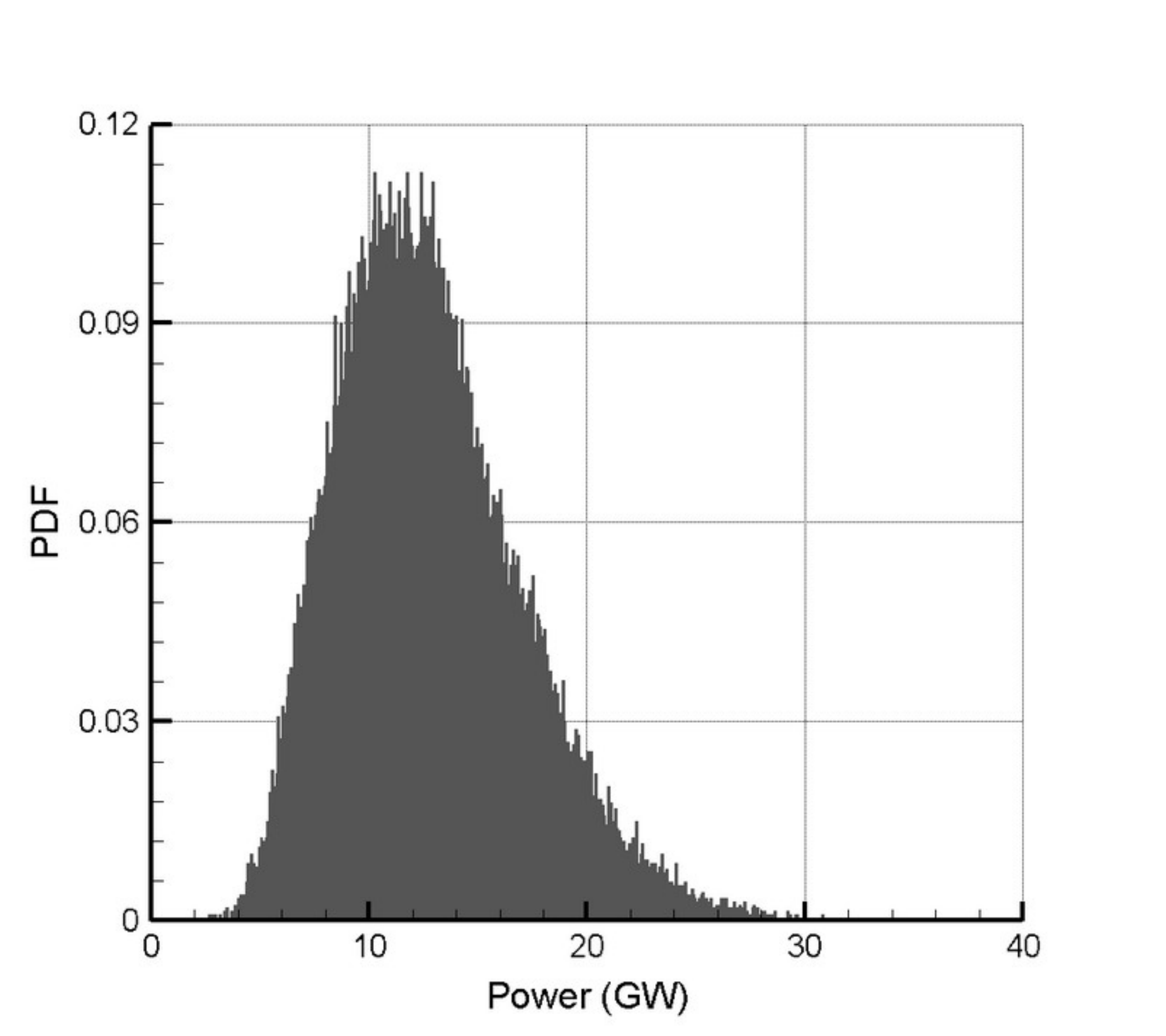}}
}
\caption{Probability density function (PDF) of the extracted power from the ocean Basin III.}
\label{f:pdf-p-3}
\end{figure}

\begin{figure}
\centering
\mbox{
\subfigure[$\epsilon = 0.05$, $\sigma_0 = 0.001$]{\includegraphics[width=0.33\textwidth]{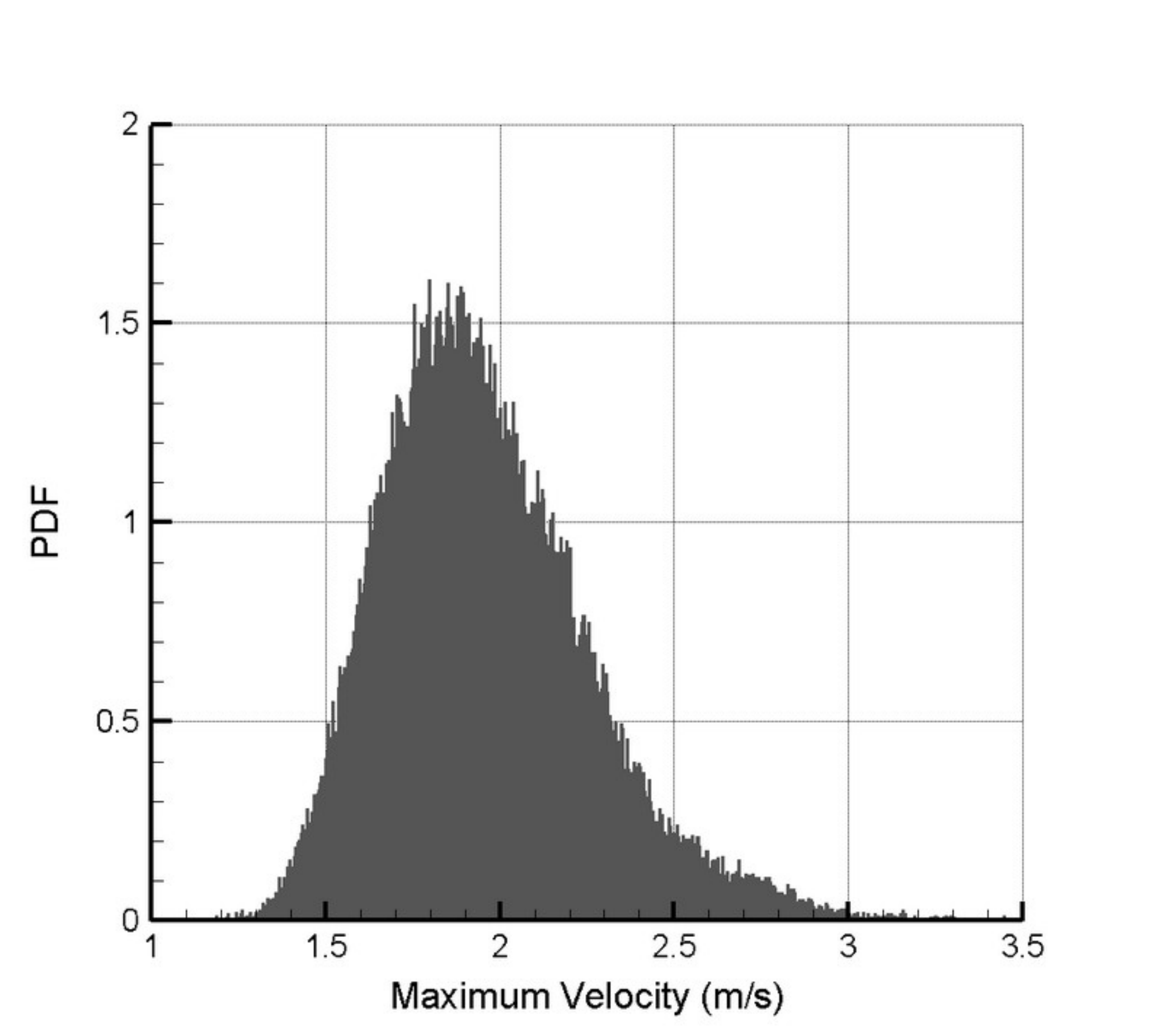}}
\subfigure[$\epsilon = 0.1$, $\sigma_0 = 0.001$]{\includegraphics[width=0.33\textwidth]{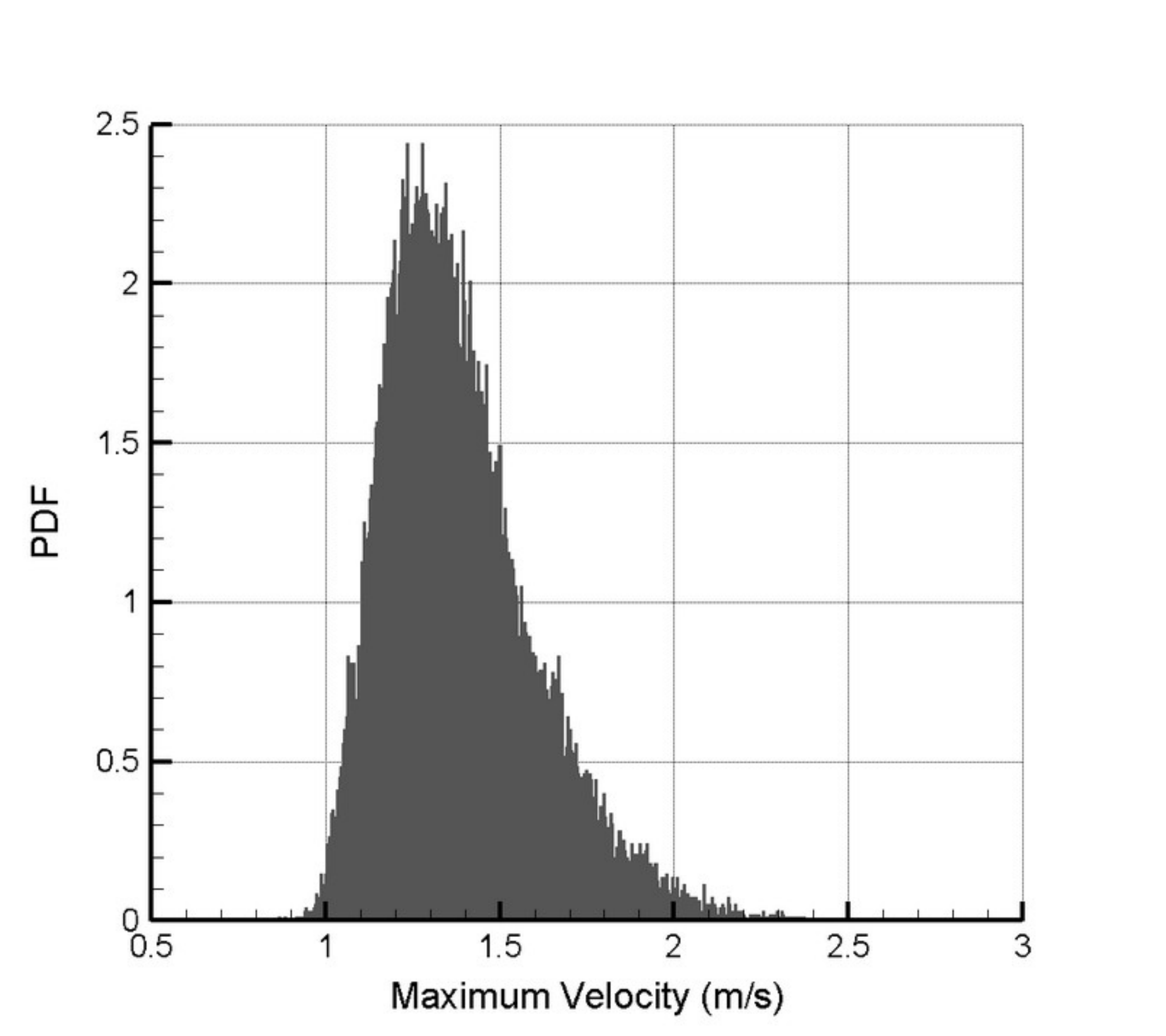}}
\subfigure[$\epsilon = 0.2$, $\sigma_0 = 0.001$]{\includegraphics[width=0.33\textwidth]{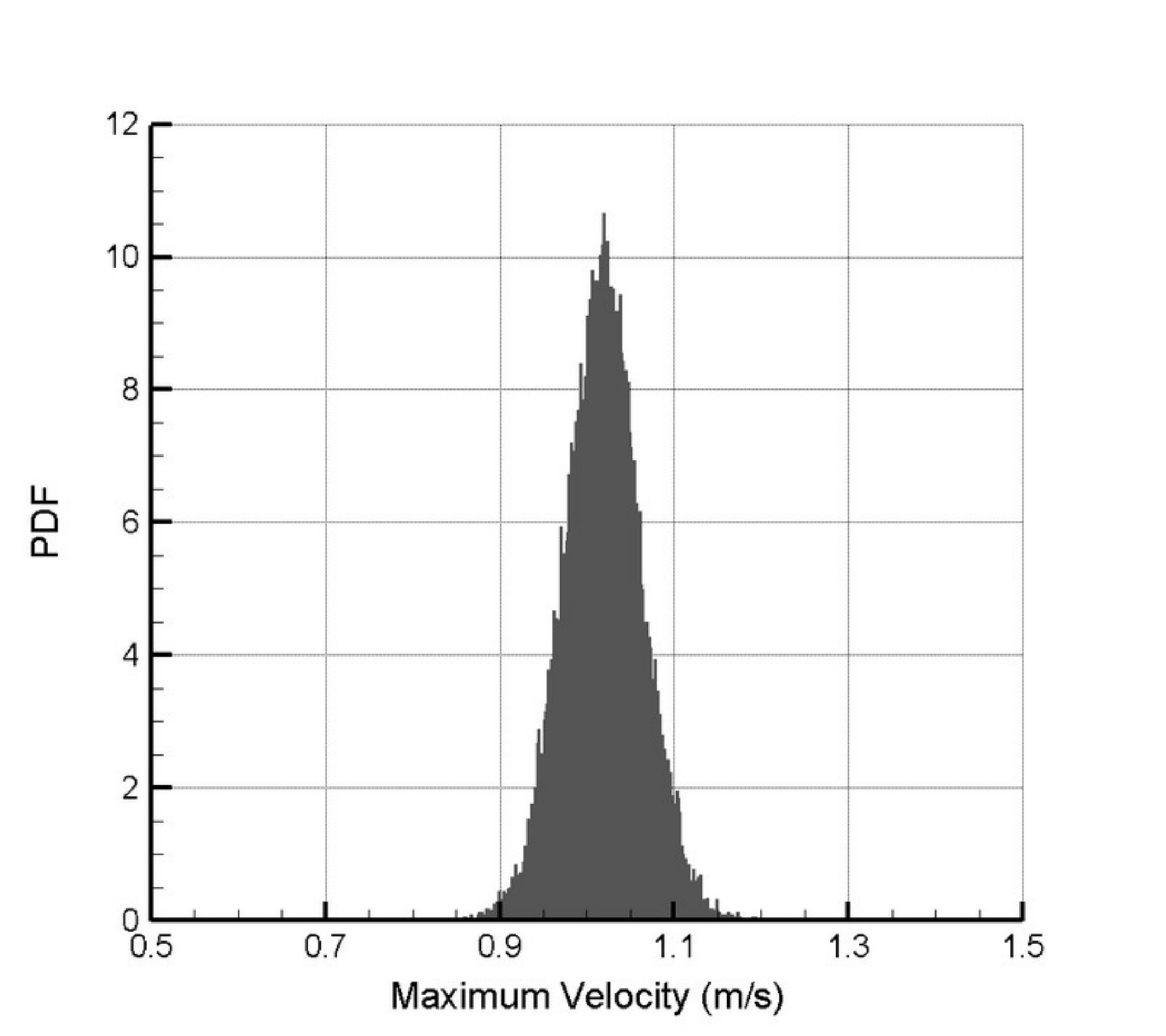}}
}\\
\mbox{
\subfigure[$\epsilon = 0.05$, $\sigma_0 = 0.0008$]{\includegraphics[width=0.33\textwidth]{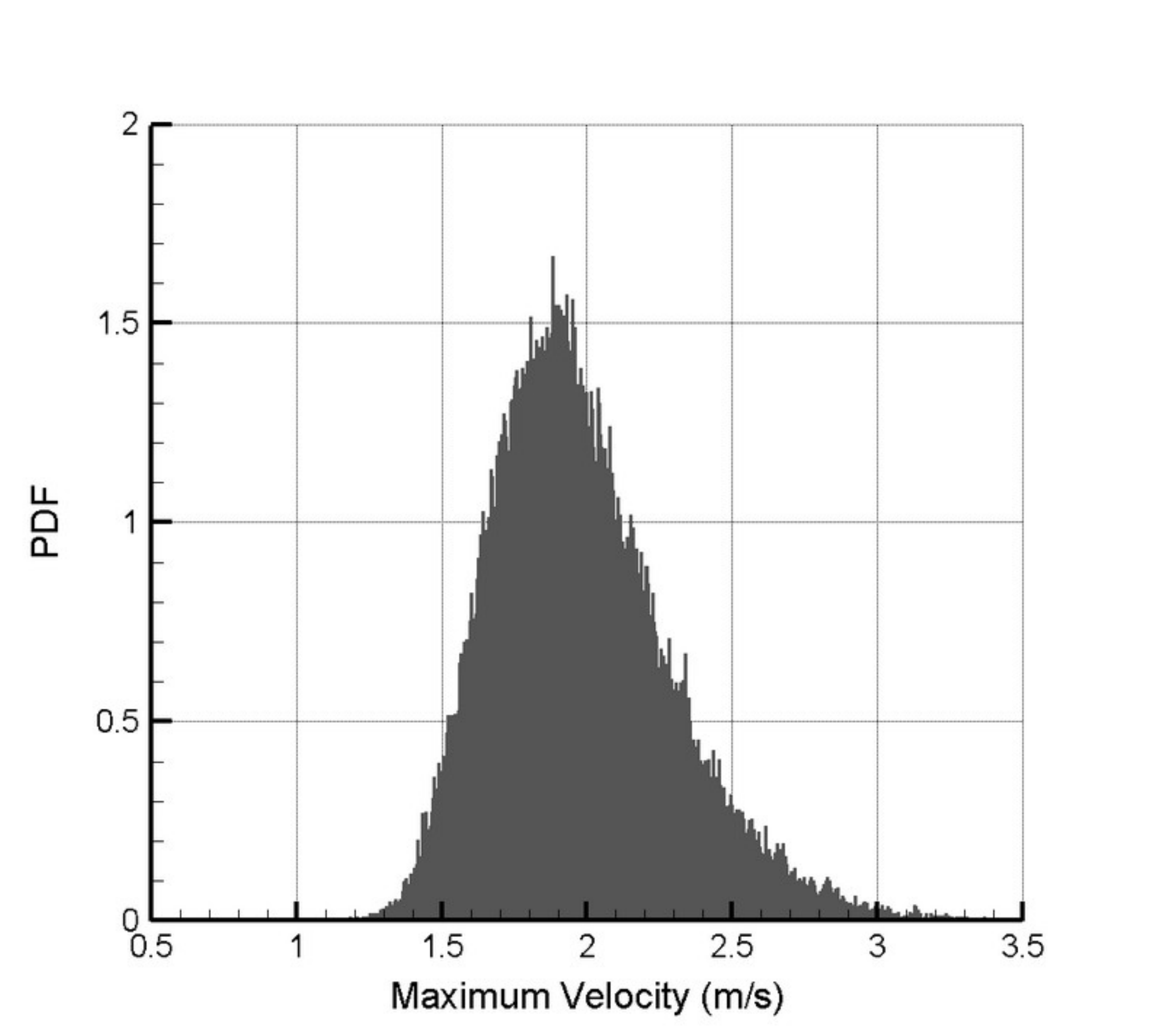}}
\subfigure[$\epsilon = 0.05$, $\sigma_0 = 0.0012$]{\includegraphics[width=0.33\textwidth]{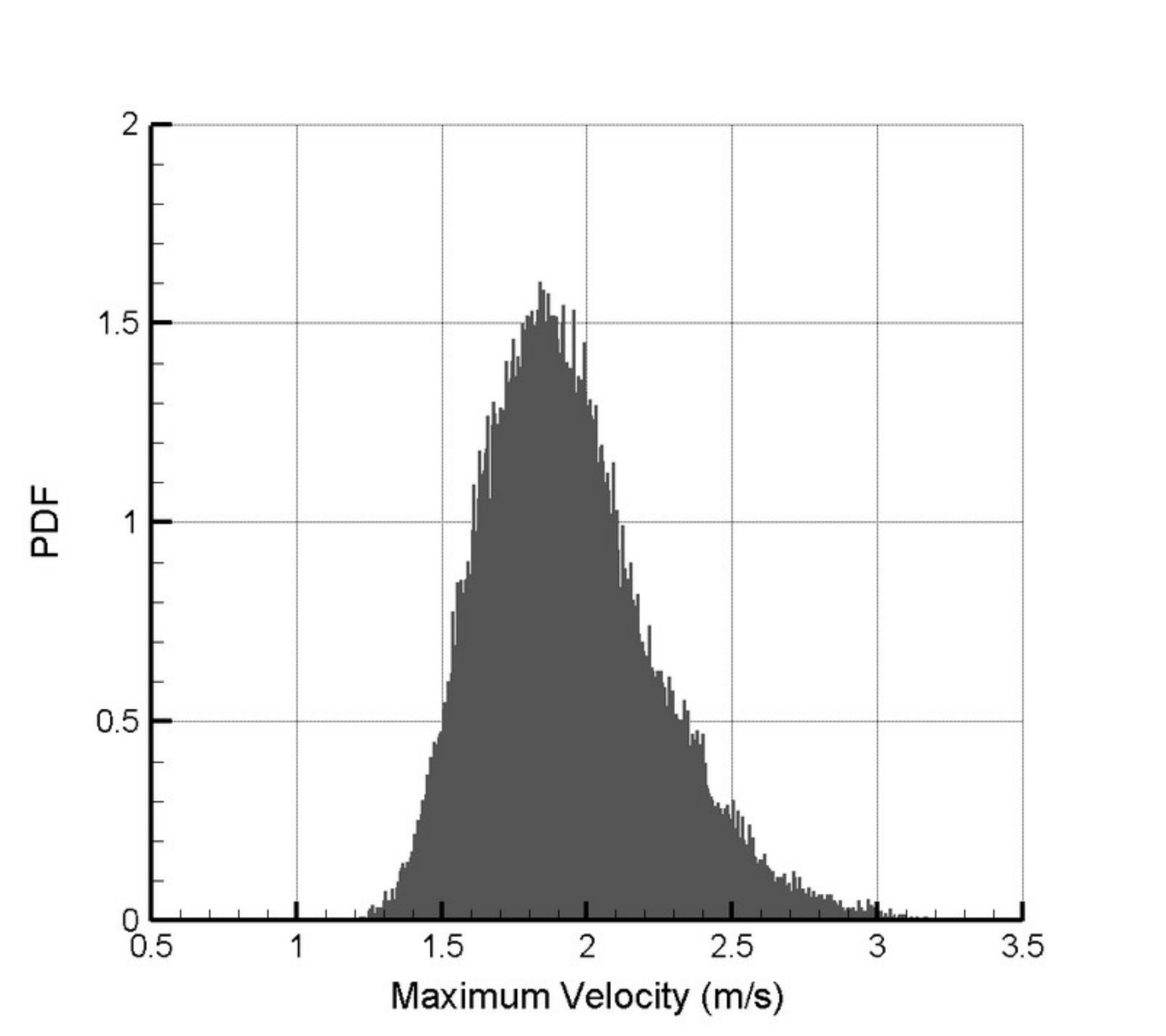}}
\subfigure[$\epsilon = 0.05$, $\sigma_0 = 0.0016$]{\includegraphics[width=0.33\textwidth]{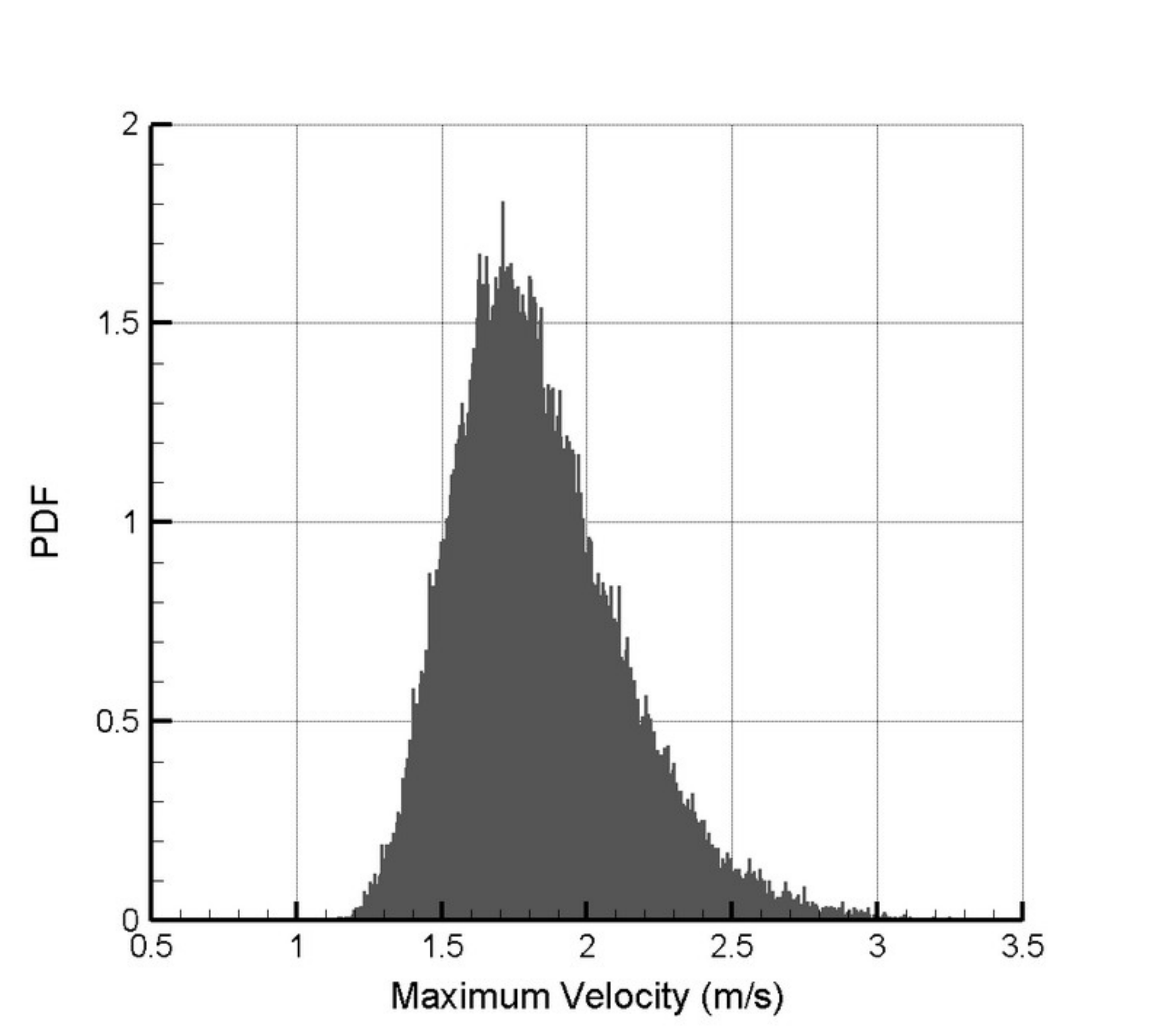}}
}
\caption{Probability density function (PDF) of the maximum flow velocity in the ocean Basin I.}
\label{f:pdf-v-1}
\end{figure}

\begin{figure}
\centering
\mbox{
\subfigure[$\epsilon = 0.0625$, $\sigma_0 = 0.001$]{\includegraphics[width=0.33\textwidth]{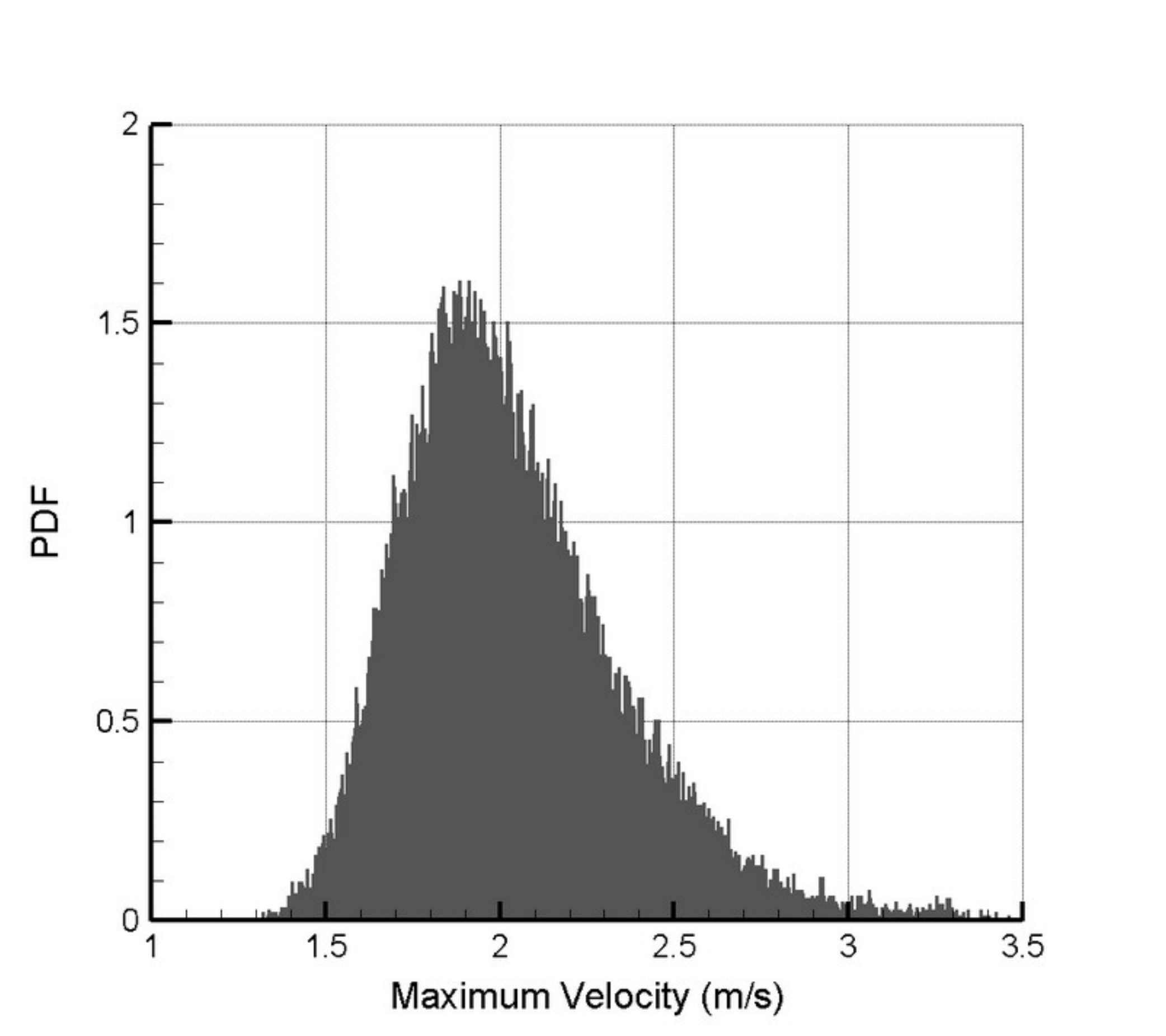}}
\subfigure[$\epsilon = 0.125$, $\sigma_0 = 0.001$]{\includegraphics[width=0.33\textwidth]{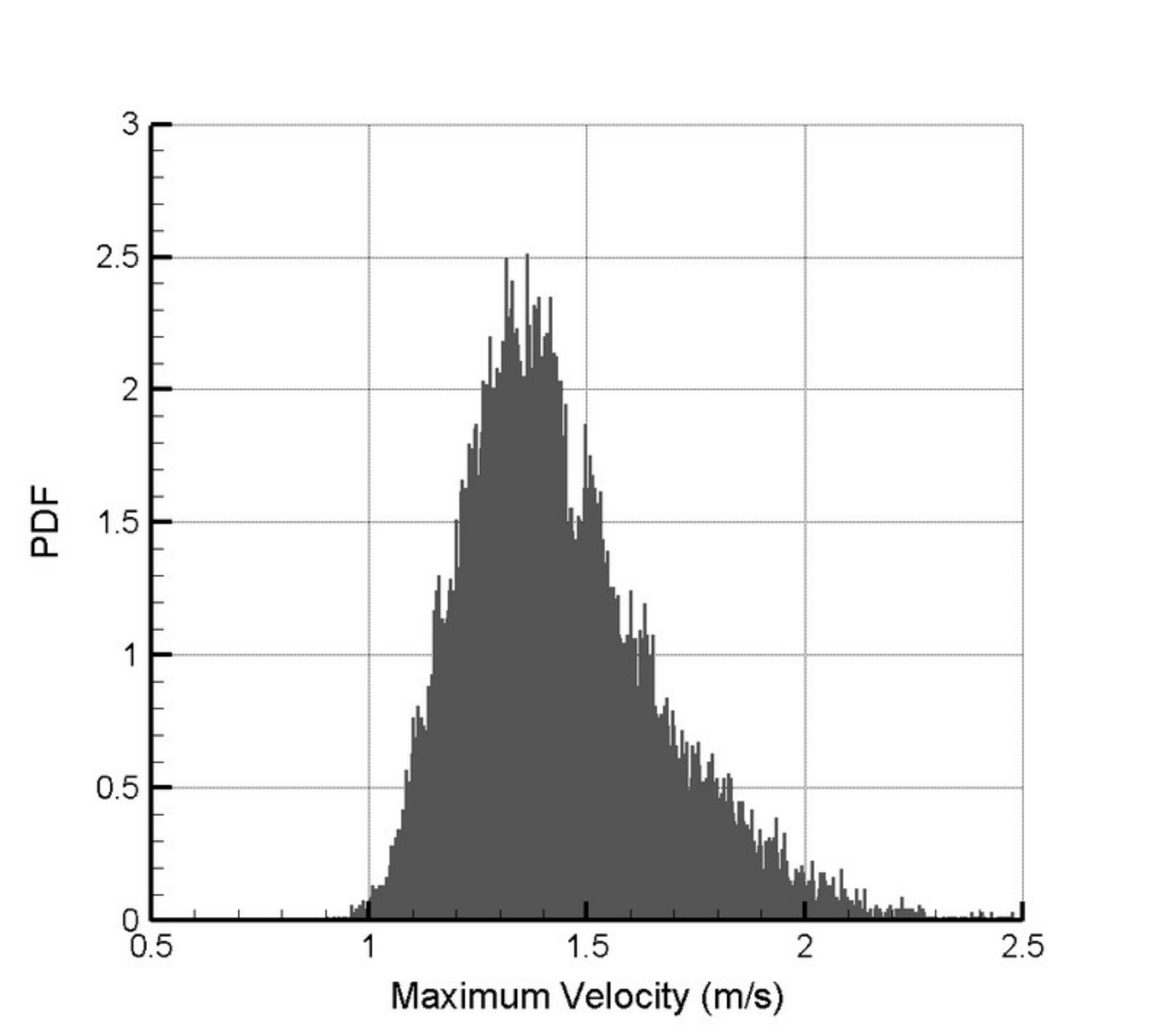}}
\subfigure[$\epsilon = 0.25$, $\sigma_0 = 0.001$]{\includegraphics[width=0.33\textwidth]{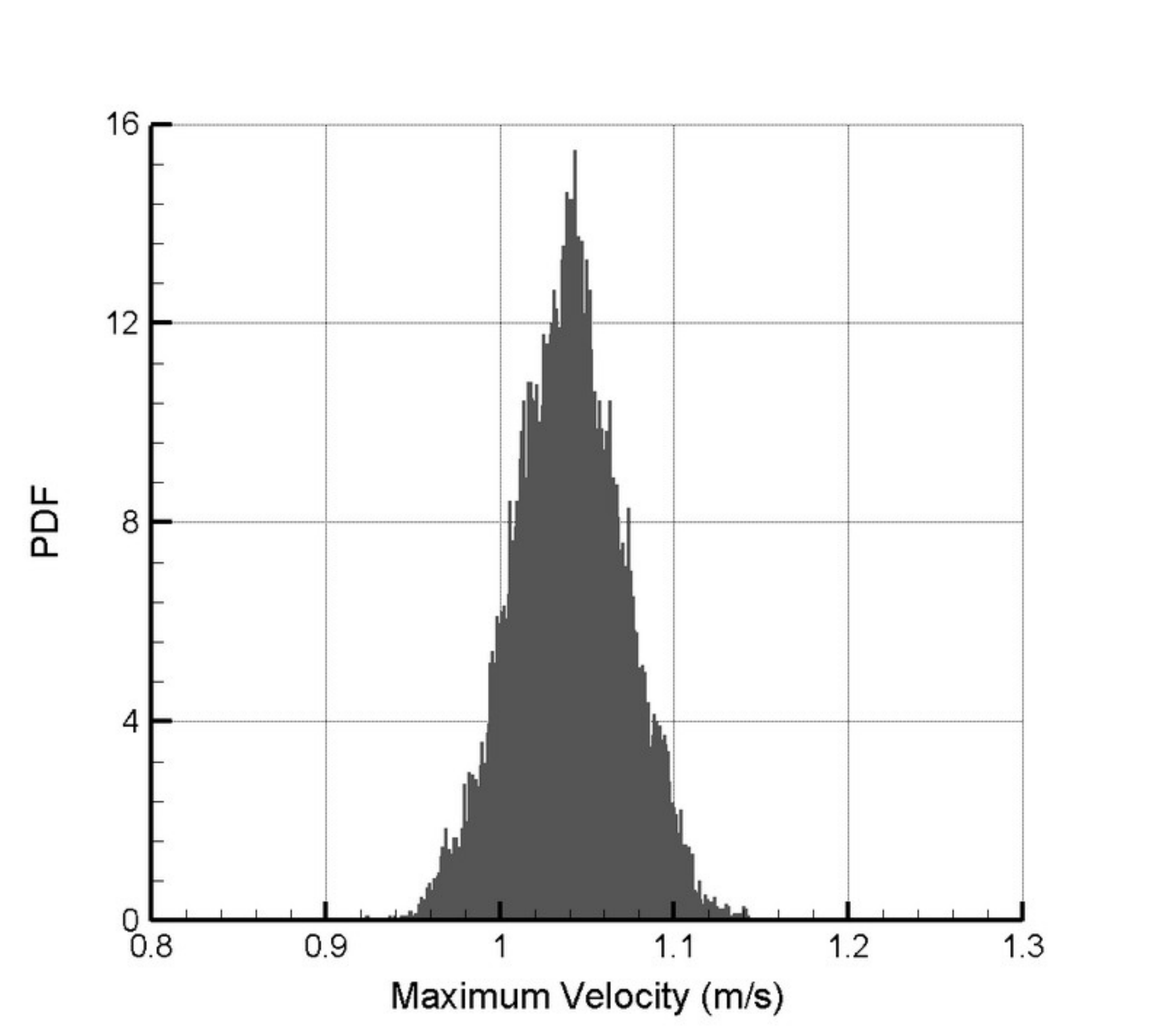}}
}\\
\mbox{
\subfigure[$\epsilon = 0.0625$, $\sigma_0 = 0.0008$]{\includegraphics[width=0.33\textwidth]{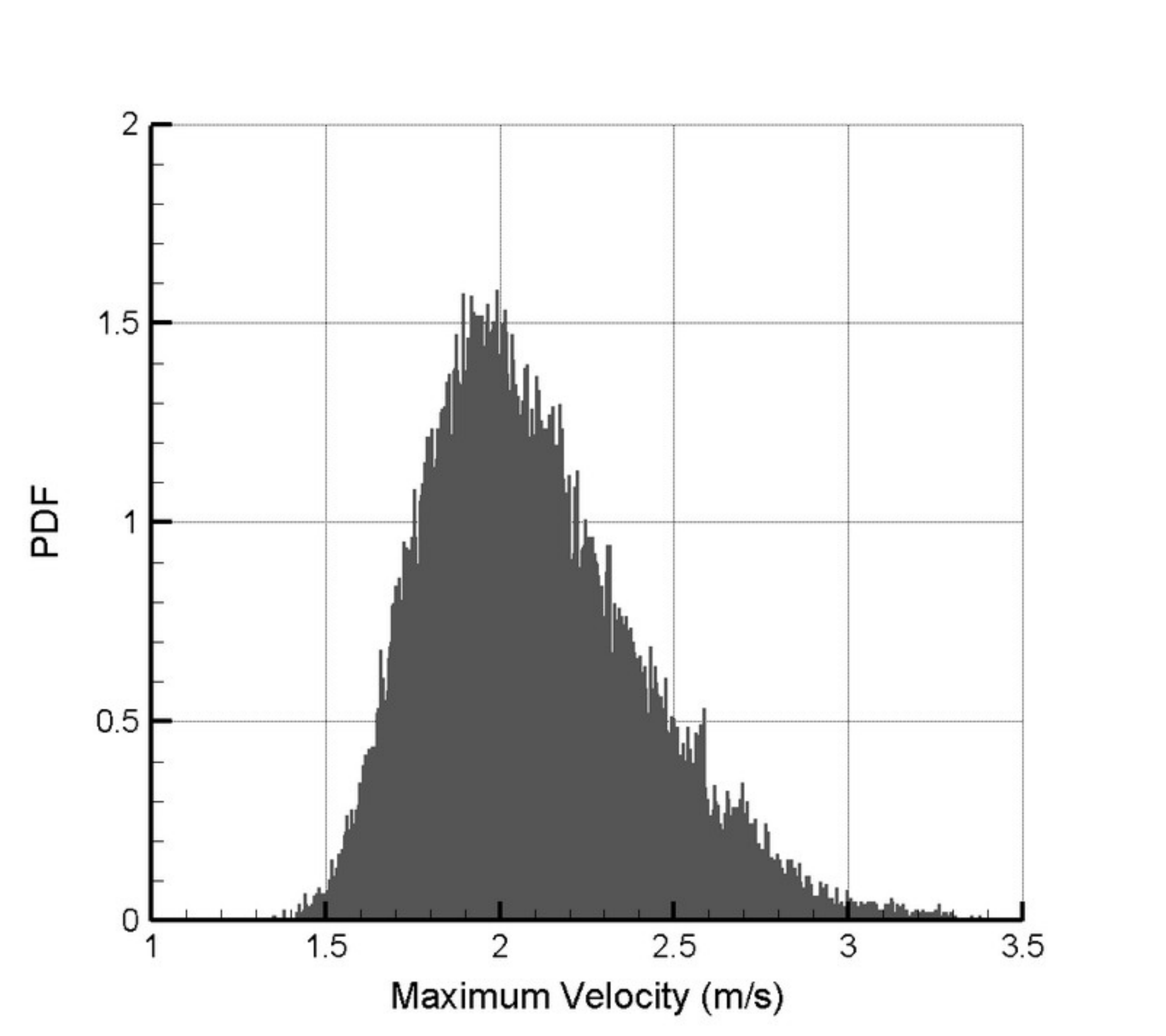}}
\subfigure[$\epsilon = 0.0625$, $\sigma_0 = 0.0012$]{\includegraphics[width=0.33\textwidth]{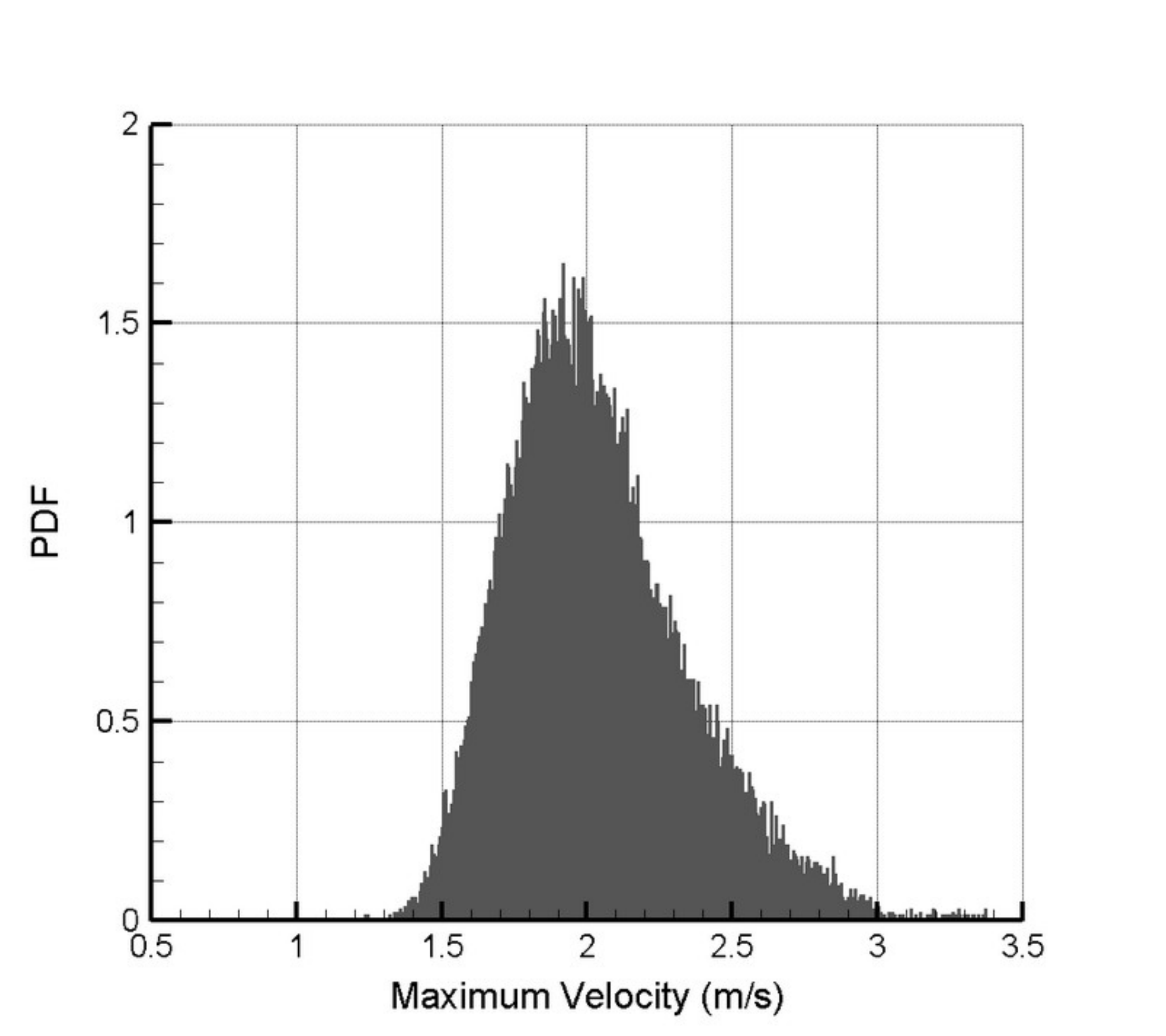}}
\subfigure[$\epsilon = 0.0625$, $\sigma_0 = 0.0016$]{\includegraphics[width=0.33\textwidth]{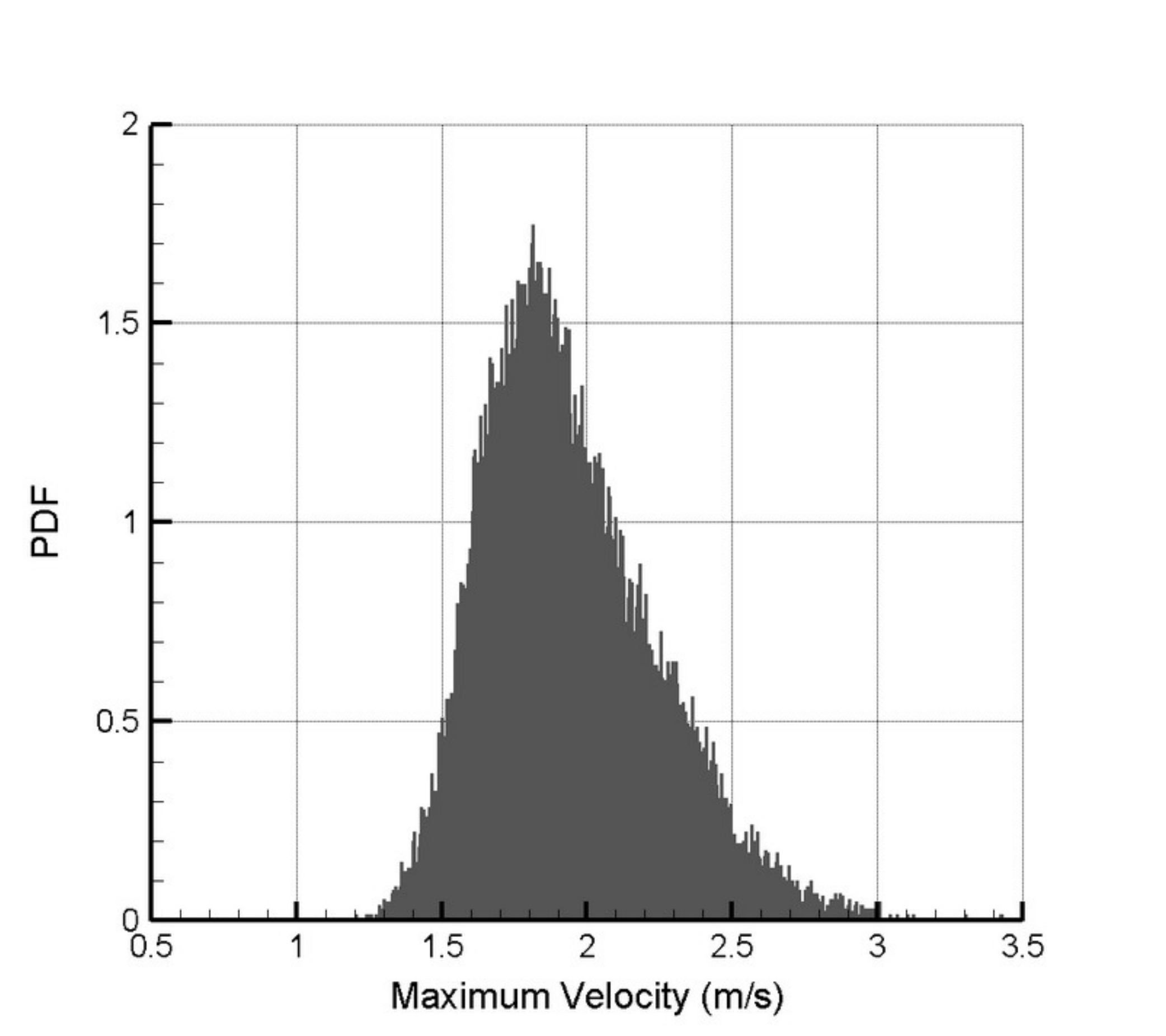}}
}
\caption{Probability density function (PDF) of the maximum flow velocity in the ocean Basin II.}
\label{f:pdf-v-2}
\end{figure}

\begin{figure}
\centering
\mbox{
\subfigure[$\epsilon = 0.05$, $\sigma_0 = 0.001$]{\includegraphics[width=0.33\textwidth]{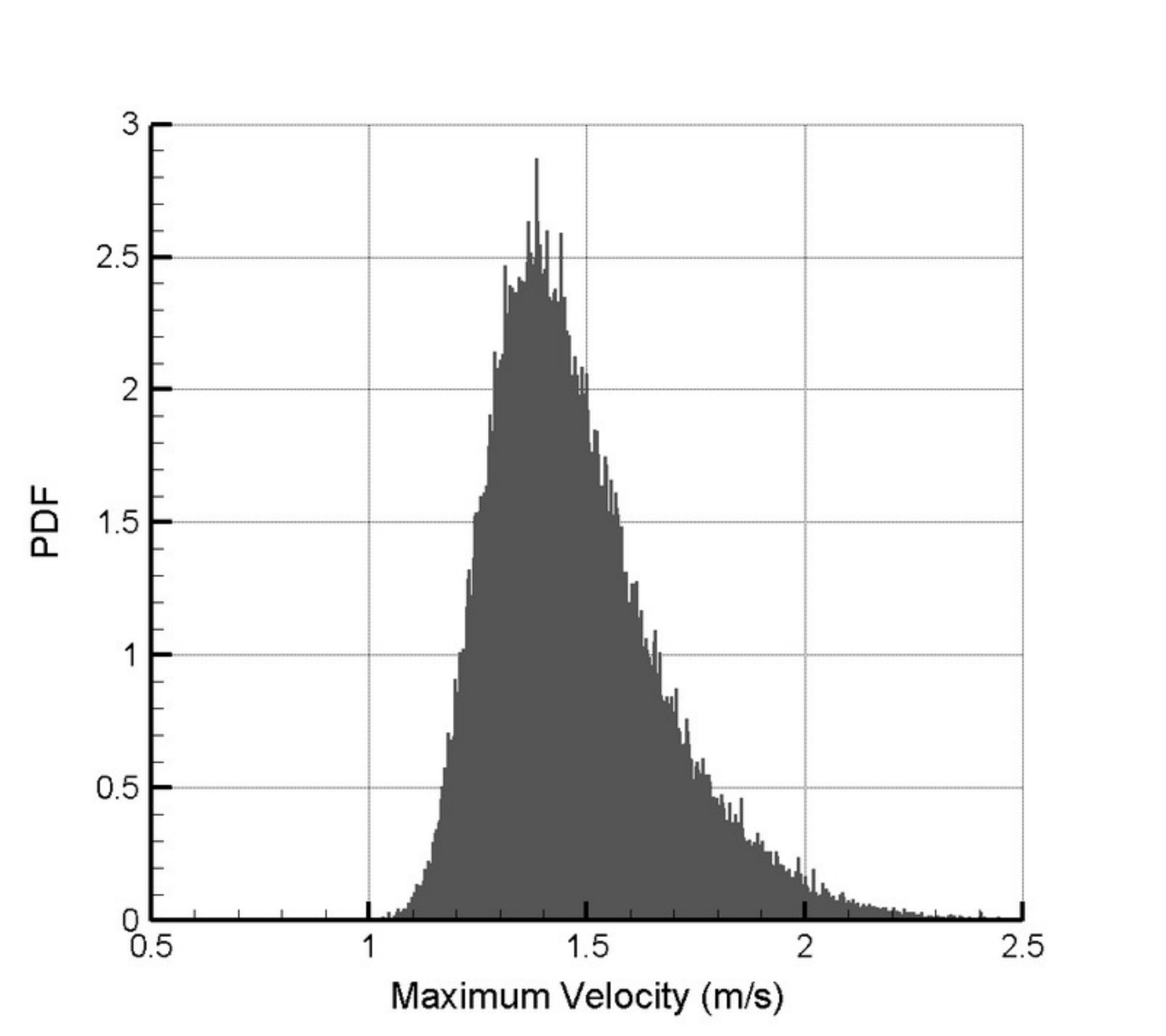}}
\subfigure[$\epsilon = 0.1$, $\sigma_0 = 0.001$]{\includegraphics[width=0.33\textwidth]{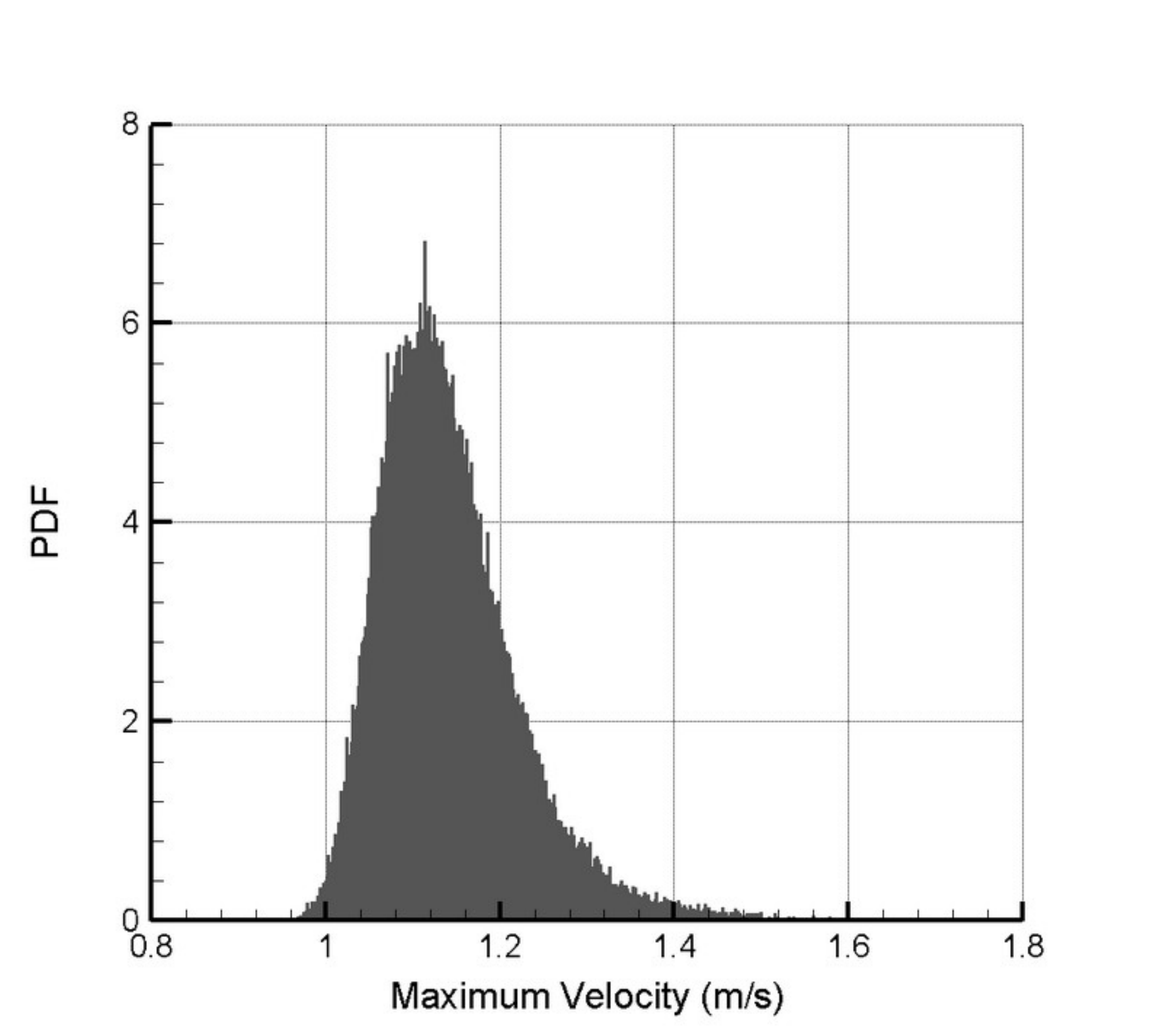}}
\subfigure[$\epsilon = 0.2$, $\sigma_0 = 0.001$]{\includegraphics[width=0.33\textwidth]{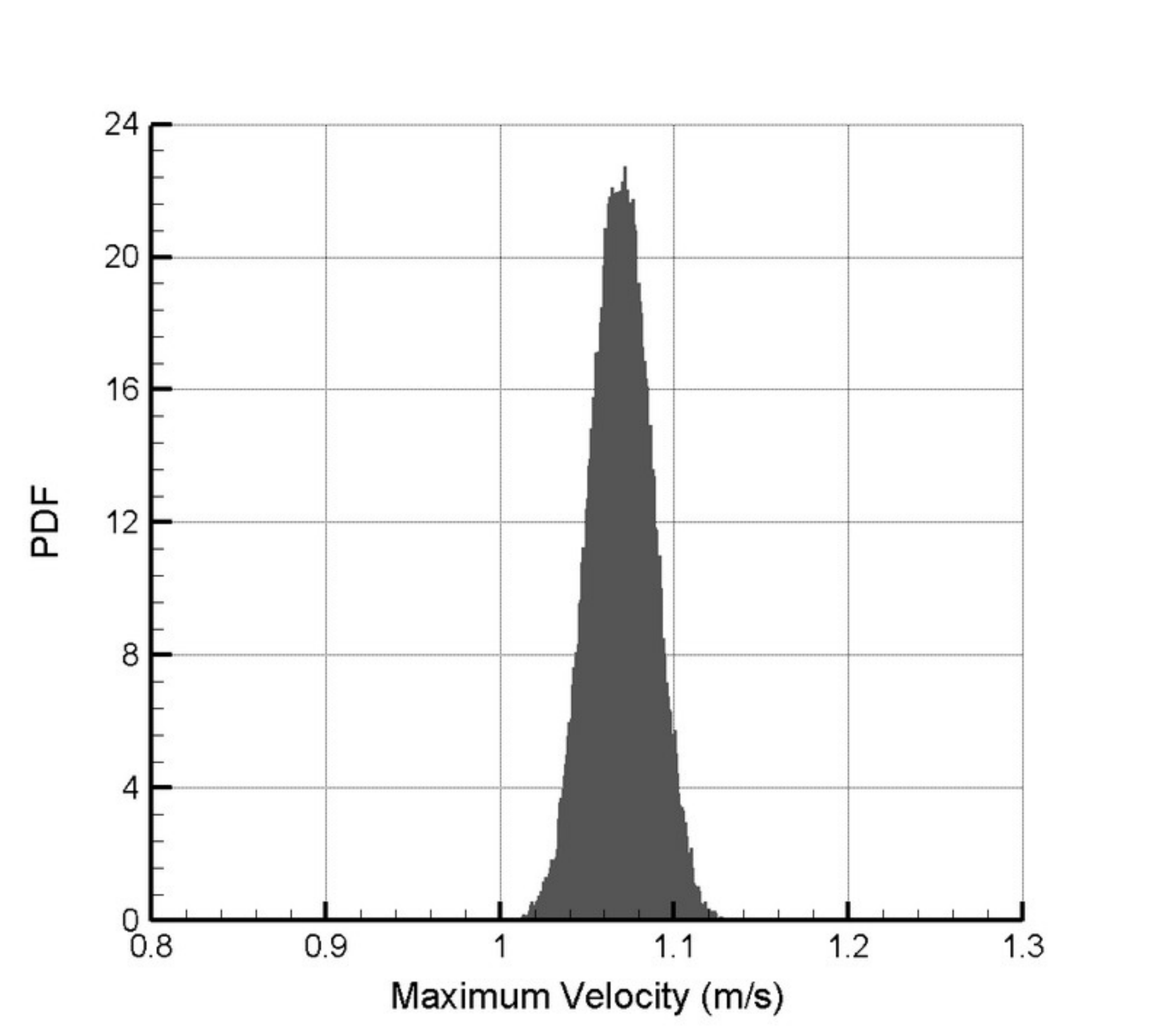}}
}\\
\mbox{
\subfigure[$\epsilon = 0.05$, $\sigma_0 = 0.0008$]{\includegraphics[width=0.33\textwidth]{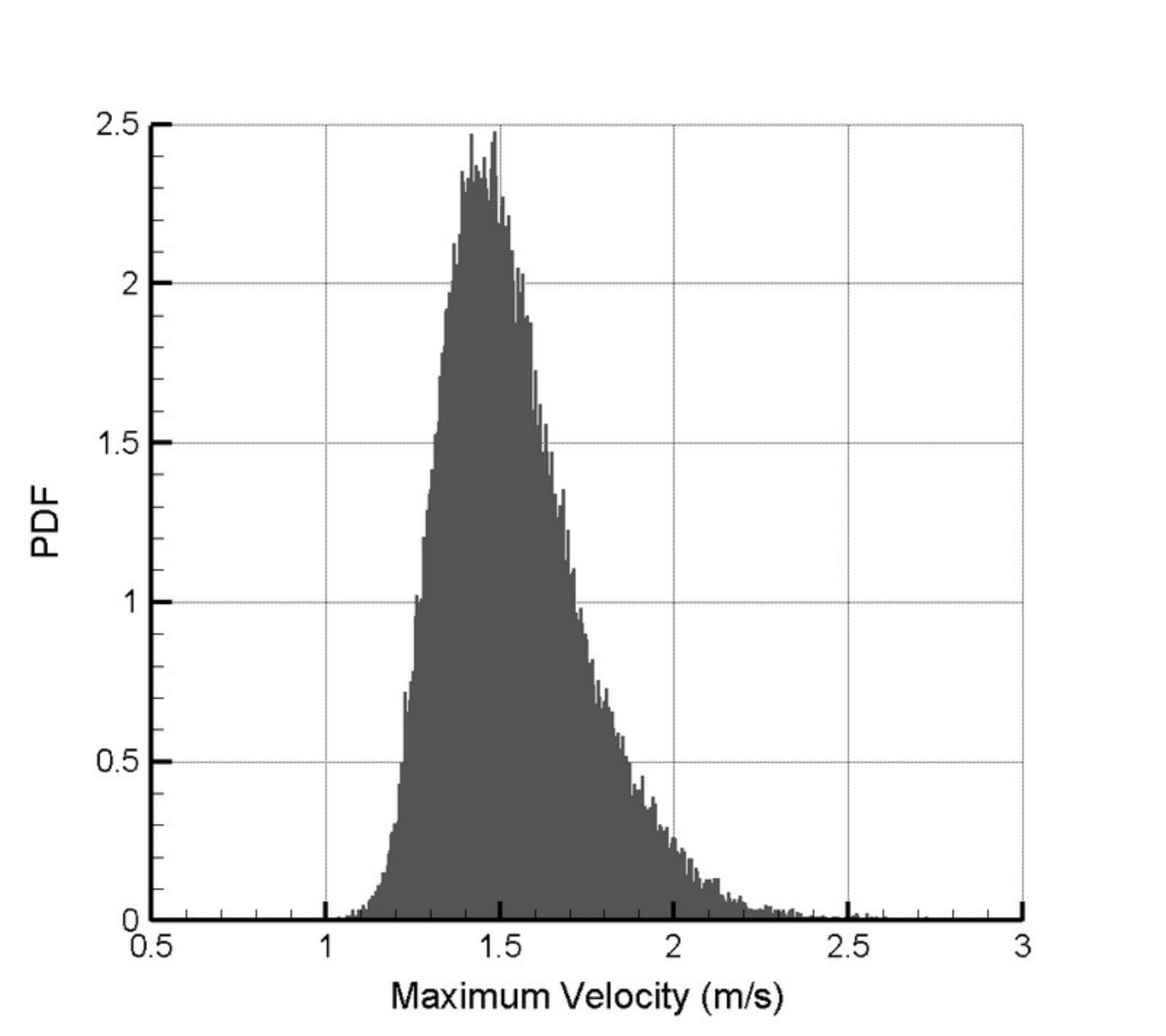}}
\subfigure[$\epsilon = 0.05$, $\sigma_0 = 0.0012$]{\includegraphics[width=0.33\textwidth]{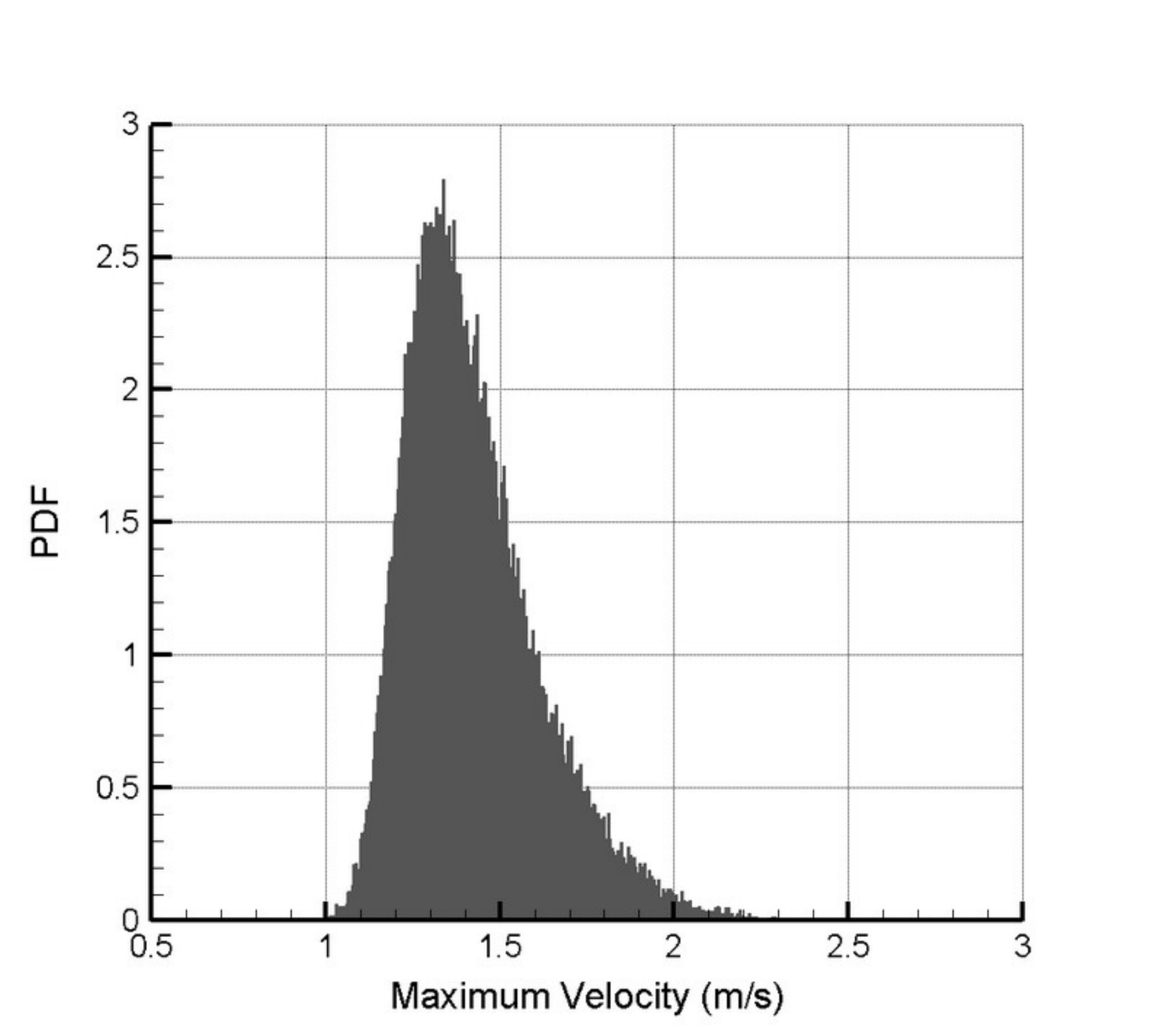}}
\subfigure[$\epsilon = 0.05$, $\sigma_0 = 0.0016$]{\includegraphics[width=0.33\textwidth]{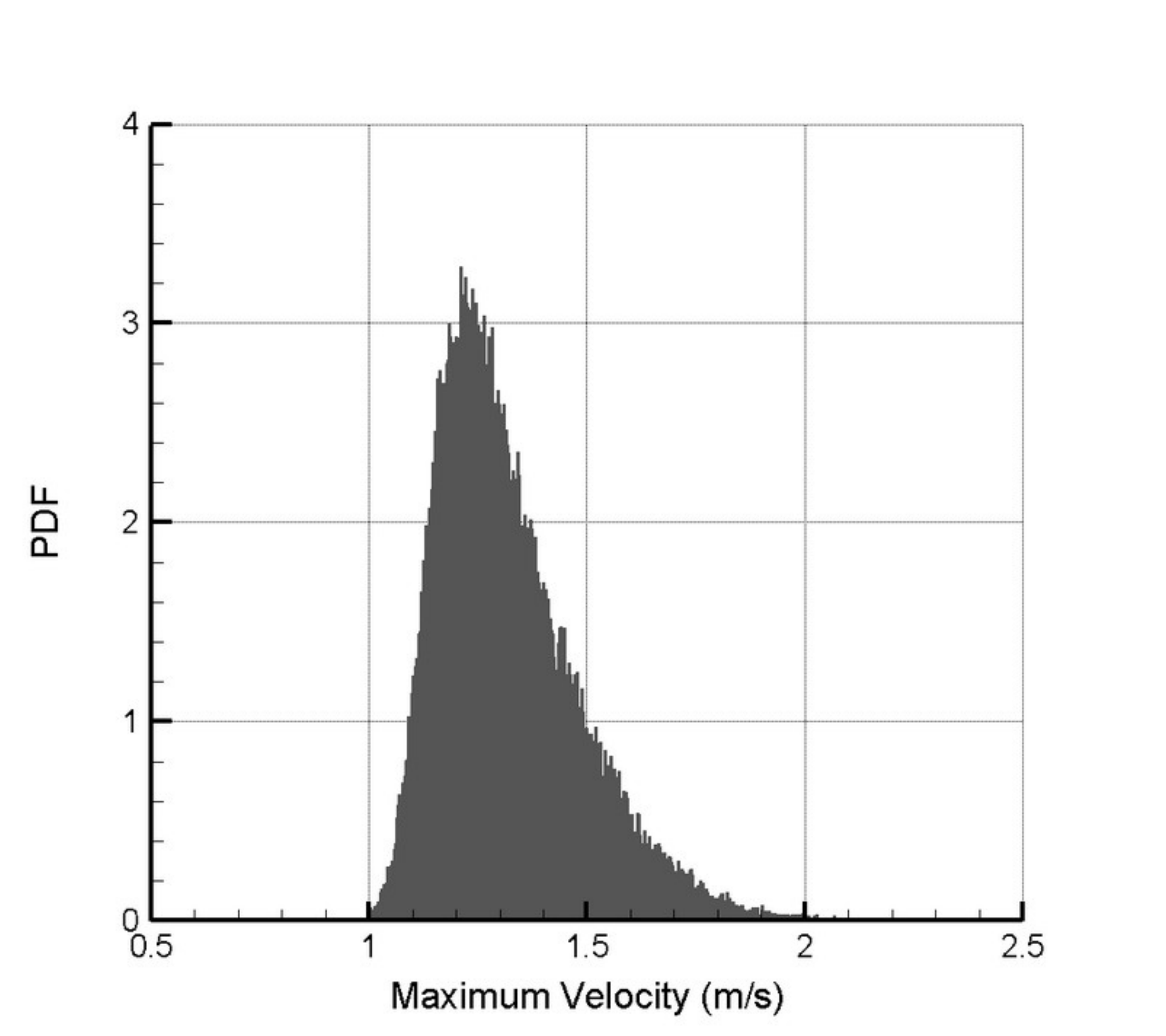}}
}
\caption{Probability density function (PDF) of the maximum flow velocity in the ocean Basin III.}
\label{f:pdf-v-3}
\end{figure}

\begin{figure}
\centering
\mbox{
\subfigure[varying $\epsilon$ ($\epsilon=\varepsilon$, $2\varepsilon$, $4\varepsilon$)]{\includegraphics[width=0.5\textwidth]{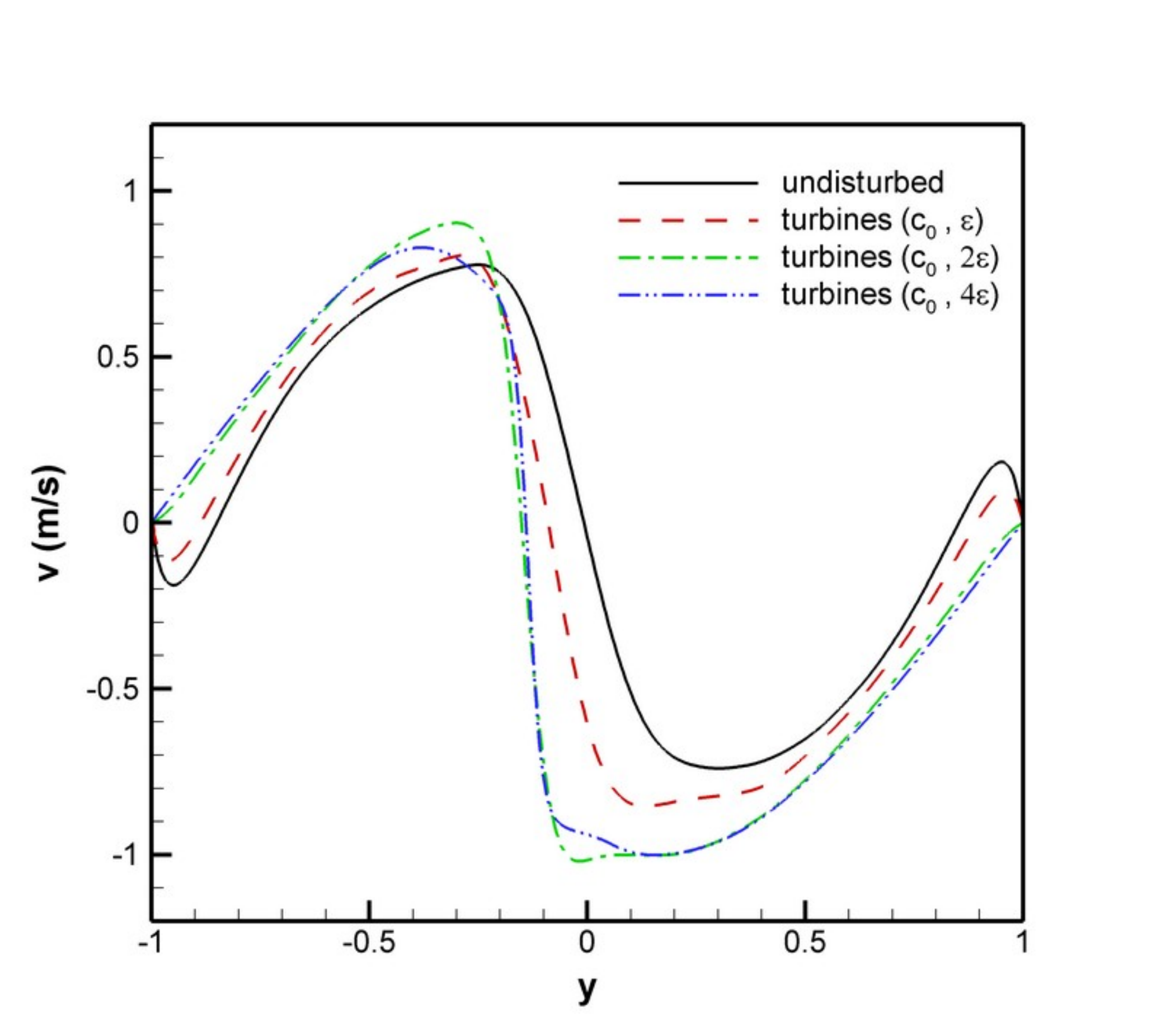}}
\subfigure[varying $C_0$ ($C_0=0.8c_0$, $1.2c_0$, $1.6c_0$)]{\includegraphics[width=0.5\textwidth]{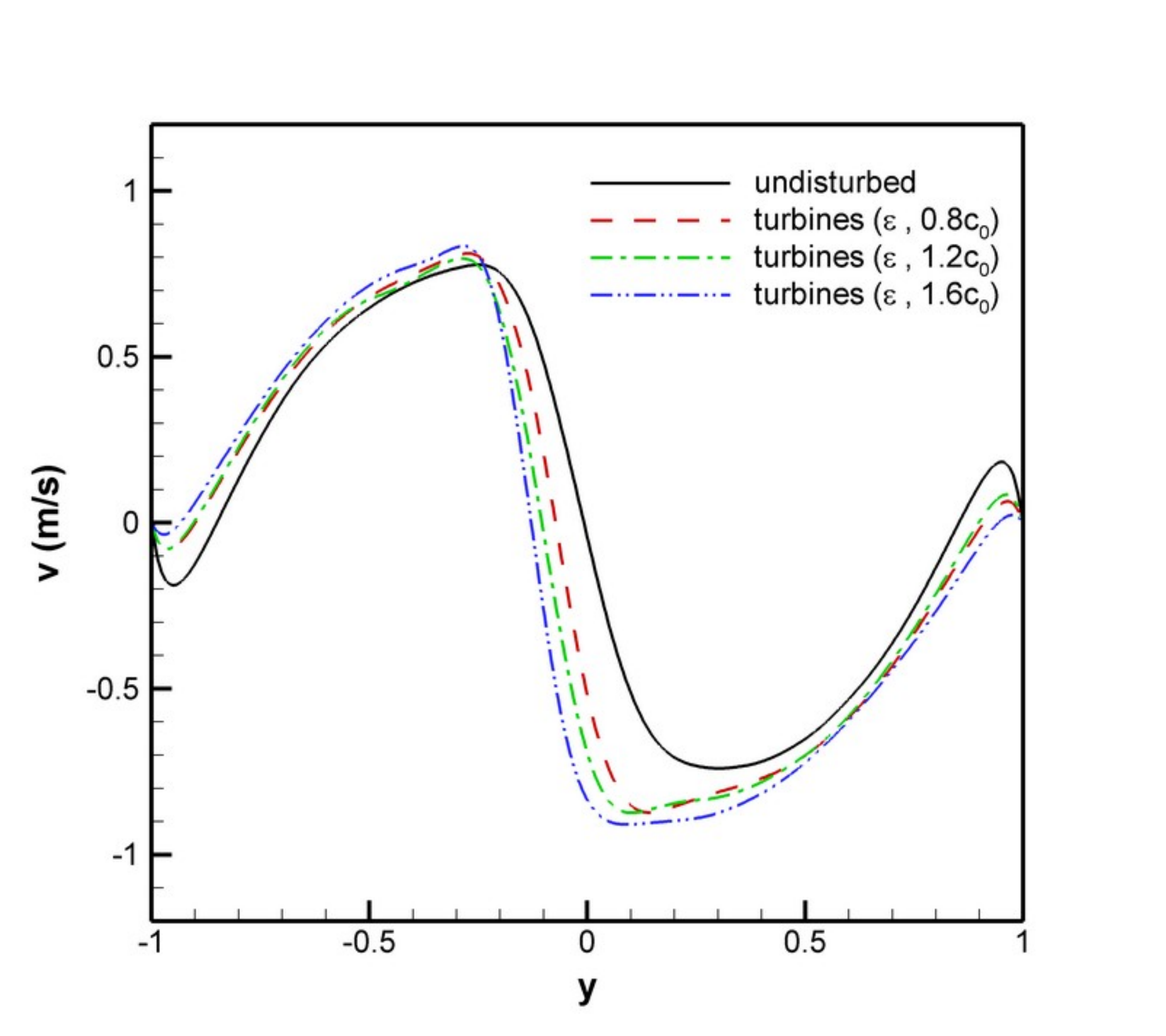}}
}
\caption{Comparison of mean meridional flow velocities along the western boundary layer ($x=0$) for the ocean Basin I, where the added turbines are parameterized by $\varepsilon = 0.05$ and $c_0 =0.001$ m/s.}
\label{f:wbc-1}
\end{figure}

\begin{figure}
\centering
\mbox{
\subfigure[varying $\epsilon$ ($\epsilon=\varepsilon$, $2\varepsilon$, $4\varepsilon$)]{\includegraphics[width=0.5\textwidth]{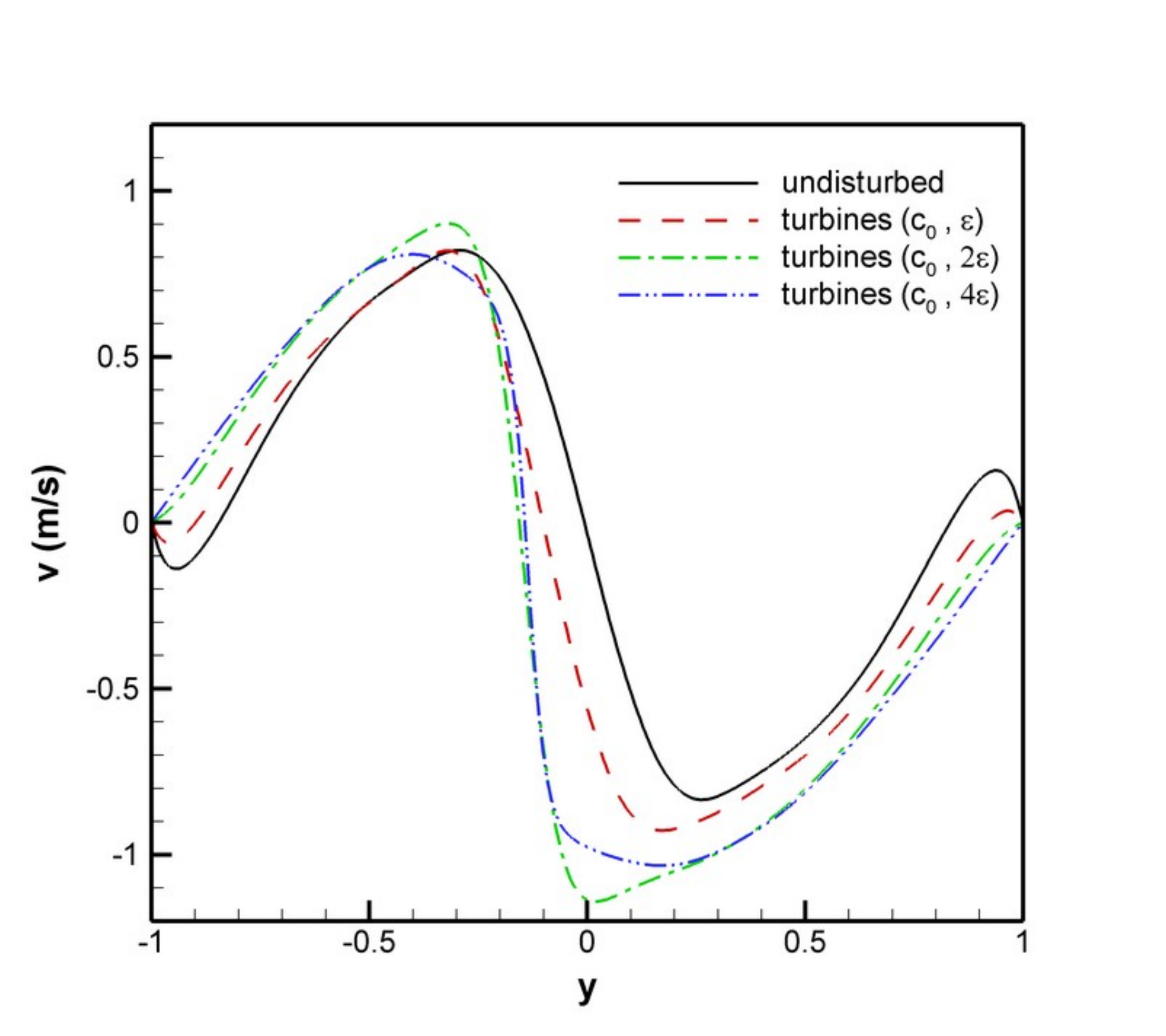}}
\subfigure[varying $C_0$ ($C_0=0.8c_0$, $1.2c_0$, $1.6c_0$)]{\includegraphics[width=0.5\textwidth]{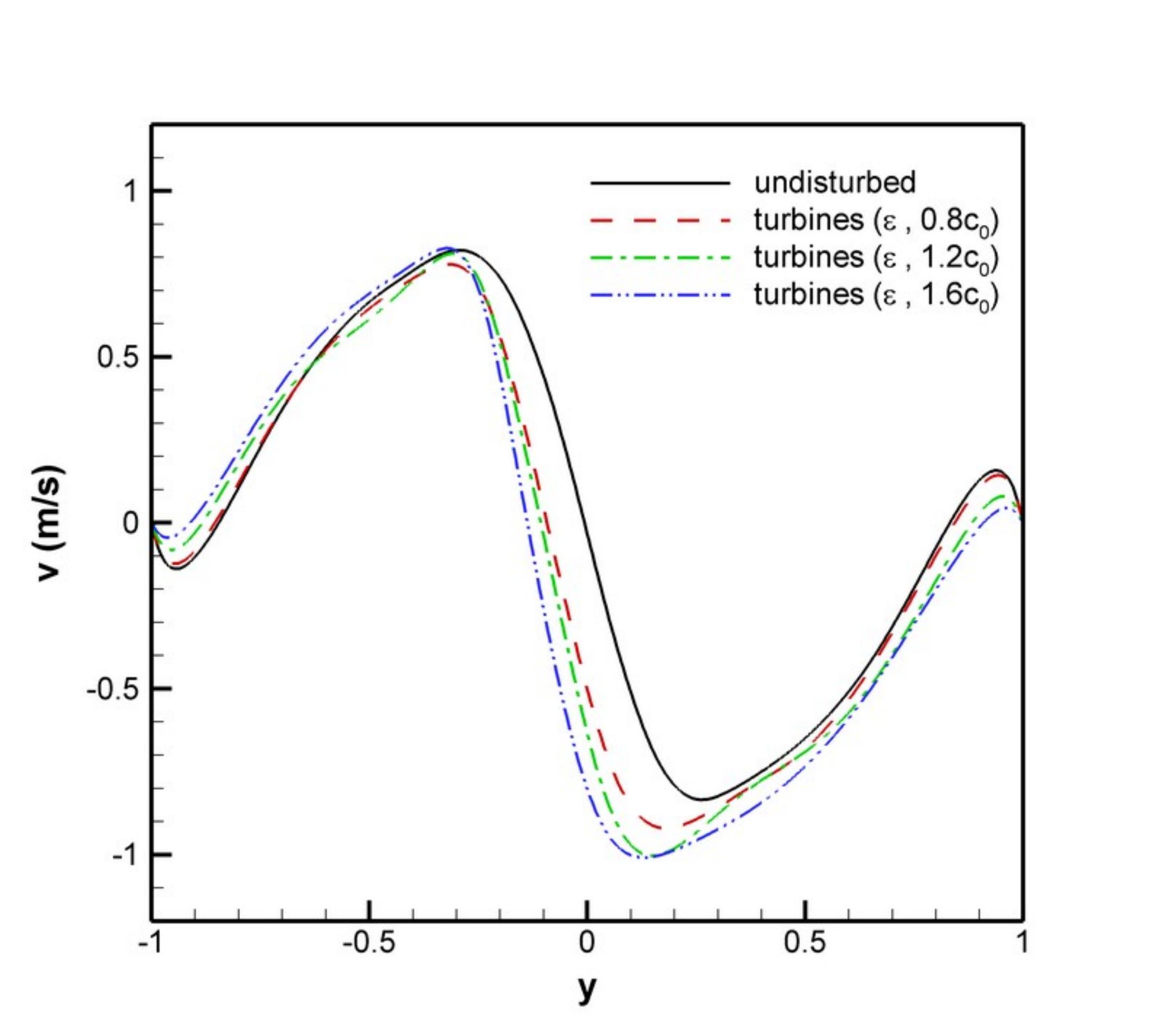}}
}
\caption{Comparison of mean meridional flow velocities along the western boundary layer ($x=0$) for the ocean Basin II, where the added turbines are parameterized by $\varepsilon = 0.0625$ and $c_0 =0.001$ m/s.}
\label{f:wbc-2}
\end{figure}

\begin{figure}
\centering
\mbox{
\subfigure[varying $\epsilon$ ($\epsilon=\varepsilon$, $2\varepsilon$, $4\varepsilon$)]{\includegraphics[width=0.5\textwidth]{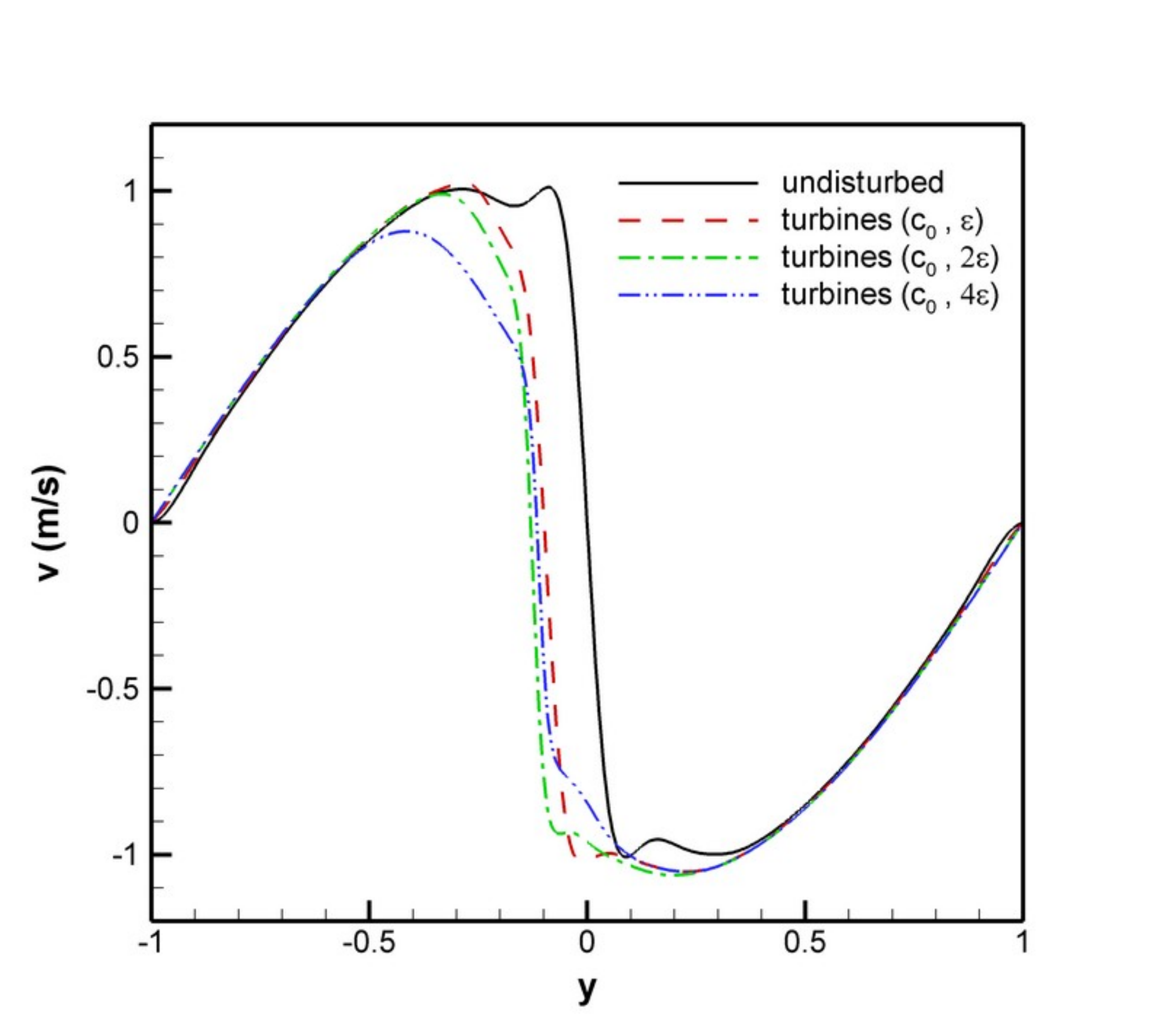}}
\subfigure[varying $C_0$ ($C_0=0.8c_0$, $1.2c_0$, $1.6c_0$)]{\includegraphics[width=0.5\textwidth]{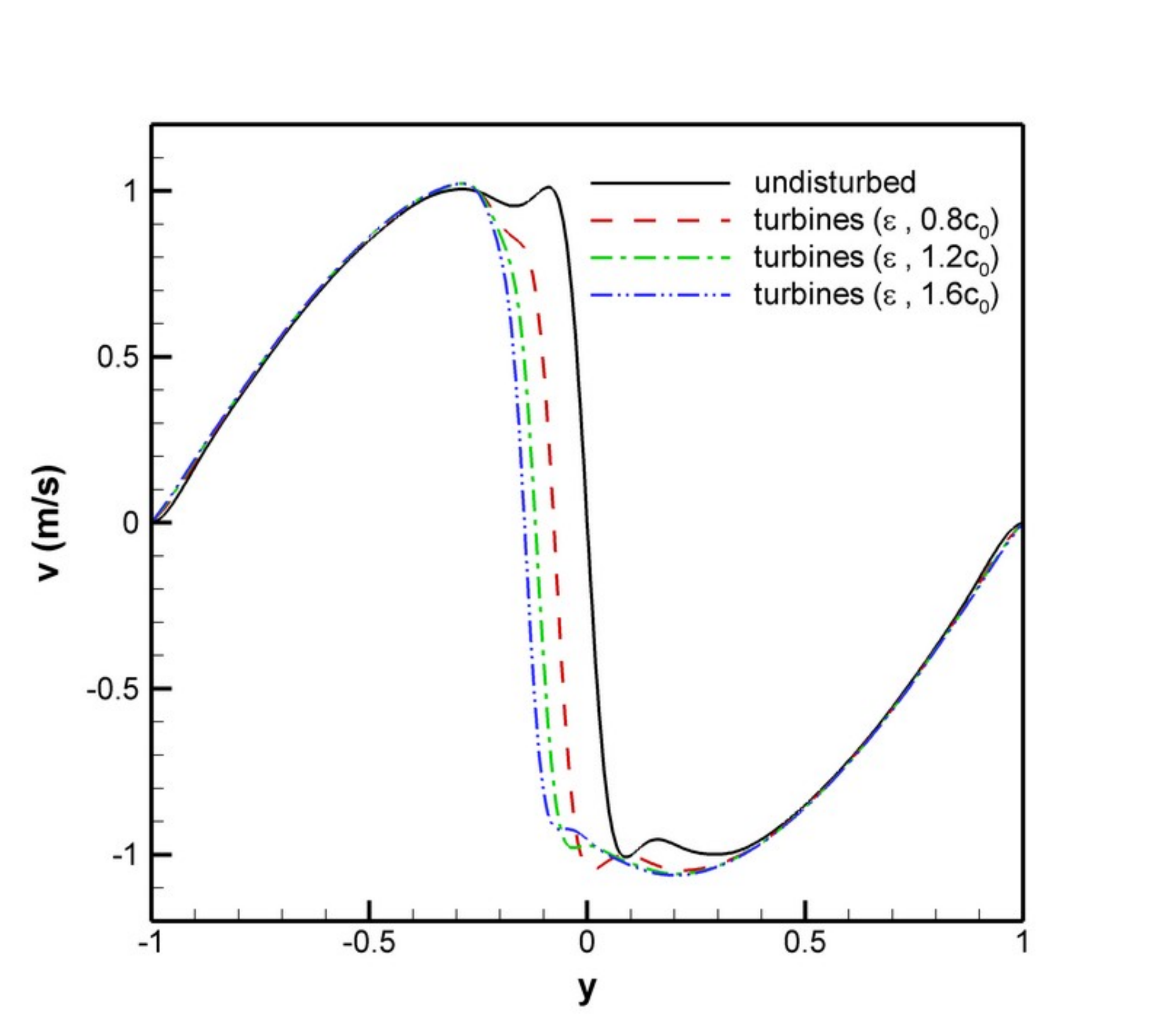}}
}
\caption{Comparison of mean meridional flow velocities along the western boundary layer ($x=0$) for the ocean Basin III, where the added turbines are parameterized by $\varepsilon = 0.05$ and $c_0 =0.001$ m/s.}
\label{f:wbc-3}
\end{figure}

Our next step is to analyze available energy in the QG model due to the added turbines. The time history of the total power dissipation, $P(t)$, from the turbine drag force is evaluated by
\begin{equation}\label{eq:power}
    \frac{P(t)}{C_0 \rho V^2 L^2} = \frac{1}{2}\int\int \exp(-\frac{x^2 + (y+0.2)^2}{\epsilon^{2}}) \Big(\Big(\frac{\partial \psi}{\partial x}\Big)^2 + \Big(\frac{\partial \psi}{\partial y}\Big)^2 \Big) dxdy,
\end{equation}
in each case. Then, we compute the probability density function (PDF) distribution of the available power from the data collected between $t=20$ and $t=40$. Fig.~\ref{f:pdf-p-1} demonstrates normalized PDF distributions for a set of numerical experiments in Basin I varying $\epsilon$ and $C_0$ parameters. An increase in $\epsilon$ results in an increase in the available power due to a dominance of relatively large area with turbines. It can be seen that maximum probabilities are 4 GW, 14 GW and 17 GW for $\epsilon$ values of $0.05$, $0.1$, $0.2$, respectively. Mean values are documented in Table~\ref{tab:b1} showing approximately 10 GW power can be extracted from Basin I for $\epsilon=0.05$ and $C_0 = 0.001$ m/s. This corresponds an 8.8 \% percent reduction for mean value of the maximum basin velocity. Similar analyses are shown in Fig.~\ref{f:pdf-p-2} for Basin II and in Fig.~\ref{f:pdf-p-3} for Basin III. Using the turbine parameter $\epsilon > 0.1$, the model predicts a maximum zonal velocity smaller than the limiting value observed in the Gulf Stream system. Summary of the mean results are also documented in Table~\ref{tab:b2} and Table~\ref{tab:b3} for the Basin II and Basin III, respectively. The PDF distributions of the maximum basin velocity are plotted in Figs.~\ref{f:pdf-v-1}-\ref{f:pdf-v-3} for the three ocean basins confirming that maximum speed is typically between 1.5 m/s and 2.5 m/s with a mean value close to 2 m/s. This agrees well with the observations of Gulf Stream currents \cite{richardson1985average,fratantoni2001north,dijkstra2005nonlinear}. Maximum speed drastically reduces to a lower value for larger $\epsilon$ due to excessive dissipation in the system. Figs.~\ref{f:wbc-1}-\ref{f:wbc-3} compare mean values of the meridional velocity along the western boundary layer ($x=0$) in response to the localized turbine arrays centered at $y=-0.2$ for the same set of modeling parameters. Although the maximum mean velocity decreases by adding turbines, it can be seen that the flow speed increases along the western boundary due to the implantation of the turbines. This corresponds an increase of the mean flow speed on coastal sites. It is also clear that change of the flow speed exceeds 10 \% for the northern sites of the turbine arrays. Therefore computational assessments indicate that the implantation of turbine arrays covering large areas (e.g., $L_t$ is greater than 100 km) may change ocean dynamics drastically. It should be noted that, using an idealized QG model, this study simulates the effects of additional localized energy dissipation due to turbines on the dynamics and circulation patterns. Although an explicit relationship between turbine modeling parameters and the positioning of turbines is not established, this study provides a systematic intercomparison with respect to the turbine parameters.

\begin{table}[!t]
\centering
\caption{Mean data for available power potential and resulting maximum velocity in Basin I. }
\label{tab:b1}       
\begin{tabular}{lcllcc}
\noalign{\smallskip}\hline\noalign{\smallskip}
$\epsilon$ & $L_t$ (km)  & $\sigma_0$ & $C_0$ (m/s) & Mean Max Velocity (m/s) & Mean Power (GW) \\
\noalign{\smallskip}\hline\noalign{\smallskip}
\multicolumn{3}{l}{undisturbed} & 0.0 & 2.15 & N/A \\
  0.05   & 100 & 106.2 & 0.001        & 1.96 & 9.98 \\
  0.1    & 200 & 106.2 & 0.001        & 1.38 & 14.99 \\
  0.2    & 400 & 106.2 & 0.001        & 1.02 & 17.98 \\
  0.05   & 100 & 85 & 0.0008          & 1.98 & 8.13 \\
  0.05   & 100 & 127.5 & 0.0012       & 1.93 & 11.38 \\
  0.05   & 100 & 170 & 0.0016         & 1.83 & 13.32 \\
\noalign{\smallskip}\hline
\end{tabular}
\end{table}

\begin{table}[!t]
\centering
\caption{Mean data for available power potential and resulting maximum velocity in Basin II. }
\label{tab:b2}       
\begin{tabular}{lcllcc}
\noalign{\smallskip}\hline\noalign{\smallskip}
$\epsilon$ & $L_t$ (km)  & $\sigma_0$ & $C_0$ (m/s) & Mean Max Velocity (m/s) & Mean Power (GW) \\
\noalign{\smallskip}\hline\noalign{\smallskip}
\multicolumn{3}{l}{undisturbed} & 0.0 & 2.19 & N/A \\
  0.0625   & 100 & 68 & 0.001         & 2.04 & 12.42 \\
  0.125    & 200 & 68 & 0.001         & 1.44 & 19.02 \\
  0.25     & 400 & 68 & 0.001         & 1.04 & 19.63 \\
  0.0625   & 100 & 54.4    & 0.0008   & 2.10 & 10.60 \\
  0.0625   & 100 & 81.6 & 0.0012      & 2.03 & 15.02 \\
  0.0625   & 100 & 108.8   & 0.0016   & 1.94 & 16.90 \\
\noalign{\smallskip}\hline
\end{tabular}
\end{table}

\begin{table}[!t]
\centering
\caption{Mean data for available power potential and resulting maximum velocity in Basin III. }
\label{tab:b3}       
\begin{tabular}{lcllcc}
\noalign{\smallskip}\hline\noalign{\smallskip}
$\epsilon$ & $L_t$ (km)  & $\sigma_0$ & $C_0$ (m/s) & Mean Max Velocity (m/s) & Mean Power (GW) \\
\noalign{\smallskip}\hline\noalign{\smallskip}
\multicolumn{3}{l}{undisturbed} & 0.0 & 1.84 & N/A \\
  0.05   & 150 & 239 & 0.001          & 1.48 & 9.51 \\
  0.1    & 300 & 239 & 0.001          & 1.14 & 16.62 \\
  0.2    & 600 & 239 & 0.001          & 1.07 & 22.26 \\
  0.05   & 150 & 191.2 & 0.0008       & 1.54 & 8.37 \\
  0.05   & 150 & 286.8 & 0.0012       & 1.42 & 10.62 \\
  0.05   & 150 & 382.4 & 0.0016       & 1.32 & 12.67 \\
\noalign{\smallskip}\hline
\end{tabular}
\end{table}

\section{Summary and Conclusions}
\label{}

A quasi-geostrophic (QG) ocean circulation model is developed to evaluate the potential of ocean energy harnessed by using ocean current turbines distributed in the Gulf Stream system. The ocean turbines are included in the model as a localized Gaussian type Rayleigh friction term parameterized by $\epsilon$ controling the effective turbine area and $C_0$, which is the effective drag coefficient. Numerical assessments for various mid-latitude ocean basins are systematically performed by using a huge set of turbine parameters. Although there is no attempt to characterize the relationship between the Gaussian drag coefficient profile and turbine arrays, numerical assessments presented in this study clearly demonstrate trends and theoretical upper bounds considering the long term inter-decade variability of ocean dynamics in quasi-stationary turbulent flow regime. Potential impacts on mean circulation patterns and mean flow speeds on coastal regions are investigated. Results are compared against the undisturbed QG simulations for the three representative Cartesian ocean basins utilizing the double-gyre wind forcing. This forcing yields a four-gyre circulation pattern in statistically steady state and represents an ideal test problem to assess whether the added turbines alter the four-gyre circulation pattern. Performing Munk layer resolving high-resolution computations, it is shown that the four-gyre pattern is recovered for small $\epsilon$ values (e.g., $\epsilon=0.05$) providing approximately 10 GW mean available power for a turbine region covering an area of approximately $100^2$ km$^2$. This corresponds an approximately 4 GW peak power value in its probability density function distribution and 10 \% reduction of the maximum flow velocity over the entire basin. It is interesting to note that higher power estimations can be obtained by tuning the turbine modeling parameters (i.e., mean power estimation between 8 GW and 22 GW). It is observed higher available power estimations for larger $\epsilon$ values resulting in drastic changes in mean circulation patterns due to excessive dissipation arising from the turbines. It is clear that the model predicts underestimated maximum zonal velocities when the turbine area parameter $\epsilon > 0.1$. Therefore, the present analysis concludes that a theoretical estimate of 10 GW mean power corresponding to an average mean value of 88 TWh/yr can be extracted from the Gulf Stream currents with slightly altering the mean flow structures and zonal velocities.




\section*{References}

  \bibliographystyle{elsarticle-num}
  \bibliography{ref}






\end{document}